%% file: paper.tex
\documentclass[12pt]{article}
\usepackage{cite,epsfig,amssymb,amsmath,feynarts}

\newenvironment{comment}[1]{}{}

\newcommand{\Eq}[1]{Eq.\,(\ref{#1})}
\newcommand{\Eqs}[1]{Eqs.\,(\ref{#1})}
\newcommand{\Eqsand}[2]{Eqs.\,(\ref{#1}) and (\ref{#2})}

\newcommand{\eq}[1]{(\ref{#1})}

\newcommand{\Fig}[1]{Figure \ref{#1}}

\newcommand{\App}[1]{Appendix \ref{#1}}
\newcommand{\Appsand}[2]{Appendices \ref{#1} and \ref{#2}}
\newcommand{\Sec}[1]{Section \ref{#1}}

\newcommand{\beq}{\begin{equation}}                
\newcommand{\eeq}{\end{equation}}        
\newcommand{\bea}{\begin{eqnarray}}               
\newcommand{\eea}{\end{eqnarray}}        
\newcommand{\bdm}{\begin{displaymath}}                 
\newcommand{\edm}{\end{displaymath}}                      
  
\newcommand{\non}{\nonumber}

\newcommand{\smallW}{{\scriptscriptstyle W}}

\newcommand{\ord}{{O}}

\newcommand{\mb}{m_b}        
\newcommand{\mt}{m_t}

\newcommand{\MW}{M_\smallW}

\newcommand{\mub}{\mu_b}
\newcommand{\mul}{\mu}  
\newcommand{\muh}{\mu_0}

\newcommand{\MSbar}{ \overline{\rm MS} }

\newcommand{\gs }{ g }
\newcommand{\as }{ \alpha_s }         
   
\newcommand{\f}{\frac}

\newcommand{\Dsl}{D \hspace{-.70em} / \hspace{.25em}}

\newcommand{\Gsl}{G \hspace{-.70em} / \hspace{.25em}}

\newcommand{\Heff}{{\cal H}_{\rm eff}}

\newcommand{\Aeff}{{\cal A}_{\rm eff}}
\newcommand{\Afull}{{\cal A}_{\rm full}}

\newcommand{\SUC}{SU (3)_C}

\newcommand{\GF}{G_F}

\newcommand{\dimension}{n}
\newcommand{\ep}{\epsilon}

\newcommand{\btosgluon}{b \to s g}

\newcommand{\BtoXsgamma}{\bar B \to X_s \gamma}

\newcommand{\BtoXslpluslminus}{\bar B \to X_s \ell^+ \ell^-}
\newcommand{\btosccbar}{b \to s q \bar{q}}
\newcommand{\btos}{b \to s}
\newcommand{\btosgluongluon}{b \to s g g}

\newcommand{\btosgluongluongluon}{b \to s g g g}
\newcommand{\btosghostantighost}{b \to s u^a \bar{u}^a}

\newcommand{\eps}{\epsilon}

\newcommand{\zetathree}{\zeta_3}

\newcommand{\QQ}{\Qmisiak \hspace{-0.0mm} \Qmisiak}
\newcommand{\EE}{\Emisiak \hspace{-0.0mm} \Emisiak}
\newcommand{\QE}{\Qmisiak \hspace{-0.2mm} \Emisiak}
\newcommand{\EQ}{\Emisiak \hspace{-0.0mm} \Qmisiak}

\newcommand{\betazero}{\beta_0}
\newcommand{\betaone}{\beta_1}
\newcommand{\betatwo}{\beta_2}

\newcommand{\nf}{f}

\newcommand{\Qmisiak}{Q}
\newcommand{\Qburas}{Q'}
\newcommand{\Emisiak}{E}
\newcommand{\Eburas}{E'}

\newcommand{\ad}{anomalous dimension}
\newcommand{\ads}{anomalous dimensions}

\newcommand{\etal}{{\it et al}.}

\newcommand{\gammadown}[1]{\gamma_{#1}}
\newcommand{\gammaup}[1]{\gamma^{#1}}

\newcommand{\cl}{{\rm Cl}_2}
\newcommand{\li}{{\rm Li}_2}

\newcommand{\deltaBRST}{s}

\newcommand{\CP}{C \hspace{-0.5mm} P}
\newcommand{\epsovereps}{\epsilon^\prime/\epsilon}

\newcommand{\PL}{P_L}

\begin{document}

\allowdisplaybreaks


\thispagestyle{empty}
\rightline{IPPP/04/66}
\rightline{DCPT/04/132}
\rightline{FERMILAB-PUB-04-281-T}
\rightline{hep-ph/0411071}
\vspace*{1.2truecm}
\bigskip

\centerline{\LARGE \bf Effective Hamiltonian for Non-Leptonic} 
\vspace{0.2cm}%
\centerline{\LARGE \bf \boldmath $| \Delta F | = 1$ Decays at NNLO in
QCD}     

\vskip1truecm
\centerline{\large \bf Martin Gorbahn$^{a}$ and Ulrich Haisch$^{b}$} 
\bigskip
\begin{center}{
{\em $^a$ IPPP, Physics Department, University of Durham, \\ DH1 3LE,
Durham, UK} \\
\vspace{.3cm}
{\em $^b$ Theoretical Physics Department, Fermilab, \\ Batavia, IL
60510, USA}    
}\end{center}
\vspace{1.cm}

\centerline{\bf Abstract}

\vspace{0.5cm}
We compute the effective hamiltonian for non-leptonic $| \Delta F | = 
1$ decays in the standard model including next-to-next-to-leading
order QCD corrections. In particular, we present the complete
three-loop \ad \ matrix describing the mixing of current-current and
QCD penguin operators. The calculation is performed in an operator
basis which allows to consistently use fully anticommuting $\gamma_5$
in dimensional regularization at an arbitrary number of loops. The
renormalization scheme dependences and their cancellation in physical
quantities is discussed in detail. Furthermore, we demonstrate how our
results are transformed to a different basis of effective operators
which is frequently adopted in phenomenological applications. We give
all necessary two-loop constant terms which allow to obtain the
three-loop \ads \ and the corresponding initial conditions of the
two-loop Wilson coefficients in the latter scheme. Finally, we solve
the renormalization group equation and give the analytic expressions
for the low-energy Wilson coefficients relevant for non-leptonic $B$
meson decays beyond next-to-leading order in both renormalization
schemes.    

\vspace*{1.8cm}

\newpage


\section{Introduction}
\label{sec:introduction}

Perturbative QCD effects have an important impact on the structure of
the effective hamiltonian for non-leptonic $| \Delta F | = 1$
processes with $F = S, C$ or $B$, which describes the weak decay of  
the corresponding mesons and hadrons. Most notably, they can lead to a
sizable enhancement of the $\Delta I = 1/2$ transitions, of the
$\CP$-violating ratio $\epsovereps$, and of the QCD penguin 
contributions to rare and radiative $B$ decays within the Standard
Model (SM) \cite{Buras:1998ra} and some of its innumerous extensions
\cite{Buchalla:1990fu, MFVSSM, Buras:2003mk, Jang:2000rk, GSSM,
others}.      

In all cases, these short-distance QCD effects can be systematically
calculated using an effective field theory framework, which allows to
resum large QCD logarithms of the form $L \equiv \ln \mu/\MW$ by
solving the Renormalization Group Equation (RGE) that governs the 
scale dependence of the Wilson coefficient functions of the relevant
$| \Delta F | = 1$ local operators built out of the light and massless
SM fields. After the pioneering Leading Order (LO) calculation of the
$\ord (\as^n L^n)$ contributions \cite{LO}, the resummation of the
$\ord (\as^n L^{n - 1})$ logarithms has been completed more than ten
years ago and subsequently confirmed by several groups. The main
components of the perturbative Next-to-Leading Order (NLO) calculation
are $i )$ the one-loop $\ord (\as)$ corrections to the relevant Wilson
coefficient functions \cite{Buras:1991jm, Ciuchini:1993vr,
Chetyrkin:1997gb} and $ii )$ the two-loop $\ord (\as^2)$ Anomalous
Dimension Matrix (ADM) describing the mixing of the associated
physical operators \cite{Altarelli:1980fi, Buras:1989xd, Buras:1992tc,
Ciuchini:1993vr, Chetyrkin:1997gb, Gambino:2003zm}.    

To improve on the present NLO calculation, one needs to include
one more order in the strong coupling expansion, aiming at a
resummation of all the $\ord (\as^n L^{n - 2})$ logarithmic enhanced  
corrections. The completion of this Next-to-Next-to-Leading Order
(NNLO) computation constitutes the core of this work. Since the
two-loop $\ord (\as^2)$ matching corrections to the relevant Wilson 
coefficients are already known from \cite{Bobeth:1999mk} the only
missing ingredient to perform this task is the knowledge of the
three-loop $\ord (\as^3)$ ADM describing the mixing of the
current-current and QCD penguin operators. In this paper we will close
this gap by employing standard techniques \cite{Gambino:2003zm,
Misiak:1994zw, Chetyrkin:1997fm} to carry out a direct calculation of 
the required ADM adopting the renormalization scheme introduced in
\cite{Chetyrkin:1997gb}. Since in this scheme the QCD penguin
operators are defined in such a way that traces with $\gamma_5$ do not
occur to all orders in perturbation theory, we are allowed to
consistently use dimensional regularization with a naive anticommuting
$\gamma_5$. This feature is very welcome, as it makes the actual
three-loop calculation completely automatic and rather
straightforward.               

The NNLO ADM we have computed can be used in analyses of new physics
models as well, provided they do not introduce new operators with
respect to the SM. This applies, for example, to the case of the two
Higgs doublet models \cite{Buchalla:1990fu}, to some supersymmetric
scenarios with minimal flavor violation \cite{MFVSSM}, and to specific
models of universal extra dimensions \cite{Buras:2003mk}. On the other
hand, in left-right-symmetric models \cite{Jang:2000rk} and in the
general supersymmetric SM \cite{GSSM}, additional operators with
different chirality structures arise \cite{Buras:2000if}. In many
cases one can exploit the chiral invariance of QCD and use the same
ADM, but in general an extended basis is required.    

Another strong motivation to write the article at hand was that the 
three-loop ADM computed here is part and parcel of the complete   
NNLO analysis of rare semi-leptonic $\BtoXslpluslminus$ decays
presented recently by us in collaboration with Christoph Bobeth and 
Paolo Gambino \cite{Bobeth:2003at}. Furthermore, it constitutes a    
integral part of the NNLO calculation of radiative $\BtoXsgamma$
decays, admittedly a very ambitious enterprise, which nevertheless has 
already aroused the interest of some theorists
\cite{NNLOBtoXsgamma}. Whereas our previous work dealt exclusively 
with the phenomenological application of our result, we will now ---
in the spirit of \cite{Buras:1991jm, Ciuchini:1993vr,
Chetyrkin:1997gb} --- focus on the more formal aspects of the
renormalization of effective field theories, such as the issue of
scheme dependences in general, their cancellation in physical
observables, and the general transformation properties of the ADM and
the Wilson coefficients under a change of scheme. In this respect we
will extend the existing NLO results \cite{Buras:1991jm,
Ciuchini:1993vr} to the next order, paying special attention to the 
conceptual features related to the renormalization of the strong
coupling constant.        
 
While much of the discussion in this paper is therefore rather
technical, our general results have important practical
applications. This will be showcased by means of a couple of
examples. In particular, we will devote a sizable part of the present 
article to derive the explicit NNLO relation between our and a 
different renormalization scheme that is commonly used in the
literature on weak decays \cite{Buras:1998ra}. In the latter scheme,
which we shall call ``traditional'' scheme from now on, the applied
form of the effective hamiltonian introduces unwanted traces
containing $\gamma_5$ by definition. These traces turn out to be
harmless at the LO, but involve a lot of technical difficulties
related to the use of Fierz symmetry arguments in $n = 4 - 2 \eps$
dimensions at the NLO \cite{Ciuchini:1993vr, Buras:1992tc}. Applying
the same scheme in a direct calculation of the ADM at the NNLO or even
beyond would thus be extremely tedious. We will not attempt such a
direct computation here, but avail the derived NNLO relation between
our and the ``traditional'' scheme to find the NNLO ADM and the
corresponding matching conditions in the latter scheme on detours. As
another exercise we solve the RGE and give the analytic expressions
for the $\Delta B = -\Delta S = 1$ low-energy Wilson coefficients
beyond NLO in both renormalization schemes. We are aware of the fact
that some of the formulas presented below are rather 
long. Nevertheless we believe that at least some of them should be 
useful to the reader interested mainly in the application of the
presented formalism to weak decays rather than in the conceptual 
subtleties, which obviously address more technical minded colleagues.

The main part of this paper is organized as follows: in 
\Sec{sec:generalstructure} we present the general structure of the  
effective hamiltonian for non-leptonic $| \Delta F | =  1$ decays at
the NNLO level. \Sec{sec:effectivehamiltonian} is devoted to the
simplest application of the general formalism, namely the $\Delta B =
-\Delta S = 1$ decays. We recall the relevant effective hamiltonian
and list all the dimension-five and six operators that will be needed
in the calculation of the three-loop ADM. In
\Sec{sec:initialconditions} we collect the results for the initial 
conditions of the relevant Wilson coefficients through NNLO 
obtained in \cite{Chetyrkin:1997gb, Bobeth:1999mk}. After a brief
description of the actual three-loop calculation the final result
for the ADM is given in \Sec{sec:anomalousdimensionmatrix}. In 
\Sec{sec:renormalizationgroupequation} we solve the RGE to find the 
explicit NNLO expressions for the low-energy Wilson coefficients 
relevant for non-leptonic $B$ meson decays. In
\Sec{sec:renormalizationschemedependences} we elaborate on the question
of scheme dependence related to the renormalization of the effective
operators as well as the strong coupling
constant. \Sec{sec:changeofoperatorbasis} starts out with a general 
discussion of the non-trivial nature of a change of the basis of
physical operators in the framework of dimensional regularization,
followed by a demonstration of how the NNLO results for the ADM and
the matching conditions are transformed to the ``traditional'' basis  
of effective operators. Finally, in \Sec{sec:summary} we summarize the
main results of this work.   

Some technical details as well as additional material has been
relegated to the appendices: in \App{app:derivationofs1ands2} we
derive the explicit form of the matrix kernels that are needed to find
the evolution matrices through NNLO, while 
\Appsand{app:traceevanescentoperators}{app:changetothetraditionaloperatorbasis}
contain all ingredients that are necessary to transform our results to
the ``traditional'' set of operators. The NNLO analytic formulas for
the low-energy Wilson coefficients relevant for non-leptonic $B$ meson
decays in the latter scheme will be given in
\App{app:wilsoncoefficientsinthetraditionaloperatorbasis}, which
concludes our paper.         

\section{General Structure} 
\label{sec:generalstructure}

The effective hamiltonian for non-leptonic $| \Delta F | = 1$ decays
has the following generic structure \cite{Buras:1998ra}  
\beq \label{eq:heffgeneric}
\Heff = -\f{4 \GF}{\sqrt{2}} V_{\rm CKM} \hspace{0.5mm} \vec{Q}^T
 \vec{C} (\mul) \, .  
\eeq
Here $\GF$ denotes the Fermi constant and $\vec{Q}^T$ is a row vector 
containing the relevant local operators $Q_i$. Explicit expressions
will be given in \Sec{sec:effectivehamiltonian}. $\vec{C} (\mul)$ is a
column vector containing the Wilson coefficients $C_i (\mul)$ that
together with the Cabibbo-Kobayashi-Maskawa (CKM) factor \cite{CKM}
$V_{\rm CKM}$ describe the strength with which a given operator enters
the hamiltonian, and $\mul$ is the renormalization scale. The decay
amplitude for a decay of a meson $M$ into a final state $F$ is simply
given by $\langle F | \Heff | M \rangle$.     

The Wilson coefficient functions evolve from the initial scale $\muh$
down to $\mul$, which in practical applications is much lower than
$\muh$, according to their RGE. Using dimensional regularization with
$n = 4 - 2 \eps$ and considering only mass independent renormalization
schemes it is given by 
\beq \label{eq:renormalizationgroupequation} 
\mu \f{d}{d \mul} \vec{C} (\mul) = \hat{\gamma}^T (\gs) \vec{C} (\mul)
\, ,          
\eeq
where $\hat{\gamma} (\gs)$ is the ADM corresponding to
$\vec{Q}$. 
Neglecting all  electromagnetic effects, the general solution of this
equation reads   
\beq \label{eq:rgesolution}
\vec{C} (\mu) = \hat{U} (\mul, \muh) \vec{C} (\muh) \, ,  
\eeq
with 
\begin{gather} 
\label{eq:evolutionmatrix}
\hat{U} (\mul, \muh) = T_\gs \exp \int^{\gs (\mul)}_{\gs (\muh)} d
g^\prime \f{\hat{\gamma}^T (g^{\prime})}{\beta (g^\prime)} \, ,
\\[2mm] 
\label{eq:gammamatrixandbetafunctionexpansion} 
\hat{\gamma} (\gs) = \sum^\infty_{i = 0} \left ( \f{\gs^2}{16 \pi^2} 
\right )^{i + 1} \hat{\gamma}^{(i)} \, , \hspace{0.5cm} \text{and}
\hspace{0.5cm} \beta (\gs) = -\gs \sum^\infty_{i = 0} \left (
\f{\gs^2}{16 \pi^2} \right )^{i + 1} \beta_i \, .   
\end{gather} 
Here $\vec{C} (\mu_0)$ are the initial conditions of the evolution and
$T_{\gs}$ denotes ordering of the coupling constants $\gs (\mu)$ in
such a way that their value increases from right to left. $\beta (\gs)$
is the QCD $\beta$ function.        

Keeping the first three terms in the expansions of $\hat{\gamma}
(\gs)$ and $\beta (\gs)$ as given in
\Eq{eq:gammamatrixandbetafunctionexpansion}, we find for the evolution
matrix $\hat{U} (\mul, \muh)$ in the NNLO approximation 
\beq \label{eq:asevolutionmatrixexpansion}
\hat{U} (\mul, \muh) = \hat{K} (\mul) \hspace{0.25mm} \hat{U}^{(0)}
(\mul, \muh) \hat{K}^{-1} (\muh) \, ,    
\eeq
where
\beq \label{eq:kmatrices}
\begin{split}
\hat{K} (\mul) & = \hat{1} + \f{\as (\mul)}{4 \pi} \hat{J}^{(1)} + 
\left ( \f{\as (\mul)}{4 \pi} \right )^2 \hat{J}^{(2)} \, , \\   
\hat{K}^{-1} (\muh) & = \hat{1} - \f{\as (\muh)}{4 \pi} \hat{J}^{(1)}
- \left ( \f{\as (\muh)}{4 \pi} \right )^2 \left ( \hat{J}^{(2)} -
\big ( \hat{J}^{(1)} \big )^2 \right ) \, ,      
\end{split}
\eeq
and 
\beq \label{eq:loevolutionmatrix}
\hat{U}^{(0)} (\mul, \muh) = \hat{V} \, {\rm diag} \left ( \f{\as
(\muh)}{\as (\mul)} \right )^{a_i} \hat{V}^{-1} \, ,
\eeq  
denotes the LO evolution matrix, which depends on the matrix $\hat{V}$ 
and the so-called magic numbers $a_i$ that are obtained via
diagonalizing $\hat{\gamma}^{(0) \, T}$:   
\beq \label{eq:magicnumbers}
\left ( \hat{V}^{-1} \hat{\gamma}^{(0) \, T} \hat{V} \right )_{i j} =
2 \betazero a_i \delta_{i j} \, .   
\eeq
In order to give the explicit expressions for the matrices
$\hat{J}^{(1)}$ and $\hat{J}^{(2)}$ we define   
\beq \label{eq:jandgmatrices}
\hat{J}^{(i)} = \hat{V} \hat{S}^{(i)} \hat{V}^{-1} \, , \hspace{0.5cm}
\text{and} \hspace{0.5cm} \hat{G}^{(i)} = \hat{V}^{-1}
\hat{\gamma}^{(i) \, T} \hat{V} \, ,  
\eeq
for $i = 1, 2$. The entries of the matrix kernels $\hat{S}^{(1)}$ and
$\hat{S}^{(2)}$ are given by   
\beq \label{eq:smatrices}
\begin{split}
S^{(1)}_{i j} & = \f{\betaone}{\betazero} a_i \delta_{i j} -
\f{G^{(1)}_{i j}}{2 \betazero \left ( 1 + a_i - a_j \right )} \, ,   
\\        
S^{(2)}_{i j} & = \f{\betatwo}{2 \betazero} a_i \delta_{i j} + \sum_k 
\f{1 + a_i - a_k}{2 + a_i - a_j} \left ( S^{(1)}_{i k} S^{(1)}_{k j} -
\f{\betaone}{\betazero} S^{(1)}_{i j} \delta_{j k} \right ) -
\f{G^{(2)}_{i j}}{2 \betazero \left ( 2 + a_i - a_j \right )} \, , 
\end{split} 
\eeq
where the first line recalls the classical NLO result
\cite{Buras:1991jm}, and the second one represents the corresponding
NNLO expression, in agreement with \cite{NNLOmatrixkernel}. The
explicit derivation of $\hat{S}^{(1)}$ and $\hat{S}^{(2)}$ is
presented in \App{app:derivationofs1ands2}.        

Let us now recall how the initial conditions of the Wilson
coefficients are obtained. The amplitude for a given non-leptonic 
quark decay is calculated perturbatively in the full theory including
all possible diagrams such as $W$-boson exchange, QCD penguin and box 
diagrams as well as gluon corrections to all these building
blocks. The result up to the NNLO is given schematically by 
\beq \label{eq:afull}
\Afull = \langle \vec{Q} \rangle^{(0) \, T} \left ( \vec{A}^{(0)} +
\f{\as (\muh)}{4 \pi} \vec{A}^{(1)} + \left ( \f{\as (\muh)}{4 \pi}
\right )^2 \vec{A}^{(2)} \right ) \, , 
\eeq  
where $\langle \vec{Q} \rangle^{(0)}$ denotes the tree-level matrix
elements of $\vec{Q}$. 

A second step involves the calculation of the decay amplitude in the
QCD effective theory. It generally requires the computation of the
operator insertions into current-current and QCD penguin diagrams of 
the effective theory together with gluon corrections to these
insertions. Including QCD corrections up to the NNLO one finds
\beq \label{eq:aeff}
\Aeff = \langle \vec{Q} \rangle^{(0) \, T} \left ( \hat{1} + \f{\as
(\muh)}{4 \pi} \hat{r}^{(1) \, T} + \left ( \f{\as (\muh)}{4 \pi}
\right )^2 \hat{r}^{(2) \, T} \right ) \vec{C} (\muh) \, ,  
\eeq  
where the matrices $\hat{r}^{(1)}$ and $\hat{r}^{(2)}$ codify the
one- and two-loop matrix elements of $\vec{Q}$, respectively. 

The matching procedure between full and effective theory establishes
the initial conditions $\vec{C} (\muh)$ for the Wilson
coefficients. Equating $\Afull$ and $\Aeff$ in
\Eqsand{eq:afull}{eq:aeff} at a scale $\muh$ translates into the
following identity  
\cite{Buras:1999st} 
\beq \label{eq:initialcondition}
\begin{split}
\vec{C} (\muh) & = \vec{A}^{(0)} + \f{\as (\muh)}{4 \pi} \left ( 
\vec{A}^{(1)} - \hat{r}^{(1) \, T} \vec{A}^{(0)} \right ) \\
& + \left ( \f{\as (\muh)}{4 \pi} \right )^2 \left ( \vec{A}^{(2)} -
\hat{r}^{(1) \, T} \left [ \vec{A}^{(1)} - \hat{r}^{(1) \, T}
\vec{A}^{(0)} \right ] - \hat{r}^{(2) \, T} \vec{A}^{(0)} \right ) \,
. 
\end{split}
\eeq
Combining \Eqs{eq:renormalizationgroupequation},
\eq{eq:asevolutionmatrixexpansion}, \eq{eq:kmatrices} and
\eq{eq:initialcondition}, we finally obtain   
\beq \label{eq:wilsoncoefficient}
\begin{split}
\vec{C} (\mul) & = \hat{K} (\mul) \hspace{0.25mm} \hat{U}^{(0)} (\mul,
\muh) \Bigg ( \vec{A}^{(0)} + \f{\as (\muh)}{4 \pi} \left [
\vec{A}^{(1)} - \hat{R}^{(1)} \vec{A}^{(0)} \right ] \\   
& + \left ( \f{\as (\muh)}{4 \pi} \right )^2 \left [ \vec{A}^{(2)} -
\hat{R}^{(1)} \vec{A}^{(1)} - \left ( \hat{R}^{(2)} - \big (
\hat{R}^{(1)} \big )^2 \right ) \vec{A}^{(0)} \right  ] \Bigg ) \, ,
\end{split}
\eeq
where 
\beq \label{eq:rmatrices}
\hat{R}^{(1)} = \hat{r}^{(1) \, T} + \hat{J}^{(1)} \, , \hspace{0.5cm}
\text{and} \hspace{0.5cm} \hat{R}^{(2)} = \hat{r}^{(2) \, T} +
\hat{J}^{(2)} + \hat{r}^{(1) \, T} \hat{J}^{(1)} \, , 
\eeq
are certain combinations of $\hat{r}^{(1) \, T}$, $\hat{r}^{(2) \,
T}$, $\hat{J}^{(1)}$ and $\hat{J}^{(2)}$ which will play a special
role in \Sec{sec:renormalizationschemedependences} where we will
discuss the issue of renormalization scheme dependences in detail. 

\section{Effective Hamiltonian for \boldmath $\Delta B = -\Delta S =
1$ Decays} 
\label{sec:effectivehamiltonian}

The simplest application of the general formalism outlined in the
previous section is the case of non-leptonic $B$ meson decays
governed by the $\btos$ transition. For definiteness we will therefore
give explicit formulas for the $\Delta B = -\Delta S = 1$ decays
only. However, it is straightforward to transform them to the other $|
\Delta F | = 1$ cases. Neglecting contributions proportional to the
small CKM factor $V^\ast_{us} V_{ub}$ which are irrelevant here, the
corresponding effective off-shell hamiltonian is given by     
\beq \label{eq:heff}
\Heff = -\f{4 \GF}{\sqrt{2}} V^\ast_{ts} V_{tb} \left ( \vec{Q}^T
\vec{C} (\mul) + \vec{N}^T \vec{C}_N (\mul) + \vec{B}^T \vec{C}_B
(\mul) + \vec{E}^T \vec{C}_E (\mul) \right ) \, . 
\eeq
In addition to the gauge-invariant operators $\vec{Q}$, non-physical
operators arise as counterterms in the renormalization of higher loop
One-Particle-Irreducible (1PI) off-shell Green's functions 
with insertions of the operators $\vec{Q}$. These non-physical
operators can in general be divided into three different classes
\cite{Buras:1989xd, collins, Simma:1993ky, evanescent}: $i )$
operators $\vec{N}$ that vanish by use of the QCD Equations Of Motion
(EOM), $ii )$ non-physical counterterms $\vec{B}$ that can be 
written as a Becchi-Rouet-Stora-Tyutin (BRST) variation \cite{BRST} of
some other operators --- so-called BRST-exact operators --- and $iii
)$ evanescent operators $\vec{E}$ that vanish algebraically in $n = 4$
dimensions.   

The set of physical operators $\vec{Q}$ consists of six dimension-six
operators, which can be chosen as \cite{Chetyrkin:1997gb,
Chetyrkin:1996vx}     
\beq \label{eq:dimensionsixphysicaloperatorsmisiak} 
\begin{split}
\Qmisiak_1 & = (\bar{s}_L \gammadown{\mu_1} T^a c_L) (\bar{c}_L
\gammaup{\mu_1} T^a b_L) \, , \\ 
\Qmisiak_2 & = (\bar{s}_L \gammadown{\mu_1} c_L) (\bar{c}_L
\gammaup{\mu_1} b_L) \, , \\     
\Qmisiak_3 & = (\bar{s}_L \gammadown{\mu_1} b_L)
\sum\nolimits_q (\bar{q} 
\gammaup{\mu_1} q) \, , \\  
\Qmisiak_4 & = (\bar{s}_L \gammadown{\mu_1} T^a b_L)
\sum\nolimits_q (\bar{q} \gammaup{\mu_1} T^a q) \, ,
\\  
\Qmisiak_5 & = (\bar{s}_L \gammadown{\mu_1 \mu_2 \mu_3} b_L)
\sum\nolimits_q (\bar{q} \gammaup{\mu_1 \mu_2 \mu_3}
q) \, , \\      
\Qmisiak_6 & = (\bar{s}_L \gammadown{\mu_1 \mu_2 \mu_3} T^a b_L) 
\sum\nolimits_q (\bar{q} \gammaup{\mu_1 \mu_2 \mu_3}
T^a q) \, ,  
\end{split}
\eeq
and one dimension-five operator  
\beq \label{eq:dimensionfivephysicaloperatorsmisiak} 
\Qmisiak_8 = \f{1}{\gs} \mb ( \bar{s}_L \sigma^{\mu_1 \mu_2} T^a b_R )
G^a_{\mu_1 \mu_2} \, . 
\eeq
Here we have used the definitions $\gammadown{\mu_1 \cdots 
\mu_m} \equiv \gamma_{\mu_1} \cdots \gamma_{\mu_m}$, $\gammaup{\mu_1 
\cdots \mu_m} \equiv \gamma^{\mu_1} \cdots \gamma^{\mu_m}$ and
$\sigma^{\mu_1 \mu_2} \equiv i \, [ \gamma^{\mu_1}, \gamma^{\mu_2}
]/2$, and the sum over $q$ extends over all light quark flavors. $\gs$
is the strong coupling constant, $q_L$ and $q_R$ are the chiral quark 
fields, $G_{\mu_1 \mu_2}^a$ is the gluonic field strength tensor, and
$T^a$ are the generators of $\SUC$, normalized so that $\mbox{Tr} (T^a
T^b) \equiv \delta^{ab}/2$. 

{%
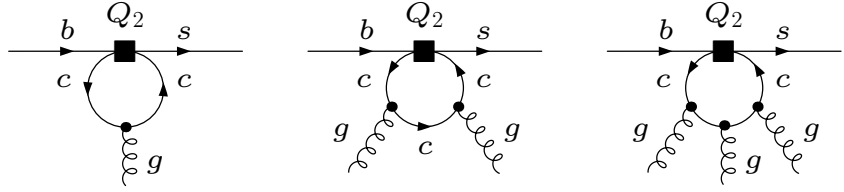
\begin{figure}[t]
\begin{center}
\scalebox{1.8}{\begin{picture}(55,45)(0,-5)\input{oneloopeom1.tex}\end{picture}}%
\hspace{5mm}%
\scalebox{1.8}{\begin{picture}(55,45)(0,-5)\input{oneloopeom2.tex}\end{picture}}%
\hspace{5mm}%
\scalebox{1.8}{\begin{picture}(55,45)(0,-5)\input{oneloopeom3.tex}\end{picture}}%
\hspace{5mm}%
\caption{The one-loop 1PI diagrams which mix $\Qmisiak_2$ into $N^{(1)}_1$.}     
\label{fig:oneloopeom}
\end{center}
\end{figure}
}%

The physical operators given in
\Eqsand{eq:dimensionsixphysicaloperatorsmisiak}{eq:dimensionfivephysicaloperatorsmisiak}
include the current-current operators $\Qmisiak_1$ and $\Qmisiak_2$,
the QCD penguin operators $\Qmisiak_3$--$\Qmisiak_6$ and the
chromomagnetic moment operator $\Qmisiak_8$. Notice that we have
defined $\Qmisiak_1$--$\Qmisiak_{6}$ in such a way that problems
connected with the treatment of $\gamma_5$ in $\dimension = 4 - 2 \ep$
dimensions do not arise \cite{Chetyrkin:1997gb}. Consequently, we are
allowed to consistently use a fully anticommuting $\gamma_5$ in
dimensional regularization throughout the calculation.

As far as the EOM-vanishing operators are concerned, the specific
structure of only one of them \cite{Chetyrkin:1997gb}
\beq \label{eq:oneloopeomvanishingoperatormisiak}  
N^{(1)}_1 = \f{1}{\gs} \bar{s}_L \gamma^{\mu_1} T^a b_L D^{\mu_2}
G_{\mu_1 \mu_2}^a + \Qmisiak_4 \, ,  
\eeq
is relevant in finding the one- and two-loop mixing of the four-quark
operators $\Qmisiak_1$--$\Qmisiak_6$. The corresponding divergent
one-loop 1PI diagrams are shown in \Fig{fig:oneloopeom}.  

In order to remove the ultraviolet (UV) divergences related to the
two-loop subdiagrams with insertions of $\Qmisiak_1$--$\Qmisiak_6$
depicted in \Fig{fig:twoloopeom}, another ten EOM-vanishing operators
need to be considered \cite{Gambino:2003zm}     
\begin{align} \label{eq:twoloopeomvanishingoperatormisiak}  
N^{(2)}_1 & = \f{1}{\gs^2} \mb \bar{s}_L \Dsl \Dsl b_R \, , \non \\
N^{(2)}_2 & = \f{i}{\gs^2} \bar{s}_L \Dsl \Dsl \Dsl b_L \, , \non \\
N^{(2)}_3 & = \f{i}{\gs} \left [ \bar{s}_L \stackrel{\leftarrow}{\Dsl}
  \sigma^{\mu_1 \mu_2} T^a b_L G_{\mu_1 \mu_2}^a - G_{\mu_1 \mu_2}^a
\bar{s}_L T^a \sigma^{\mu_1 \mu_2} \Dsl b_L \right ] + Q_8 \, , \non
\\ 
N^{(2)}_4 & = \f{i}{\gs} \mb \bar{s}_L \left [
\stackrel{\leftarrow}{\Dsl} \Gsl - \Gsl \Dsl \right ] b_R \, , \non \\
N^{(2)}_5 & = i \left [ \bar{s}_L \left ( \stackrel{\leftarrow}{\Dsl} 
  \Gsl \Gsl - \Gsl \Gsl \Dsl \right ) b_L - i \mb \bar{s}_L \Gsl \Gsl
  b_R \right ] \, , \non \\
N^{(2)}_6 & = \f{1}{\gs} \left [ \bar{s}_L \left (
  \stackrel{\leftarrow}{\Dsl} \stackrel{\leftarrow}{\Dsl} \Gsl + \Gsl 
  \Dsl \Dsl \right ) b_L + i \mb \bar{s}_L \Gsl \Dsl 
  b_R \right ] \, , \non \\
N^{(2)}_7 & = i \left [ \bar{s}_L \left ( \stackrel{\leftarrow}{\Dsl}
  G^a_{\mu_1} G^{a \mu_1} - G^a_{\mu_1} G^{a \mu_1} \Dsl \right ) b_L
- i \mb \bar{s}_L G^a_{\mu_1} G^{a \mu_1} b_R \right ] \, , \non \\
N^{(2)}_8 & = \f{1}{\gs} \left [ \bar{s}_L \left (
  \stackrel{\leftarrow}{\Dsl} \stackrel{\leftarrow
  \hspace{1.5mm}}{D_{\mu_1}} G^{\mu_1} + G_{\mu_1} D^{\mu_1} \Dsl
\right ) b_L + i \mb \bar{s}_L G_{\mu_1} D^{\mu_1} b_R \right ] \, ,
\non \\  
N^{(2)}_9 & = \f{1}{\gs} \left [ \bar{s}_L \stackrel{\leftarrow}{\Dsl}
\Gsl \Dsl b_L + i \mb \bar{s}_L \stackrel{\leftarrow}{\Dsl} \Gsl b_R
\right ] \, , \non \\ 
N^{(2)}_{10} & = d^{abc} \left [ \bar{s}_L \left (
\stackrel{\leftarrow}{\Dsl} T^a - T^a \Dsl \right ) b_L - i \mb 
  \bar{s}_L T^a b_R \right ] G^b_{\mu_1} G^{c \mu_1} \, ,
\end{align}
where $D_\mu \equiv \partial_\mu + i \gs G_\mu$ and $\stackrel{\!
\leftarrow}{D_\mu} \hspace{1.5mm} \equiv \hspace{1.5mm} \stackrel{\!
\leftarrow}{\partial_\mu} \hspace{-0.5mm} - \hspace{1.0mm} i \gs
G_\mu$ denotes the covariant derivative of the gauge group $\SUC$
acting on the fields to the right and left, respectively. $G^a_\mu$
denotes the gluon field, and we have used the definitions $G_\mu
\equiv G^a_\mu T^a$ and $d^{abc} \equiv 2 \mbox{Tr} ( \{ T^a, T^b \}
T^c )$.     

{%
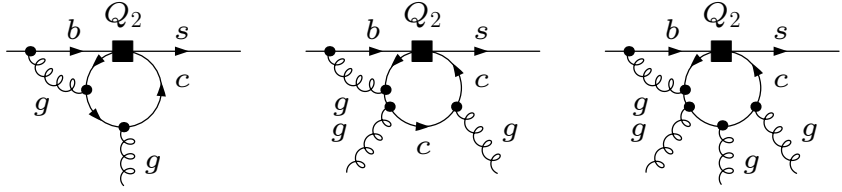
\begin{figure}[t]
\begin{center}
\scalebox{1.8}{\begin{picture}(55,45)(0,-5)\input{twoloopeom1.tex}\end{picture}}%
\hspace{5mm}%
\scalebox{1.8}{\begin{picture}(55,45)(0,-5)\input{twoloopeom2.tex}\end{picture}}%
\hspace{5mm}%
\scalebox{1.8}{\begin{picture}(55,45)(0,-5)\input{twoloopeom3.tex}\end{picture}}%
\hspace{5mm}%
\caption{Some of the two-loop 1PI diagrams which mix $\Qmisiak_2$ into
$N^{(1)}_1$ and $N^{(2)}_1$--$N^{(2)}_{10}$.}   
\label{fig:twoloopeom}
\end{center}
\end{figure}
}%

It is important to remark that the EOM-vanishing operators introduced
in
\Eqsand{eq:oneloopeomvanishingoperatormisiak}{eq:twoloopeomvanishingoperatormisiak}
arise as counterterms independently of what kind of infrared (IR)
regularization is adopted in the computation of the \ads \ of 
$\Qmisiak_1$--$\Qmisiak_6$. However, if the regularization respects 
the underlying symmetry, and all the diagrams are calculated on-shell,
non-physical operators have vanishing matrix elements \cite{collins,
Simma:1993ky, evanescent, Politzer:1980me}. In this case the
EOM-vanishing operators given in
\Eqsand{eq:oneloopeomvanishingoperatormisiak}{eq:twoloopeomvanishingoperatormisiak}
play no role in the calculation of the mixing of physical operators. 
If the gauge symmetry is broken this is no longer the case, as
diagrams with insertions of non-physical operators will generally have
non-vanishing projection on the physical operators. As we will discuss
in \Sec{sec:anomalousdimensionmatrix}, our IR regularization implies a
massive gluon propagator, and therefore non-physical counterterms play
a crucial role at intermediate stages of the \ads \ calculation.

{%
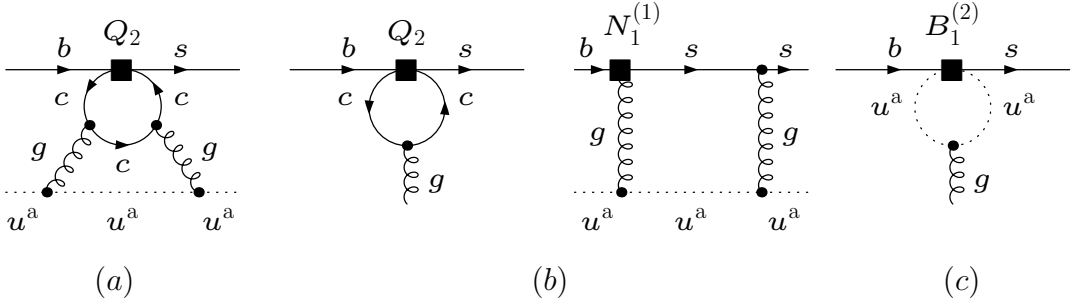
\begin{figure}[t]
\begin{center}
\vspace{-0.5cm}%
\scalebox{1.8}{\begin{picture}(55,45)(0,-5)\input{brstexact1.tex}\end{picture}}%
\hspace{3mm}%
\scalebox{1.8}{\begin{picture}(55,45)(0,-5)\input{brstexact2.tex}\end{picture}}%
\hspace{3mm}%
\scalebox{1.8}{\begin{picture}(55,45)(0,-5)\input{brstexact3.tex}\end{picture}}%
\vspace{3mm}%
\scalebox{1.8}{\begin{picture}(55,45)(0,-5)\input{brstexact4.tex}\end{picture}}%
\vspace{1mm}%

\hspace{0mm} $(a)$ \hspace{50mm} $(b)$ \hspace{47mm} $(c)$
\vspace{-0.5mm}%
\caption{$(a)$ A typical example of a divergent two-loop 1PI diagram
which potentially could introduce a mixing of $\Qmisiak_2$ into
$B_1^{(2)}$. $(b)$ A typical example of a counterterm contribution
needed to renormalize the corresponding two-loop 1PI diagrams. $(c)$
The contribution to the one-loop matrix element of $B_1^{(2)}$
containing an effective $\btosghostantighost$ vertex, which has a
non-vanishing on-shell projection on $\Qmisiak_4$ if a non-zero ghost
mass is used in the calculation. The QCD penguin and box contributions
to the matrix element that contain an effective $\btosgluon$ vertex
are not shown.}  
\label{fig:brstexact}
\end{center}
\end{figure}
}%

In contrast to the case of the two-loop mixing of the magnetic
operators considered in \cite{Gambino:2003zm, Misiak:1994zw}, 
it is a priori not clear if BRST-exact operators do arise as
counterterms of $\Qmisiak_1$--$\Qmisiak_6$. Since the BRST variation 
raises both ghost number and mass dimension by one unit, it is evident
that any BRST-exact operator that potentially could mix with
$\Qmisiak_1$--$\Qmisiak_6$ has to be a BRST variation of a
dimension-five operator containing a single anti-ghost field. The only
possibility for the latter operator having the correct chirality
structure is given in the $R_\xi$ gauge by 
\cite{Simma:1993ky}    
\bea \label{eq:brstexactoperatormisiak}  
B_1^{(2)} = \deltaBRST \left [ \f{1}{\gs} \left (\partial_{\mu_1}
\bar{u}^a \right ) \left ( \bar{s}_L \gamma^{\mu_1} T^a b_L \right )
\right ] \! = -\f{1}{\gs} \left [ \f{1}{\xi} \partial_{\mu_1}
\partial^{\mu_2} G^a_{\mu_2} - \gs f^{abc} \left (\partial_{\mu_1}
\bar{u}^b \right ) u^c \right ] \left ( \bar{s}_L \gamma^{\mu_1} T^a
b_L \right ) \hspace{-0.25mm},
\hspace{1.75mm}
\eea
where $\deltaBRST$ denotes the BRST operator, $u^a$ and $\bar{u}^a$
are the ghost and anti-ghost fields, $f^{abc}$ are the totally
antisymmetric structure constants of $\SUC$ and $\xi$ is the covariant
gauge-parameter. 

Although there is no obvious reason why $B_1^{(2)}$ should not appear
as a counterterm of $\Qmisiak_1$--$\Qmisiak_6$, it turns out that up
to three loops $B_1^{(2)}$ does not play a role in the mixing of
physical operators considered in the paper at hand. The key
observation thereby is that the overall contribution from the two-loop
1PI diagrams depicted in \Fig{fig:brstexact} $(a)$ is canceled by the
corresponding counterterm contribution as shown in \Fig{fig:brstexact}
$(b)$, so that the associated renormalization constant is exactly zero
at $\ord (\as^2)$. Therefore $B_1^{(2)}$ does not contribute to the
mixing of $\Qmisiak_1$--$\Qmisiak_6$ into $\Qmisiak_4$, although its
one-loop $\ord (\as)$ matrix element displayed in \Fig{fig:brstexact}
$(c)$ does not vanish if it is computed using non-vanishing gluon and
ghost masses to regulate IR divergences.       

In order to remove the divergences of all possible 1PI Green's
functions with single insertion of $\Qmisiak_1$--$\Qmisiak_6$ we have
to introduce some evanescent operators $\vec{E}$ as well. At the
one-loop level one encounters four evanescent operators, which can be
chosen to be \cite{Chetyrkin:1997gb, Chetyrkin:1996vx} 
\beq \label{eq:oneloopevanescentoperatorsmisiak} 
\begin{split}
\Emisiak^{(1)}_1 & = (\bar{s}_L \gammadown{\mu_1 \mu_2 \mu_3} T^a c_L)
(\bar{c}_L \gammaup{\mu_1 \mu_2 \mu_3} T^a b_L) - 16 \Qmisiak_1 \, ,
\\   
\Emisiak^{(1)}_2 & = (\bar{s}_L \gammadown{\mu_1 \mu_2 \mu_3} c_L)
(\bar{c}_L \gammaup{\mu_1 \mu_2 \mu_3} b_L) - 16 \Qmisiak_2 \, , \\  
\Emisiak^{(1)}_3 & = (\bar{s}_L \gammadown{\mu_1 \mu_2 \mu_3 \mu_4
\mu_5} b_L) \sum\nolimits_q (\bar{q} \gammaup{\mu_1
\mu_2 \mu_3 \mu_4 \mu_5} q) + 64 \Qmisiak_3 - 20 \Qmisiak_5 \, , \\  
\Emisiak^{(1)}_4 & = (\bar{s}_L \gammadown{\mu_1 \mu_2 \mu_3 \mu_4
\mu_5} T^a b_L) \sum\nolimits_q (\bar{q}
\gammaup{\mu_1 \mu_2 \mu_3 \mu_4 \mu_5} T^a q) + 64 \Qmisiak_4 - 20 
\Qmisiak_6 \, . 
\end{split}
\eeq
At the two-loop level four more evanescent operators do arise, that
can be defined as \cite{Chetyrkin:1997gb, Chetyrkin:1996vx} 
\begin{align} \label{eq:twoloopevanescentoperatorsmisiak} 
\Emisiak^{(2)}_1 & = (\bar{s}_L \gammadown{\mu_1 \mu_2 \mu_3 \mu_4
\mu_5} T^a c_L) (\bar{c}_L \gammaup{\mu_1 \mu_2 \mu_3 \mu_4 \mu_5} T^a
b_L) - 256 \Qmisiak_1 - 20 \Emisiak^{(1)}_1 \, , \non \\   
\Emisiak^{(2)}_2 & = (\bar{s}_L \gammadown{\mu_1 \mu_2 \mu_3 \mu_4
\mu_5} c_L) (\bar{c}_L \gammaup{\mu_1 \mu_2 \mu_3 \mu_4 \mu_5} b_L) -
256 \Qmisiak_2 - 20 \Emisiak^{(1)}_2 \, , \non \\    
\Emisiak^{(2)}_3 & = (\bar{s}_L \gammadown{\mu_1 \mu_2 \mu_3 \mu_4
\mu_5 \mu_6 \mu_7} b_L) \sum\nolimits_q (\bar{q}
\gammaup{\mu_1 \mu_2 \mu_3 \mu_4 \mu_5 \mu_6 \mu_7} q) + 1280 
\Qmisiak_3 - 336 \Qmisiak_5 \, , \non \\  
\Emisiak^{(2)}_4 & = (\bar{s}_L \gammadown{\mu_1 \mu_2 \mu_3 \mu_4
\mu_5 \mu_6 \mu_7} T^a b_L) \sum\nolimits_q (\bar{q} 
\gammaup{\mu_1 \mu_2 \mu_3 \mu_4 \mu_5 \mu_6 \mu_7}  T^a q) + 1280
\Qmisiak_4 - 336 \Qmisiak_6 \, . 
\end{align}
Finally, at the three-loop level another four evanescent operators are
needed. We define them in the following way: 
\bea \label{eq:threeloopevanescentoperatorsmisiak} 
\begin{split}
\Emisiak^{(3)}_1 & = (\bar{s}_L \gammadown{\mu_1 \mu_2 \mu_3 \mu_4 
\mu_5 \mu_6 \mu_7} T^a c_L) (\bar{c}_L \gammaup{\mu_1 \mu_2 \mu_3
\mu_4 \mu_5 \mu_6 \mu_7} T^a b_L) - 4096 \Qmisiak_1 - 336
\Emisiak^{(1)}_1 \, , \\ 
\Emisiak^{(3)}_2 & = (\bar{s}_L \gammadown{\mu_1 \mu_2 \mu_3 \mu_4
\mu_5 \mu_6 \mu_7} c_L) (\bar{c}_L \gammaup{\mu_1 \mu_2 \mu_3 \mu_4
\mu_5 \mu_6 \mu_7} b_L) - 4096 \Qmisiak_2 - 336 \Emisiak^{(1)}_2 \, ,
\\      
\Emisiak^{(3)}_3 & = (\bar{s}_L \gammadown{\mu_1 \mu_2 \mu_3 \mu_4 
\mu_5 \mu_6 \mu_7 \mu_8 \mu_9} b_L) \sum\nolimits_q 
(\bar{q} \gammaup{\mu_1 \mu_2 \mu_3 \mu_4 \mu_5 \mu_6 \mu_7 \mu_8
\mu_9} q) + 21504 \Qmisiak_3 - 5440 \Qmisiak_5 \, , \\    
\Emisiak^{(3)}_4 & = (\bar{s}_L \gammadown{\mu_1 \mu_2 \mu_3 \mu_4 
\mu_5 \mu_6 \mu_7 \mu_8 \mu_9} T^a b_L)
\sum\nolimits_q (\bar{q} \gammaup{\mu_1 \mu_2 \mu_3
\mu_4 \mu_5 \mu_6 \mu_7 \mu_8 \mu_9} T^a q) + 21504 \Qmisiak_4 - 5440
\Qmisiak_6 \, . 
\end{split}
\eea
Needless to say, the above choice of evanescent operators
$\Emisiak^{(3)}_1$--$\Emisiak^{(3)}_4$ is not unique, in the sense 
that their particular structure can be changed quite a lot without
affecting the three-loop ADM of the four-quark operators
$\Qmisiak_1$--$\Qmisiak_6$. For instance, adding any multiple of 
$\eps$ times any physical operator to them leaves the ADM unchanged up to
$\ord (\as^3)$. This is in contrast to what happens if such a
redefinition is applied to the one- and two-loop evanescent operators
given in
\Eqsand{eq:oneloopevanescentoperatorsmisiak}{eq:twoloopevanescentoperatorsmisiak}.

\section{Initial Conditions of the Wilson Coefficients} 
\label{sec:initialconditions}

Let us now turn to the initial conditions $\vec{C} (\muh)$ of the
Wilson coefficients. Their values are found by matching the full to
the effective theory amplitudes perturbatively in $\as$. The NLO
and NNLO approximation requires the calculation of one- and two-loop  
diagrams both in the SM and the low-energy effective theory. Some of
the SM two-loop 1PI diagrams one has to consider in order to find the
$\ord (\as^2)$ corrections to $\vec{C} (\muh)$ are displayed in 
\Fig{fig:matching}. Restricting ourselves to the physical on-shell
operators $\Qmisiak_1$--$\Qmisiak_6$ and setting $\muh = \MW$, one
obtains using dimensional regularization with a naive anticommuting
$\gamma_5$ \cite{Chetyrkin:1997gb, Bobeth:1999mk}:            
\begin{align} \label{eq:wilsoncoefficientsmisiak} 
  C_1 (\MW) & = 15 \f{\as (\MW)}{4 \pi} + \left( \f{\as (\MW)}{4 \pi}
  \right )^2 \left ( \f{7987}{72} + \f{17}{3} \pi^2 - \widetilde{T}_0
    (x_t) \right ) \, , \non \\
  C_2 (\MW) & = 1 + \left( \f{\as (\MW)}{4 \pi} \right )^2 \left (
    \f{127}{18} + \f{4}{3} \pi^2 \right ) \, , \non \\
  C_3 (\MW) & = \left( \f{\as (\MW)}{4 \pi} \right )^2 \widetilde{G}_1
  (x_t) \, , \non \\
  C_4 (\MW) & = \f{\as (\MW)}{4 \pi} \widetilde{E}_0 (x_t) + \left(
    \f{\as (\MW)}{4 \pi} \right )^2 \widetilde{E}_1 (x_t) \, , \non \\
  C_5 (\MW) & = \left( \f{\as (\MW)}{4 \pi} \right )^2 \left (
    \f{14}{135} + \f{2}{15} \widetilde{E}_0 (x_t) - \f{1}{10}
    \widetilde{G}_1 (x_t) \right ) \, , \non \\
  C_6 (\MW) & = \left( \f{\as (\MW)}{4 \pi} \right )^2 \left (
    \f{7}{36} + \f{1}{4} \widetilde{E}_0 (x_t) - \f{3}{16}
    \widetilde{G}_1 (x_t) \right ) \, ,
\end{align}
where $x_t = \mt^2/\MW^2$. The one-loop Inami-Lim \cite{Inami:1980fz}
function $\widetilde{E}_0 (x_t)$ characterizing the effective
off-shell vertex involving a gluon reads       
\beq \label{eq:inamilimoneloop}
\widetilde{E}_0 (x_t) = \f{8 - 42 x_t + 35 x_t^2 - 7 x_t^3}{12 (x_t -
1)^3} - \f{4 - 16 x_t + 9 x_t^2}{6 (x_t - 1)^4} \ln x_t \, . 
\eeq

{%
\begin{figure}[t]
\begin{center}
\scalebox{1.8}{\begin{picture}(55,45)(0,-5)\input{matching1.tex}\end{picture}}%
\hspace{3mm}%
\scalebox{1.8}{\begin{picture}(55,45)(0,-5)\input{matching2.tex}\end{picture}}%
\hspace{3mm}%
\scalebox{1.8}{\begin{picture}(55,45)(0,-5)\input{matching3.tex}\end{picture}}%
\hspace{3mm}%
\scalebox{1.8}{\begin{picture}(55,45)(0,-5)\input{matching4.tex}\end{picture}}%
\vspace{-0.5cm}%
\caption{Some of the SM two-loop 1PI diagrams one has to calculate in 
order to find the Wilson coefficients of the four-quark operators
$\Qmisiak_1$--$\Qmisiak_6$ at $\ord (\as^2)$.}  
\label{fig:matching}
\end{center}
\end{figure}
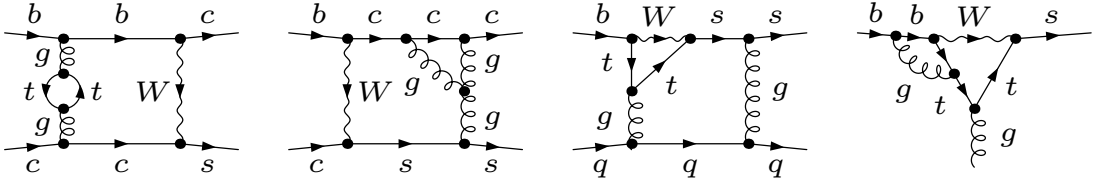
}%

\noindent The one-loop function $\widetilde{T}_0 (x_t)$ originates
from diagrams like the first one shown in
\Fig{fig:matching}. Subtracting the corresponding terms in the gluon
propagator in the so-called momentum space subtraction scheme at $q^2 
= 0$, which guarantees that $\as$ has the same numerical value on the
full and effective side at the matching scale through NNLO, one finds
\cite{Bobeth:1999mk}        
\beq \label{eq:oneloopselfenergy}
\begin{split}
\widetilde{T}_0 (x_t) & = \f{112}{9} + 32 x_t + \left ( \f{20}{3} + 16
x_t \right ) \ln x_t \\ 
& - \left ( 8 + 16 x_t \right ) \sqrt{4 x_t - 1} \, \cl \left ( 2
\arcsin \left ( \f{1}{2 \sqrt{x_t}} \right ) \right ) \, , 
\end{split}
\eeq
with $\cl (x) = {\rm Im} [ \li ( e^{i x} ) ]$ and $\li (x) = -\int^x_0
dt \, \ln (1 - t)/t$. The remaining two-loop functions
$\widetilde{E}_1 (x_t)$ and $\widetilde{G}_1 (x_t)$ take the following
form \cite{Bobeth:1999mk}  
\begin{align} \label{eq:inamilimtwoloop} 
  \widetilde{E}_1 (x_t) & = -\f{1120 - 12044 x_t - 5121 x_t^2 - 5068
    x_t^3 + 7289 x_t^4}{648 (x_t - 1)^4} \non \\
  & - \f{380 - 7324 x_t + 17702 x_t^2 + 2002 x_t^3 - 5981 x_t^4 + 133
    x_t^5}{324 (x_t - 1)^5} \ln x_t \non \\
  & - \f{112 - 530 x_t - 3479 x_t^2 + 2783 x_t^3 - 1129 x_t^4 + 515
    x_t^5}{108 (x_t - 1)^5} \ln^2 x_t \non \\
  & - \f{40 - 190 x_t - 81 x_t^2 - 614 x_t^3 + 515 x_t^4}{54 (x_t -
    1)^4} \, \li (1 - x_t) + \f{10}{81} \pi^2 \, , \non \\[2mm]
  \widetilde{G}_1 (x_t) & = \f{554 - 2523 x_t + 2919 x_t^2 - 662 x_t^3
    {243 (x_t - 1)^3}} \non \\
  & + \f{88 - 142 x_t - 357 x_t^2 + 100 x_t^3 + 35 x_t^4}{81 (x_t -
    1)^4} \ln x_t \
  - \f{20 - 40 x_t + 5 x_t^2}{27 (x_t - 1)^2} \ln^2 x_t \non \\
  & + \f{40 - 160 x_t - 30 x_t^2 + 100 x_t^3 - 10 x_t^4}{27 (x_t -
    1)^4} \li (1 - x_t) - \f{20}{81} \pi^2 \, .
\end{align}

\section{Anomalous Dimension Matrix} 
\label{sec:anomalousdimensionmatrix}

Before presenting our results for the \ads \ describing the mixing  
of the four-quark operators $Q_1$--$Q_6$ up to $\ord (\as^3)$ let us
recall some definitions that will turn out to be useful in the rest of
the paper. 

Upon renormalization the bare Wilson coefficients $\vec{C}_B (\mul)$
of \Eq{eq:heffgeneric} transform as          
\beq \label{eq:renormalization}
\vec{C}_B (\mul) = \hat{Z}^T \vec{C} (\mul) \, .  
\eeq
In terms of the renormalization constant matrix $\hat{Z}$ the ADM of
\Eq{eq:renormalizationgroupequation} is then given by  
\beq \label{eq:definitionanomalousdimensions}
\hat{\gamma} (\gs) = \hat{Z} \hspace{0.5mm} \mul \f{d}{d \mul} 
\hat{Z}^{-1} \, .        
\eeq
The renormalization constants $Z_{ij}$ of the operator $Q_j$ can be
expanded in powers of $\gs$ in the following way  
\beq \label{eq:zfactors}
Z_{ij} = \delta_{ij} + \sum^\infty_{k = 1} \left ( \f{\gs^2}{16 \pi^2}
\right )^k Z_{ij}^{(k)} \, , \hspace{5mm} \text{with} \hspace{5mm}
Z_{ij}^{(k)} = \sum_{l = 0}^k \frac{1}{\eps^l} Z_{ij}^{(k,l)} \, .    
\eeq
Following the standard $\MSbar$ scheme prescription, $Z_{ij}$ is given
by pure $1/\eps^l$ poles, except when $i$ corresponds to an evanescent
operator and $j$ does not. In the latter case, the renormalization
constant is finite, to make sure that the matrix elements of the
evanescent operators vanish in $n = 4$ dimensions \cite{Buras:1989xd,
evanescent}.  

In a mass independent renormalization scheme the only $\mu$-dependence
of $Z_{ij}$ resides in the coupling constant. In consequence, we might
rewrite \Eq{eq:definitionanomalousdimensions} as 
\beq \label{eq:gammaintermsofbeta}
\gamma_{ij} = \beta (\eps, \gs) Z_{ik} \f{d}{d \gs} Z^{-1}_{kj} \, ,
\eeq
where $\beta (\eps, \gs) $ is related to the usual QCD $\beta$
function via
\beq \label{eq:betaprimeeps}
\beta (\eps, \gs) = -\eps \, \gs + \beta (\gs) \, .    
\eeq
The finite parts of \Eq{eq:gammaintermsofbeta} in the limit of $\eps$
going to zero give the \ads. Inserting the expansions of $\hat{\gamma}
(\gs)$ and $\beta (\gs)$ in powers of $\gs$, as given in
\Eq{eq:gammamatrixandbetafunctionexpansion}, one immediately finds   
\cite{Gambino:2003zm, Chetyrkin:1997fm} for the \ads \ governing the
evolution of physical operators up to third order in the strong
coupling parameter:
\beq \label{eq:physicalgammaexpansion}
\begin{split}
\hat{\gamma}^{(0)} & = 2 \hat{Z}^{(1,1)} \, , \\
\hat{\gamma}^{(1)} & = 4 \hat{Z}^{(2,1)} - 2 \hat{Z}^{(1,1)}
\hat{Z}^{(1,0)} \, , \\
\hat{\gamma}^{(2)} & = 6 \hat{Z}^{(3,1)} - 4 \hat{Z}^{(2,1)}
\hat{Z}^{(1,0)} - 2 \hat{Z}^{(1,1)} \hat{Z}^{(2,0)} \, .  
\end{split}
\eeq

{%
\begin{figure}[t]
\begin{center}
\vspace{-0.5cm}%
\scalebox{1.8}{\begin{picture}(55,45)(0,-5)\input{mixing1.tex}\end{picture}}%
\hspace{1mm}%
\scalebox{1.8}{\begin{picture}(55,45)(0,-5)\input{mixing2.tex}\end{picture}}%
\hspace{3mm}%
\scalebox{1.8}{\begin{picture}(55,45)(0,-5)\input{mixing3.tex}\end{picture}}%
\hspace{3mm}%
\scalebox{1.8}{\begin{picture}(55,45)(0,-5)\input{mixing4.tex}\end{picture}}%
\vspace{0.5cm}%
\caption{Some of the three-loop 1PI diagrams we had to calculate in
order to find the mixing among the four-quark operators
$\Qmisiak_1$--$\Qmisiak_6$ at $\ord (\as^3)$.}    
\label{fig:mixing}
\end{center}
\end{figure}
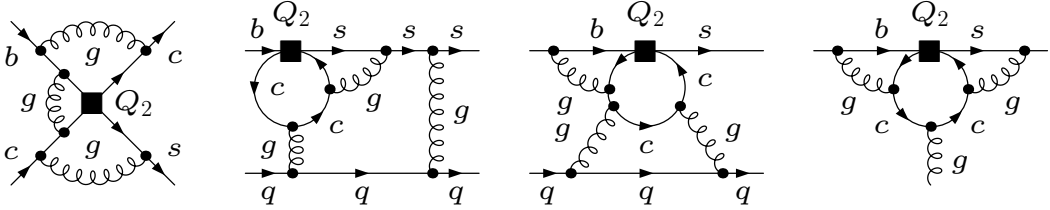
}%

The matrices $\hat{Z}^{(1,0)}$, $\hat{Z}^{(1,1)}$, $\hat{Z}^{(2,0)}$
and $\hat{Z}^{(2,1)}$ are found by calculating various one- and
two-loop diagrams with a single insertion of
$\Qmisiak_1$--$\Qmisiak_6$, $E^{(1)}_1$--$E^{(1)}_4$ and
$E^{(2)}_1$--$E^{(2)}_4$, whereas the matrix $\hat{Z}^{(3,1)}$
requires the computation of three-loop diagrams with insertions of
$\Qmisiak_1$--$\Qmisiak_6$ as shown in \Fig{fig:mixing}. The pole and
finite parts of these one-, two- and three-loop diagrams are evaluated
using the method we have described together with Paolo Gambino in
detail in \cite{Gambino:2003zm}: We perform the calculation off-shell 
in an arbitrary $R_\xi$ gauge which allows us to explicitly check the
gauge-parameter independence of the mixing among physical
operators. To distinguish between IR and UV divergences we follow
\cite{Misiak:1994zw, Chetyrkin:1997fm} and introduce a common mass $M$
for all fields, expanding all loop integrals in inverse powers of
$M$. This makes the calculation of the UV divergences possible even at
three loops, as $M$ becomes the only relevant internal scale and
three-loop tadpole integrals with a single non-zero mass are known
\cite{Chetyrkin:1997fm, threeloop}. On the other hand, this procedure
requires to take into account insertions of the non-physical operators
$N^{(1)}_1$ and $N^{(2)}_1$--$N^{(2)}_{10}$, as well as of appropriate
counterterms of dimension-three and four, some of which explicitly
break gauge invariance. A comprehensive discussion of the technical
details of the renormalization of the effective theory and the actual
calculation of the operator mixing is given in \cite{Gambino:2003zm}. 

Having summarized the general formalism and our method, we will now
present our results for an arbitrary number of quark flavors denoted
by $\nf$. For completeness we start with the regularization- and
renormalization-scheme independent matrix $\hat{\gamma}^{(0)}$, which 
is given by     
\beq \label{eq:gamma0new}
\hat{\gamma}^{(0)} = 
\left (
\begin{array}{cccccc}
-4 & {\scriptstyle \f{8}{3}} & 0 & {\scriptstyle -\f{2}{9}} & 0 & 0 \\  
12 & 0 & 0 & {\scriptstyle \f{4}{3}} & 0 & 0 \\  
0 & 0 & 0 & {\scriptstyle -\f{52}{3}} & 0 & 2 \\  
0 & 0 & {\scriptstyle -\f{40}{9}} & {\scriptstyle -\f{160}{9}}
{\scriptstyle +\f{4}{3} \nf} & {\scriptstyle \f{4}{9}} & {\scriptstyle
\f{5}{6}} \\ 
0 & 0 & 0 & {\scriptstyle -\f{256}{3}} & 0 & 20 \\
0 & 0 & {\scriptstyle -\f{256}{9}} & {\scriptstyle -\f{544}{9}}
{\scriptstyle +\f{40}{3} \nf} & {\scriptstyle \f{40}{9}} &
{\scriptstyle -\f{2}{3}} 
\end{array}
\right ) \, .    
\eeq
While the matrix $\hat{\gamma}^{(0)}$ is renormalization-scheme
independent, $\hat{\gamma}^{(1)}$ and $\hat{\gamma}^{(2)}$ are not. In
the $\MSbar$ scheme supplemented by the definition of evanescent
operators given in \Eqs{eq:oneloopevanescentoperatorsmisiak},
\eq{eq:twoloopevanescentoperatorsmisiak} and
\eq{eq:threeloopevanescentoperatorsmisiak} we obtain         
\beq \label{eq:gamma1new}
\hat{\gamma}^{(1)} = 
\left (
\begin{array}{cccccc}
{\scriptstyle -\f{145}{3}} {\scriptstyle +\f{16}{9} \nf} &
{\scriptstyle -26} {\scriptstyle +\f{40}{27} \nf} & {\scriptstyle
-\f{1412}{243}} & {\scriptstyle -\f{1369}{243}} & {\scriptstyle
\f{134}{243}} & {\scriptstyle -\f{35}{162}} \\ 
{\scriptstyle -45}  {\scriptstyle +\f{20}{3} \nf} & {\scriptstyle
-\f{28}{3}} & {\scriptstyle -\f{416}{81}} & {\scriptstyle
\f{1280}{81}} & {\scriptstyle \f{56}{81}} & {\scriptstyle \f{35}{27}}
\\ 
0 & 0 & {\scriptstyle -\f{4468}{81}} & {\scriptstyle -\f{29129}{81}}
{\scriptstyle -\f{52}{9} \nf} & {\scriptstyle \f{400}{81}} &
{\scriptstyle \f{3493}{108}} {\scriptstyle -\f{2}{9} \nf} \\
0 & 0 & {\scriptstyle -\f{13678}{243}} {\scriptstyle +\f{368}{81} \nf}
& {\scriptstyle -\f{79409}{243}} {\scriptstyle +\f{1334}{81} \nf} &
{\scriptstyle \f{509}{486}} {\scriptstyle -\f{8}{81} \nf} &
{\scriptstyle \f{13499}{648}} {\scriptstyle -\f{5}{27} \nf} \\
0 & 0 & {\scriptstyle -\f{244480}{81}} {\scriptstyle -\f{160}{9} \nf}
& {\scriptstyle -\f{29648}{81}} {\scriptstyle -\f{2200}{9} \nf} &
{\scriptstyle \f{23116}{81}} {\scriptstyle +\f{16}{9} \nf} &
{\scriptstyle \f{3886}{27}} {\scriptstyle +\f{148}{9} \nf} \\
0 & 0 & {\scriptstyle \f{77600}{243}} {\scriptstyle -\f{1264}{81} \nf}
& {\scriptstyle -\f{28808}{243}} {\scriptstyle +\f{164}{81} \nf} &
{\scriptstyle -\f{20324}{243}} {\scriptstyle +\f{400}{81} \nf} &
{\scriptstyle -\f{21211}{162}} {\scriptstyle +\f{622}{27} \nf}
\end{array}
\right ) \, ,  
\eeq
and 
\bea \label{eq:gamma2new}
\begin{split}
\hat{\gamma}^{(2)} & = 
\left ( 
\begin{array}{ccccccc}
{\scriptstyle -\f{1927}{2} + \f{257}{9} \nf  + \f{40}{9} \nf^2 + \left
( 224 + \f{160}{3} \nf \right ) \zetathree} & & & & & &
\scriptstyle{\f{475}{9} + \f{362}{27} \nf - \f{40}{27} \nf^2 - \left (
\f{896}{3} + \f{320}{9} \nf \right ) \zetathree} \\  
{\scriptstyle \f{307}{2} + \f{361}{3} \nf - \f{20}{3} \nf^2 - \left (
1344 + 160 \nf \right ) \zetathree} & & & & & & 
{\scriptstyle \f{1298}{3} - \f{76}{3} \nf - 224 \zetathree} \\ 
0 & & & & & & 0 \\
0 & & & & & & 0 \\
0 & & & & & & 0 \\
0 & & & & & & 0 
\end{array}
\right. \\[2mm]
&  
\begin{array}{cc}
{\scriptstyle \f{269107}{13122} - \f{2288}{729} \nf - \f{1360}{81}
\zetathree} & {\scriptstyle -\f{2425817}{13122} + \f{30815}{4374} \nf 
- \f{776}{81} \zetathree} \\ 
{\scriptstyle \f{69797}{2187} + \f{904}{243} \nf + \f{2720}{27}
\zetathree} & {\scriptstyle \f{1457549}{8748} - \f{22067}{729} \nf -
\f{2768}{27} \zetathree} \\ 
{\scriptstyle -\f{4203068}{2187} + \f{14012}{243} \nf - \f{608}{27}
\zetathree} & {\scriptstyle -\f{18422762}{2187} + \f{888605}{2916} \nf
+ \f{272}{27} \nf^2 + \left ( \f{39824}{27} + 160 \nf \right )
\zetathree} \\  
{\scriptstyle -\f{5875184}{6561} + \f{217892}{2187} \nf + \f{472}{81}
\nf^2 + \left ( \f{27520}{81} + \f{1360}{9} \nf \right ) \zetathree} &
{\scriptstyle -\f{70274587}{13122} + \f{8860733}{17496} \nf -
\f{4010}{729} \nf^2 + \left ( \f{16592}{81} + \f{2512}{27} \nf \right
) \zetathree} \\
{\scriptstyle -\f{194951552}{2187} + \f{358672}{81}\nf - \f{2144}{81}
\nf^2 + \f{87040}{27} \zetathree} & {\scriptstyle -\f{130500332}{2187}
- \f{2949616}{729} \nf + \f{3088}{27} \nf^2 + \left ( \f{238016}{27} +
640 \nf \right ) \zetathree} \\
{\scriptstyle \f{162733912}{6561} - \f{2535466}{2187} \nf +
\f{17920}{243} \nf^2 + \left ( \f{174208}{81} + \f{12160}{9} \nf
\right ) \zetathree} & {\scriptstyle \f{13286236}{6561} -
\f{1826023}{4374} \nf - \f{159548}{729} \nf^2 - \left ( \f{24832}{81}
+ \f{9440}{27} \nf \right ) \zetathree}
\end{array} \\[2mm]
& \left. 
\begin{array}{cc} 
{\scriptstyle -\f{343783}{52488} + \f{392}{729} \nf + \f{124}{81}
\zetathree} & {\scriptstyle -\f{37573}{69984} + \f{35}{972} \nf +
\f{100}{27} \zetathree} \\
{\scriptstyle -\f{37889}{8748} - \f{28}{243} \nf - \f{248}{27}
\zetathree} & {\scriptstyle \f{366919}{11664} - \f{35}{162} \nf -
\f{110}{9} \zetathree} \\
{\scriptstyle \f{674281}{4374} - \f{1352}{243} \nf - \f{496}{27}
\zetathree} & {\scriptstyle \f{9284531}{11664} - \f{2798}{81} \nf -
\f{26}{27} \nf^2 - \left ( \f{1921}{9} + 20 \nf \right ) \zetathree}
\\
{\scriptstyle \f{2951809}{52488} - \f{31175}{8748} \nf - \f{52}{81}
\nf^2 - \left ( \f{3154}{81} + \f{136}{9} \nf \right ) \zetathree} & 
{\scriptstyle \f{3227801}{8748} - \f{105293}{11664} \nf - \f{65}{54}
\nf^2 + \left ( \f{200}{27} - \f{220}{9} \nf \right ) \zetathree} \\
{\scriptstyle \f{14732222}{2187} - \f{27428}{81} \nf + \f{272}{81}
\nf^2 - \f{13984}{27} \zetathree} & {\scriptstyle \f{16521659}{2916} +
\f{8081}{54} \nf - \f{316}{27} \nf^2 - \left ( \f{22420}{9} + 200 \nf
\right ) \zetathree} \\
{\scriptstyle -\f{22191107}{13122} + \f{395783}{4374} \nf -
\f{1720}{243} \nf^2 - \left ( \f{33832}{81} + \f{1360}{9} \nf \right )
\zetathree} & {\scriptstyle -\f{32043361}{8748} + \f{3353393}{5832}
\nf - \f{533}{81} \nf^2 + \left ( \f{9248}{27} - \f{1120}{9} \nf
\right ) \zetathree}
\end{array}
\right ) \, . 
\end{split}
\eea
As far as the one- and two-loop mixing of the four-quark operators
$\Qmisiak_1$--$\Qmisiak_6$, namely $\hat{\gamma}^{(0)}$ and
$\hat{\gamma}^{(1)}$ are concerned, our results agree with those of 
\cite{Chetyrkin:1997gb}. Furthermore, they also agree with the results
obtained in \cite{Ciuchini:1993vr, Buras:1992tc} after a
transformation to the ``traditional'' operator basis. This will be
shown in \Sec{sec:changeofoperatorbasis} by an explicit
calculation. On the other hand, the three-loop mixing of
$\Qmisiak_1$--$\Qmisiak_6$ described by $\hat{\gamma}^{(2)}$, is 
entirely new and has never been given before. As it is characteristic
for three-loop \ads \ the entries of $\hat{\gamma}^{(2)}$, contain
terms proportional to the Riemann zeta function $\zetathree$.       

\section{Renormalization Group Evolution}
\label{sec:renormalizationgroupequation}

In this section we shall use the obtained ADM to find the explicit
NNLO expressions for the Wilson coefficients   
\beq \label{eq:asexpansionwilsoncoefficients}
C_i (\mub) = C^{(0)}_i (\mub) + \f{\as (\mub)}{4 \pi} C^{(1)}_i (\mub)
+ \left ( \f{\as (\mub)}{4 \pi} \right )^2 C^{(2)}_i (\mub) \, , 
\eeq 
with $i = 1$--$6$, at the low-energy scale $\mub = \ord (\mb)$, which
is appropriate for studying non-leptonic $B$ meson decays. Using the
general solution of the RGE given in
\Eq{eq:asevolutionmatrixexpansion}, we arrive at 
\beq 
\begin{split}
C^{(0)}_i (\mub) & = \sum^6_{j = 1} c^{(0)}_{0, i j} \eta^{a_j} \, ,
\\[2mm]  
C^{(1)}_i (\mub) & = \sum^6_{j = 1} \left ( c^{(1)}_{0, i j} +
c^{(1)}_{1, i j} \eta + e^{(1)}_{1, i j} \eta \widetilde{E}_0 (x_t)
\right ) \eta^{a_j} \, , \\[2mm]  
C^{(2)}_i (\mub) & = \sum^6_{j = 1} \bigg ( c^{(2)}_{0, i j} + 
c^{(2)}_{1, i j} \eta + c^{(2)}_{2, i j} \eta^2 + \left [ e^{(2)}_{1,
i j} \eta + e^{(2)}_{2, i j} \eta^2 \right ] \widetilde{E}_0 (x_t) \\
& + t^{(2)}_{2, i j} \eta^2 \widetilde{T}_0 (x_t) + e^{(1)}_{1, i j} 
\eta^2 \widetilde{E}_1 (x_t) + g^{(2)}_{2, i j} \eta^2 \widetilde{G}_1
(x_t) \bigg ) \eta^{a_j} \, ,       
\end{split}
\eeq
where $\eta = \as (\MW)/\as (\mub)$ and 
\begin{align} 
\label{eq:avector}
\vec{a}^T & =
\left ( 
\begin{array}{cccccc}
\f{6}{23} & -\f{12}{23} & 0.4086 & -0.4230 & -0.8994 & 0.1456 
\end{array} 
\right )
\, , \\[2mm]
\label{eq:c00matrix}
\hat{c}^{(0)}_0 & = \left ( 
\begin{array}{cccccc}
1 & -1 & 0 & 0 & 0 & 0 \\[1mm]
\f{2}{3} & \f{1}{3} & 0 & 0 & 0 & 0 \\[1mm]
\f{2}{63} & -\f{1}{27} & -0.0659 & 0.0595 & -0.0218 & 0.0335 \\[1mm]
\f{1}{21} & \f{1}{9} & 0.0237 & -0.0173 & -0.1336 & -0.0316 \\[1mm]
-\f{1}{126} & \f{1}{108} & 0.0094 & -0.0100 & 0.0010 & -0.0017 \\[1mm]
-\f{1}{84} & -\f{1}{36} & 0.0108 & 0.0163 & 0.0103 & 0.0023 
\end{array} 
\right ) \, , \\[2mm]
\label{eq:c10matrix}
\hat{c}^{(1)}_0 & = \left ( 
\begin{array}{cccccc}
5.9606 & 1.0951 & 0 & 0 & 0 & 0 \\
1.9737 & -1.3650 & 0 & 0 & 0 & 0 \\
-0.5409 & 1.6332 & 1.6406 & -1.6702 & -0.2576 & -0.2250 \\
2.2203 & 2.0265 & -4.1830 & -0.7135 & -1.8215 & 0.7996 \\
0.0400 & -0.1861 & -0.1669 & 0.1887 & 0.0201 & 0.0304 \\
-0.2614 & -0.1918 & 0.4197 & 0.0295 & 0.1474 & -0.0640 
\end{array} 
\right ) \, , \\[2mm]
\label{eq:c11matrix}
\hat{c}^{(1)}_1 & = \left ( 
\begin{array}{cccccc}
2.0394 & 5.9049 & 0 & 0 & 0 & 0 \\
1.3596 & -1.9683 & 0 & 0 & 0 & 0 \\
0.0647 & 0.2187 & -0.4268 & -0.5165 & 0.2832 & -0.2034 \\
0.0971 & -0.6561 & 0.1534 & 0.1500 & 1.7355 & 0.1916 \\
-0.0162 & -0.0547 & 0.0606 & 0.0865 & -0.0128 & 0.0103 \\
-0.0243 & 0.1640 & 0.0700 & -0.1412 & -0.1339 & -0.0140 
\end{array} 
\right ) \, , \\[2mm]
\label{eq:e11matrix}
\hat{e}^{(1)}_1 & = \left (  
\begin{array}{cccccc}
0 & 0 & 0 & 0 & 0 & 0 \\
0 & 0 & 0 & 0 & 0 & 0 \\
0 & 0 & -0.1933 & 0.1579 & 0.1428 & -0.1074 \\
0 & 0 & 0.0695 & -0.0459 & 0.8752 & 0.1012 \\
0 & 0 & 0.0274 & -0.0264 & -0.0064 & 0.0055 \\
0 & 0 & 0.0317 & 0.0432 & -0.0675 & -0.0074 
\end{array} 
\right ) \, , \\[2mm]
\label{eq:c20matrix}
\hat{c}^{(2)}_0 & = \left ( 
\begin{array}{cccccc}
56.4723 & 22.2650 & 0 & 0 & 0 & 0 \\
14.7825 & -11.7987 & 0 & 0 & 0 & 0 \\
1.9905 & 19.2386 & -24.6846 & -12.9233 & -4.0085 & 2.0820 \\ 
8.1141 & 42.7264 & -11.7014 & -35.4784 & -14.1041 & 4.9828 \\
-0.3660 & -1.2588 & 2.7564 & 0.6168 & 0.2854 & -0.2620 \\
-2.3243 & -3.5577 & 2.9357 & 2.4965 & 1.5568 & -0.4249 
\end{array} 
\right ) \, , \\[2mm]
\label{eq:c21matrix}
\hat{c}^{(2)}_1 & = \left ( 
\begin{array}{cccccc}
12.1560 & -6.4667 & 0 & 0 & 0 & 0 \\
4.0252 & 8.0604 & 0 & 0 & 0 & 0 \\
-1.1032 & -9.6435 & 10.6219 & 14.5052 & 3.3472 & 1.3651 \\
4.5281 & -11.9660 & -27.0825 & 6.1964 & 23.6695 & -4.8514 \\
0.0816 & 1.0987 & -1.0803 & -1.6385 & -0.2612 & -0.1847 \\
-0.5332 & 1.1326 & 2.7171 & -0.2564 & -1.9148 & 0.3886 
\end{array} 
\right ) \, , \\[2mm]
\label{eq:c22matrix}
\hat{c}^{(2)}_2 & = \left ( 
\begin{array}{cccccc}
32.6228 & 49.8089 & 0 & 0 & 0 & 0 \\
21.7486 & -16.6030 & 0 & 0 & 0 & 0 \\
1.0356 & 1.8448 & -0.6250 & -6.6619 & 2.8566 & 0.7622 \\
1.5535 & -5.5343 & 0.2246 & 1.9350 & 17.5058 & -0.7181 \\
-0.2589 & -0.4612 & 0.0887 & 1.1155 & -0.1290 & -0.0387 \\ 
-0.3884 & 1.3836 & 0.1026 & -1.8214 & -1.3503 & 0.0524 
\end{array} 
\right ) \, , \\[2mm]
\label{eq:e21matrix}
\hat{e}^{(2)}_1 & = \left (  
\begin{array}{cccccc}
0 & 0 & 0 & 0 & 0 & 0 \\
0 & 0 & 0 & 0 & 0 & 0 \\
0 & 0 & 4.8111 & -4.4336 & 1.6880 & 0.7207 \\
0 & 0 & -12.2667 & -1.8940 & 11.9366 & -2.5613 \\
0 & 0 & -0.4893 & 0.5008 & -0.1317 & -0.0975 \\
0 & 0 & 1.2307 & 0.0784 & -0.9657 & 0.2051 
\end{array} 
\right ) \, , \\[2mm]
\label{eq:e22matrix}
\hat{e}^{(2)}_2 & = \left (  
\begin{array}{cccccc}
0 & 0 & 0 & 0 & 0 & 0 \\
0 & 0 & 0 & 0 & 0 & 0 \\
0 & 0 & -1.3169 & -0.7444 & 0.4827 & -1.2075 \\
0 & 0 & 0.4733 & 0.2162 & 2.9582 & 1.1376 \\
0 & 0 & 0.1869 & 0.1247 & -0.0218 & 0.0613 \\
0 & 0 & 0.2161 & -0.2035 & -0.2282 & -0.0830 
\end{array} 
\right ) \, , \\[2mm]
\label{eq:t22matrix}
\hat{t}^{(2)}_2 & = \left ( 
\begin{array}{cccccc}
-\f{1}{3} & -\f{2}{3} & 0 & 0 & 0 & 0 \\[1mm]
-\f{2}{9} & \f{2}{9} & 0 & 0 & 0 & 0 \\[1mm]
-\f{2}{189} & -\f{2}{81} & 0.0129 & 0.0497 & -0.0092 & -0.0182 \\[1mm]
-\f{1}{63} & \f{2}{27} & -0.0046 & -0.0144 & -0.0562 & 0.0171 \\[1mm]
\f{1}{378} & \f{1}{162} & -0.0018 & -0.0083 & 0.0004 & 0.0009 \\[1mm]
\f{1}{252} & -\f{1}{54} & -0.0021 & 0.0136 & 0.0043 & -0.0012 
\end{array} 
\right ) \, , \\[2mm]
\label{eq:g22matrix}
\hat{g}^{(2)}_2 & = \left (  
\begin{array}{cccccc}
0 & 0 & 0 & 0 & 0 & 0 \\
0 & 0 & 0 & 0 & 0 & 0 \\
0 & 0 & 0.7557 & -0.1643 & 0.0861 & 0.3224 \\
0 & 0 & -0.2716 & 0.0477 & 0.5277 & -0.3038 \\
0 & 0 & -0.1072 & 0.0275 & -0.0039 & -0.0164 \\
0 & 0 & -0.1240 & -0.0449 & -0.0407 & 0.0222 
\end{array} 
\right ) \, .  
\end{align}
As far as the LO and NLO corrections parameterized by 
$\hat{c}^{(0)}_0$, $\hat{c}^{(1)}_0$, $\hat{c}^{(1)}_1$ and
$\hat{e}^{(1)}_1$ are concerned our results agree perfectly with the
findings of \cite{Chetyrkin:1997gb}. Contrariwise, the resummation of
the NNLO logarithms is entirely new, and the corresponding matrices 
$\hat{c}^{(2)}_0$, $\hat{c}^{(2)}_1$, $\hat{c}^{(2)}_2$,
$\hat{e}^{(2)}_1$, $\hat{e}^{(2)}_2$, $\hat{t}^{(2)}_2$ and 
$\hat{g}^{(2)}_2$ have never been computed before.  

\section{Renormalization Scheme Dependences}
\label{sec:renormalizationschemedependences}

We would now like to elaborate on the question of renormalization
scheme dependences in explicit terms, to gain an insight on how they 
arise beyond the LO, how various quantities transform under a change
of scheme, and how the scheme dependences cancels out in physical
observables. In this respect we will not only extend the existing NLO
results \cite{Buras:1991jm, Ciuchini:1993vr} to the NNLO level, but
will also discuss the conceptual features related to the
renormalization of $\as$ that, to the best of our knowledge, have not
been studied in the context of the renormalization of effective field
theories so far.  

It is well-known that beyond LO various quantities such as the ADM or
the Wilson coefficients depend on the scheme adopted for the
renormalization of the operators present in the effective theory. This
scheme dependence arises because the requirement that all UV
divergences are removed by a suitable renormalization of parameters,
fields as well as operators, does not fix the finite parts of the
associated renormalization constants. Indeed, these constants can be
defined in different ways corresponding to distinct renormalization
schemes, which are always related by a finite renormalization. In the
framework of dimensional regularization one example of how such a
scheme dependence may occur is the treatment of $\gamma_5$ in $n = 4 -
2 \eps$ dimensions. In this context two well-known choices of scheme
are the naive dimensional regularization scheme
\cite{Chanowitz:1979zu} with $\gamma_5$ taken to be fully
anticommuting and the 't Hooft-Veltman (HV) scheme \cite{HVscheme} 
which comprises a $\gamma_5$ that does not have simple commutation
properties with respect to the other Dirac matrices. Another example
is the scheme dependence related to the exact form of the physical and
evanescent operators chosen in the effective theory. We will discuss
the latter issue in great detail in the following section.    

In order to show that physical quantities do not depend on the
renormalization scheme and on the choice of the operator basis, we
have to demonstrate how this dependence cancels out in the matrix
elements of the effective hamiltonian introduced in
\Eq{eq:heffgeneric} with $\vec{C} (\mul)$ given by
\Eq{eq:wilsoncoefficient}. First, let us denote by
$\hat{\gamma}^{(i)}_0$, $\hat{r}^{(i)}_0$ and $\hat{\gamma}^{(i)}_a$, 
$\hat{r}^{(i)}_a$ with $i = 1, 2$ the results 
for the ADM and the local parts of the matrix elements, that is, the
finite pieces without logarithms of external momenta divided by the
renormalization scale squared, obtained in two different
renormalization schemes --- see
\Eqsand{eq:gammamatrixandbetafunctionexpansion}{eq:aeff}. Furthermore,
let us assume without loss of generality that the first scheme, which
we shall call reference scheme hereafter, is distinguished from the
other ones by the subsidiary condition $\hat{r}^{(1)}_0 =
\hat{r}^{(2)}_0 = 0$.    

It should be clear that for any given scheme $a$ we can always switch
to the reference scheme by the following finite renormalization:         
\beq \label{eq:finiterenormalization}
\hat{Z}_0 = \left [ \hat{1} - \f{\as (\mul)}{4 \pi} \hat{r}^{(1)}_a -
\left ( \f{\as (\mul)}{4 \pi} \right)^2 \left ( \hat{r}^{(2)}_a - \big
( \hat{r}^{(1)}_a \big )^2 \right ) \right ] \hat{Z}_a \, . 
\eeq
The corresponding transformations of the $\ord (\as^2)$ and $\ord
(\as^3)$ \ads \ is easily obtained using
\Eq{eq:definitionanomalousdimensions}. At the NLO we reproduce the 
well-known result \cite{Buras:1991jm, Ciuchini:1993vr}   
\beq \label{eq:nlogammareference}
\hat{\gamma}^{(1)}_0 = \hat{\gamma}^{(1)}_a - \left [ \hat{r}^{(1)}_a,
\hat{\gamma}^{(0)} \right ] - 2 \betazero \hat{r}^{(1)}_a \, ,  
\eeq 
whereas at the NNLO we find
\beq \label{eq:nnlogammareference}
\hat{\gamma}^{(2)}_0 = \hat{\gamma}^{(2)}_a - \left [ \hat{r}^{(2)}_a,
\hat{\gamma}^{(0)} \right ] - \left [ \hat{r}^{(1)}_a,
\hat{\gamma}^{(1)}_a \right ] + \hat{r}^{(1)}_a \left [
\hat{r}^{(1)}_a, \hat{\gamma}^{(0)} \right ] - 4 \betazero
\hat{r}^{(2)}_a  - 2 \betaone \hat{r}^{(1)}_a + 2 \betazero \big (
\hat{r}^{(1)}_a \big )^2 \, .   
\eeq
Similarly, for the transformation of the Wilson coefficients through
NNLO we obtain
\beq \label{eq:nnlowilsonreference}
\vec{C}_0 (\mul) = \left [ \hat{1} + \f{\as (\mul)}{4 \pi}
\hat{r}^{(1)}_a + \left ( \f{\as (\mul)}{4 \pi} \right )^2
\hat{r}^{(2)}_a \right ]^T \vec{C}_a (\mul) \, , 
\eeq
where again the NLO result is known for quite some time
\cite{Buras:1993dy}. The relations connecting the ADM and the Wilson 
coefficients in two different schemes $a$ and $b$ can easily be
derived from the above equations. 

With \Eqsand{eq:nlogammareference}{eq:nnlogammareference} at hand,
it is now straightforward to show that the matrices $\hat{R}^{(1)}$
and $\hat{R}^{(2)}$ introduced in \Eq{eq:rmatrices} are independent
of the renormalization scheme and the form of the operators
considered. We start from the \ads \ in the reference scheme
$\hat{\gamma}^{(1)}_0$ and $\hat{\gamma}^{(2)}_0$. These 
matrices can be accessed from any arbitrary scheme $a$ using 
\Eqsand{eq:nlogammareference}{eq:nnlogammareference}. Let us transpose
the latter equations and eliminate $\hat{\gamma}^{(1) \, T}_a$ and
$\hat{\gamma}^{(2) \, T}_a$ by means of \Eq{eq:jcondition}. Finally,
dropping the unnecessary subscript $a$, we obtain   
\beq \label{eq:gammatransposedreference}
\begin{split}
\hat{\gamma}^{(1) \, T}_0 & = \f{\betaone}{\betazero}
\hat{\gamma}^{(0) \, T} - \left [ \hat{\gamma}^{(0) \, T},
\hat{R}^{(1)} \right ] - 2 \betazero \hat{R}^{(1)} \, , \\[2mm]
\hat{\gamma}^{(2) \, T}_0 & = \f{\betatwo}{\betazero}
\hat{\gamma}^{(0) \, T} - \left [ \hat{\gamma}^{(0) \, T},
\hat{R}^{(2)} \right ] - \f{\betaone}{\betazero} \left [
\hat{\gamma}^{(0) \, T}, \hat{R}^{(1)} \right ] + \left [
\hat{\gamma}^{(0) \, T}, \hat{R}^{(1)} \right ] \hat{R}^{(1)} \\
& - 4 \betazero \hat{R}^{(2)} - 2 \betaone \hat{R}^{(1)} + 2 \betazero
\big ( \hat{R}^{(1)} \big )^2 \, ,  
\end{split}
\eeq
which proves the scheme independence of $\hat{R}^{(1)}$ and
$\hat{R}^{(2)}$. 

Next, $\vec{A}^{(0)}$, $\vec{A}^{(1)}$ and $\vec{A}^{(2)}$, obtained
from the calculation in the full theory, clearly do not depend on the
particular choice adopted for the renormalization of operators. In
consequence, the factor to the right of $\hat{U}^{(0)} (\mul, \muh)$
in $\vec{C} (\mul)$, as given in \Eq{eq:wilsoncoefficient}, which is
related to the upper end of the evolution, is independent of the
renormalization scheme. The same is true for the LO evolution matrix
$\hat{U}^{(0)} (\mul, \muh)$. However, $\vec{C} (\mul)$ still depends
on the renormalization scheme through $\hat{K} (\mul)$ and
consequently on $\hat{J}^{(1)}$ and $\hat{J}^{(2)}$, entering the
Wilson coefficients to the left of $\hat{U}^{(0)} (\mul, \muh)$. As is
evident from \Eqsand{eq:kmatrices}{eq:aeff}, this dependence on the
lower end of the evolution is canceled by the one of the matrix
elements $\langle \vec{Q}^T (\mul) \rangle$. We have therefore
explicitly seen that the matrix elements of the effective hamiltonian
and the resulting physical amplitudes are scheme independent. 

It is important to emphasize that the renormalization scheme
dependence discussed above refers to the renormalization of operators
only, and has to be distinguished from the renormalization scheme 
dependence related to a redefinition of the strong coupling
constant. In the following we will discuss the latter issue in detail,
illustrating which effect a change in the charge renormalization has
on the miscellaneous ingredients of the renormalization group improved 
perturbation theory. Finally, we will also prove that the matrix
elements in the effective theory are invariant under a change of
coupling constant.   

The coupling parameter $\as (\mu)$ in the new scheme will be denoted 
by $\as' (\mu)$ and the finite renormalization relating the two
schemes is written as 
\beq \label{eq:alpharedefinition}
\as (\mu) = \left [ 1 + \f{\as' (\mu)}{4 \pi} c_1 + \left ( \f{\as'
(\mu)}{4 \pi} \right )^2 c_2 \right ] \as' (\mu) \, .      
\eeq
It is now easy to show that the first two terms in the strong coupling
expansion of the QCD $\beta$ function, that is, $\betazero$ and
$\betaone$, are scheme independent, while the third coefficient,
namely $\betatwo$, transforms non-trivially \cite{muta}:     
\beq \label{eq:betachange}
\betatwo' = \betatwo + c_1 \betaone - \left ( c_2 - c_1^2 \right )
\betazero \, . 
\eeq

The renormalization constant matrices $\hat{Z}$ and $\hat{Z}'$ of the
two different schemes can in general be related through a finite 
renormalization $\hat{\rho} (\gs')$ in such a way that  
\beq \label{eq:zprimealpharedefinition}
\hat{Z} = \hat{\rho} (\gs') \hspace{0.1mm} \hat{Z}' \, , \hspace{5mm}
\text{with} \hspace{5mm} \hat{\rho} (\gs') = \hat{1} + \sum^\infty_{k
= 1} \left ( \f{\gs'^2}{16 \pi^2} \right )^{k} \hat{\rho}^{(k)} \, .   
\eeq
Inserting this into \Eq{eq:definitionanomalousdimensions} we obtain
for the ADM in the primed scheme  
\beq \label{eq:gammaprimemsbar}
\hat{\gamma}' (\gs') =  \hat{\rho} (\gs')^{-1} \left ( \hat{\gamma}
(\gs) + \mul \f{d}{d \mul} \right ) \hat{\rho} (\gs') \, .      
\eeq
Comparing \Eqs{eq:definitionanomalousdimensions},
\eq{eq:finiterenormalization}, \eq{eq:zprimealpharedefinition} and 
\eq{eq:gammaprimemsbar} it should be clear that the NLO counterpart
of \Eq{eq:nlogammareference} corresponding to the latter
transformation reads    
\beq \label{eq:nlogammaalpharedefintion}
\hat{\gamma}'^{(1)} = \hat{\gamma}^{(1)} - \left [ \hat{\rho}^{(1)},
\hat{\gamma}^{(0)} \right ] - 2 \betazero \hat{\rho}^{(1)} + c_1
\hat{\gamma}^{(0)} \, , 
\eeq
whereas the NNLO analog of \Eq{eq:nnlogammareference} takes the 
following form
\beq \label{eq:nnlogammaalpharedefintion}
\begin{split}
\hat{\gamma}'^{(2)} & = \hat{\gamma}^{(2)} - \left [ \hat{\rho}^{(2)},
\hat{\gamma}^{(0)} \right ] - \left [ \hat{\rho}^{(1)},
\hat{\gamma}^{(1)} \right ] + \hat{\rho}^{(1)} \left [
\hat{\rho}^{(1)}, \hat{\gamma}^{(0)} \right ] - 4 \betazero
\hat{\rho}^{(2)} \\ 
& - 2 \betaone \hat{\rho}^{(1)} + 2 \betazero \big ( \hat{\rho}^{(1)}
\big )^2 - c_1 \big [ \hat{\rho}^{(1)}, \hat{\gamma}^{(0)} \big ] +
c_2 \hat{\gamma}^{(0)} + 2 c_1 \hat{\gamma}^{(1)} \, .      
\end{split}
\eeq 

Needless to say, that the finite operator renormalization $\hat{\rho}$
is unambiguously determined by the exact form of the change of charge
renormalization parameterized by $c_1$ and $c_2$. Yet, it is not
difficult to understand that by making use of the prior established
fact that the matrix elements of the effective hamiltonian introduced
in \Eq{eq:heffgeneric} are renormalization scheme independent one can
always remove the dependence on $\hat{\rho}^{(1)}$ and
$\hat{\rho}^{(2)}$ from 
\Eqsand{eq:nlogammaalpharedefintion}{eq:nnlogammaalpharedefintion}. 
In other words, it is always possible to choose a scheme in which the
operator renormalization constants remain pure UV poles. In
consequence, the explicit relations between the coefficients codifying
the finite charge and operator renormalization are not needed to prove
the invariance of the effective theory under a redefinition of the
strong coupling constant. Therefore we will not give this relations
here.    

It is easy to see, that the suitable change of renormalization scheme
that one has to perform in order to remove the finite renormalization
matrices $\hat{\rho}^{(1)}$ and $\hat{\rho}^{(2)}$ in
\Eqsand{eq:nlogammaalpharedefintion}{eq:nnlogammaalpharedefintion}, is
characterized through 
\beq \label{eq:raalpharedefintion}
\hat{r}^{(1)}_a = -\hat{\rho}^{(1)} \, , \hspace{5mm} \text{and}
\hspace{5mm} \hat{r}^{(2)}_a = -\hat{\rho}^{(2)} + \big (
\hat{\rho}^{(1)} \big )^2\, ,
\eeq
and implemented by
\Eqsand{eq:nlogammareference}{eq:nnlogammareference}. While the strong
coupling constant is obviously invariant under such a change of
renormalization scheme, that is,  
\beq \label{eq:doubleprimedalpha}
\as'' (\mul) = \as' (\mul) \, , 
\eeq
the \ads \ beyond LO transform non-trivially. Combining
\Eq{eq:zprimealpharedefinition} with the latter transformation we 
obtain the following NLO \ads \ in the double primed scheme  
\beq \label{eq:nlogammaalpharedefintionfinal}
\hat{\gamma}''^{(1)} = \hat{\gamma}^{(1)} + c_1 \hat{\gamma}^{(0)}
\, , 
\eeq
whereas the NNLO result reads
\beq \label{eq:nnlogammaalpharedefintionfinal}
\hat{\gamma}''^{(2)} = \hat{\gamma}^{(2)} + c_2 \hat{\gamma}^{(0)} + 2
c_1 \hat{\gamma}^{(1)} \, , 
\eeq 
and the dependence on $\hat{\rho}^{(1)}$ and $\hat{\rho}^{(2)}$ has
dropped out from the last two equations as intended.      

Next, let us write down the explicit expressions for the matrices 
$\hat{J}^{(1)}$ and $\hat{J}^{(2)}$ in the double primed
scheme. Using \Eq{eq:jcondition} and taking into account that the
expansion coefficients of the QCD $\beta$ function in the double
primed scheme coincide with those in the primed one, we find up to the
NNLO level     
\beq \label{eq:jdoubleprimes}
\begin{split}
\hat{J}''^{(1)} & = \hat{J}^{(1)} - \f{c_1}{2 \betazero}
\hat{\gamma}^{(0) \, T} \, , \\  
\hat{J}''^{(2)} & = \hat{J}^{(2)} + c_1 \hat{J}^{(1)} - \f{c_1}{2
\betazero} \hat{J}^{(1)} \hat{\gamma}^{(0) \, T} + \f{c_1^2 - 2 c_2}{4
\betazero} \hat{\gamma}^{(0) \, T} + \f{c_1^2}{8 \betazero^2} \big (
\hat{\gamma}^{(0) \, T} \big )^2 \, ,     
\end{split}
\eeq
with $\hat{J}^{(1)}$ and $\hat{J}^{(2)}$ defined as in
\Eq{eq:jandgmatrices}. 

Furthermore, in order to separate the coupling scheme from the
renormalization scheme dependence, let us assume without loss of
generality, that the local parts of the matrix elements in the
unprimed scheme fulfill the constraint $\hat r^{(1)} = \hat r^{(2)} =
0$. From \Eqsand{eq:nlogammaalpharedefintionfinal}{eq:nnlogammaalpharedefintionfinal}
it is then immediately clear that the local parts of the matrix
elements in the double primed scheme correspondingly satisfy the
relation $r''^{(1)} = \hat r''^{(2)} = 0$. These two subsidiary
conditions together with \Eq{eq:jdoubleprimes} now imply that the
matrices $\hat{R}^{(1)}$ and  $\hat{R}^{(2)}$ in the double primed
scheme are given by         
\beq \label{eq:rdoubleprimes}
\begin{split}
\hat{R}''^{(1)} & = \hat{R}^{(1)} - \f{c_1}{2 \betazero}
\hat{\gamma}^{(0) \, T} \, , \\  
\hat{R}''^{(2)} & = \hat{R}^{(2)} + c_1 \hat{R}^{(1)} - \f{c_1}{2
\betazero} \hat{R}^{(1)} \hat{\gamma}^{(0) \, T} + \f{c_1^2 - 2 c_2}{4
\betazero} \hat{\gamma}^{(0) \, T} + \f{c_1^2}{8 \betazero^2} \big (
\hat{\gamma}^{(0) \, T} \big )^2 \, , 
\end{split}
\eeq
with $\hat{R}^{(1)}$ and $\hat{R}^{(2)}$ defined as in
\Eq{eq:rmatrices}. 

For the sake of completeness let us remark that only the third
coefficient in the expansion of the amplitude calculated in the full
theory transforms non-trivially under a redefinition of the strong
coupling constant: 
\beq \label{eq:fullamplitudechange}
\vec {A}''^{(2)} = \vec A^{(2)} + c_1 \vec A^{(1)} \, .     
\eeq

To prove that the effective theory does in fact not depend on the
renormalization scheme employed for the strong coupling constant, let
us first write down the analog of \Eq{eq:wilsoncoefficient} in the
double primed scheme: 
\beq \label{eq:onedoubleprimedwilsoncoefficient}
\begin{split}
\vec{C}'' (\mul) & = \hat{K}'' (\mul) \hspace{0.25mm} \hat{U}''^{(0)} 
(\mul, \muh) \Bigg ( \vec{A}''^{(0)} + \f{\as'' (\muh)}{4 \pi} \left [ 
\vec{A}''^{(1)} - \hat{R}''^{(1)} \vec{A}''^{(0)} \right ] \\   
& + \left ( \f{\as'' (\muh)}{4 \pi} \right )^2 \left [ \vec{A}''^{(2)} -
\hat{R}''^{(1)} \vec{A}''^{(1)} - \left ( \hat{R}''^{(2)} - \big (
\hat{R}''^{(1)} \big )^2 \right ) \vec{A}''^{(0)} \right  ] \Bigg ) \, ,
\end{split}
\eeq
where 
\beq \label{eq:doubleprimedkmatrix}
\hat{K}'' (\mul) = \hat{1} + \f{\as (\mul)}{4 \pi} \hat{J}''^{(1)} +
\left ( \f{\as (\mul)}{4 \pi} \right )^2 \left ( \hat{J}''^{(2)} - c_1
\hat{J}''^{(1)} \right ) \, , 
\eeq
and 
\beq \label{eq:doubleprimedloevolutionmatrix}
\hat{U}''^{(0)} (\mul, \muh) = \hat{V} \, {\rm diag} \left ( \f{\as'' 
(\muh)}{\as'' (\mul)} \right )^{a_i} \hat{V}^{-1} \, , 
\eeq  
with $\hat{V}$ and $a_i$ defined as in \Eq{eq:magicnumbers}. 

It is now a matter of simply algebra to show that the LO evolution
matrix in the double primed scheme is related to the LO evolution
matrix in the unprimed scheme by        
\beq \label{eq:doubleprimedloevolutionmatrixexpansion}
\hat{U}''^{(0)} (\mul, \muh) = \hat{L} (\mul) \hspace{0.25mm}
\hat{U}^{(0)} (\mul, \muh) \hat{L}^{-1} (\muh) \, ,     
\eeq
where
\beq \label{eq:Lmatrices}
\begin{split}
\hat{L} (\mul) & = \hat{1} + \f{\as (\mul)}{4 \pi} \f{c_1}{2
\betazero} \hat{\gamma}^{(0) \, T} - \left ( \f{\as (\mul)}{4 \pi}
\right )^2  \left [ \f{3 c_1^2 - 2 c_2}{4 \betazero}
\hat{\gamma}^{(0) \, T} - \f{c_1^2}{8 \betazero^2} \big (
\hat{\gamma}^{(0) \, T} \big )^2 \right ] \, , \\ 
\hat{L}^{-1} (\muh) & = \hat{1} - \f{\as (\muh)}{4 \pi} \f{c_1}{2
\betazero} \hat{\gamma}^{(0) \, T} + \left ( \f{\as (\muh)}{4 \pi}
\right )^2 \left [ \f{3 c_1^2 - 2 c_2}{4 \betazero} \hat{\gamma}^{(0)
\, T} + \f{c_1^2}{8 \betazero^2} \big ( \hat{\gamma}^{(0) \, T} \big
)^2 \right ] \, .  
\end{split}
\eeq
Taking into account \Eq{eq:fullamplitudechange} and expanding the
double primed Wilson coefficient $\vec{C}'' (\mul)$ in terms of the
unprimed coupling parameter leads to  
\beq \label{eq:twodoubleprimedwilsoncoefficient}
\begin{split}
\vec{C}'' (\mul) & = \hat{K}'' (\mul) \hspace{0.25mm} \hat{L} (\mul) 
\hspace{0.25mm} \hat{U}^{(0)} (\mul, \muh) \hspace{0.25mm}
\hat{L}^{-1} (\muh) \Bigg ( \vec{A}^{(0)} + \f{\as (\muh)}{4 \pi}
\left [ \vec{A}^{(1)} - \hat{R}''^{(1)} \vec{A}^{(0)} \right ] \\    
& + \left ( \f{\as (\muh)}{4 \pi} \right )^2 \left [ \vec{A}^{(2)} -
\hat{R}''^{(1)} \left ( \vec{A}^{(1)} - c_1 \vec{A}^{(0)} \right ) -
\left (\hat{R}''^{(2)} - \big ( \hat{R}''^{(1)} \big )^2 \right )
\vec{A}^{(0)} \right  ] \Bigg ) \, . 
\end{split}
\eeq
Inserting \Eqs{eq:jdoubleprimes}, \eq{eq:rdoubleprimes},
\eq{eq:doubleprimedkmatrix} and \eq{eq:Lmatrices} into the latter 
equation we finally obtain  
\beq \label{eq:threedoubleprimedwilsoncoefficient}
\begin{split}
\vec{C}'' (\mul) & = \hat{K} (\mul) \hspace{0.25mm} \hat{U}^{(0)}
(\mul, \muh) \Bigg ( \vec{A}^{(0)} + \f{\as (\muh)}{4 \pi} \left [
\vec{A}^{(1)} - \hat{R}^{(1)} \vec{A}^{(0)} \right ] \\    
& + \left ( \f{\as (\muh)}{4 \pi} \right )^2 \left [ \vec{A}^{(2)} -
\hat{R}^{(1)} \vec{A}^{(1)} - \left ( \hat{R}^{(2)} - \big (
\hat{R}^{(1)} \big )^2 \right ) \vec{A}^{(0)} \right  ] \Bigg ) \, , 
\end{split}
\eeq
which shows that $\vec{C}'' (\mul)$ is nothing but $\vec{C}
(\mul)$. Since $\vec{C} (\mul)$ does not depend on the scheme used to 
renormalize the high scale coupling constant, the same is obviously
true for $\vec{C}'' (\mul)$. However, $\vec{C}'' (\mul)$ still
depends  on the renormalization scheme through $\hat{K} (\mul)$ 
and consequently on $\hat{J}^{(1)}$ and $\hat{J}^{(2)}$, entering the 
Wilson coefficients to the left of $\hat{U}^{(0)} (\mul, \muh)$. As is
evident from \Eqsand{eq:kmatrices}{eq:aeff}, and has already been
discussed before, this dependence on the lower end of the evolution is
canceled by the one of the matrix elements $\langle \vec{Q}^T (\mul) 
\rangle$. We have therefore explicitly seen that the matrix elements
of the effective hamiltonian and the resulting physical amplitudes are
in fact invariant under a change of coupling constant. 

\section{Change of Operator Basis}
\label{sec:changeofoperatorbasis}

In $n = 4$ dimensions, a change of the physical operators is always
equivalent to a simple linear transformation  
\beq \label{eq:lineartransformation}
\vec{\Qburas} = \hat{R} \hspace{0.4mm} \vec{\Qmisiak} \, ,  
\eeq
parameterized by a rotation matrix $\hat{R}$, which affects the
renormalization constants and the ADM in a trivial way:   
\beq \label{eq:rchange}
\hat{Z}' = \hat{R} \hspace{0.4mm} \hat{Z} \hspace{0.4mm} \hat{R}^{-1}
\, , \hspace{5mm} \text{and} \hspace{5mm} \hat{\gamma}' = \hat{R}
\hspace{0.2mm} \hat{\gamma} \hspace{0.2mm} \hat{R}^{-1} \, . 
\eeq

In the framework of dimensional regularization, the transformation 
corresponding to the change of basis turns out to be more complicated,
as it generally involves evanescent operators as well. This feature
basically reflects the fact that in order to formulate consistently
the dimensional regularization of a theory containing fermionic
degrees of freedom, the Dirac algebra has to be infinite-dimensional,
which implies that evanescent operators are necessary to form a
complete basis in $n = 4 - 2 \eps$ dimensions. In consequence,
specifying the evanescent operators is necessary to make precise the
definition of the $\MSbar$ scheme in the effective theory beyond
LO, as can been seen for instance in
\Eq{eq:physicalgammaexpansion}. Clearly, EOM-vanishing operators are 
irrelevant to the present discussion.    

As long as the change of basis does not mix physical and evanescent
operators, the ADM still changes in a trivial way. In particular, a
linear transformation of evanescent operators does not affect the
physical ADM  at all. However, when the change of basis involves
linear combinations of evanescent and physical operators, the
situation turns out to be more complicated
\cite{Chetyrkin:1997gb}. Indeed, as we will explain in a moment, the  
new ADM is still given by \Eq{eq:rchange}, but the presence of
evanescent operators induces a finite renormalization constant for the
physical operators in the new basis. In order to restore the standard
$\MSbar$ scheme definitions, a change of scheme is therefore required.

Let us first consider a change of basis that consists of adding
some evanescent operators to the physical ones, 
\beq \label{eq:wchange}
\vec{\Qburas} = \vec{\Qmisiak} + \hat{W} \vec{\Emisiak} \, ,   
\eeq
parameterized by the matrix $\hat{W}$. In this case the new ADM is
still given by \Eq{eq:rchange} because of the absence of mixing of
evanescent into physical operators in the original basis. However,
after the above transformation, the renormalization matrix
corresponding to the physical operators in the new basis will contain
a finite, non-vanishing contribution   
\beq \label{eq:wzchange}
\hat{Z}'^{(1, 0)}_{\QQ}= \hat{W} \hat{Z}^{(1, 0)}_{\EQ} \, ,  
\eeq
where the subscript $\Qmisiak$ and $\Emisiak$ denotes an element of
the physical and evanescent operators, respectively. In order to
re-impose the standard $\MSbar$ conditions, the latter contribution
must be removed by a change of scheme, implemented by
\Eq{eq:finiterenormalization}.   

The situation is very similar for a change of basis that consists of
adding multiples of $\epsilon$ times physical operators to the
evanescent ones     
\beq \label{eq:uchange}
\vec{\Eburas} = \vec{\Emisiak} + \eps \, \hat{U} \vec{\Qmisiak} \, , 
\eeq
parameterized by the matrix $\hat{U}$. In this case the ADM is
unchanged because of its finiteness. However, the renormalization
matrix of the physical operators in the new basis will contain a
finite, non-vanishing contribution as well:    
\beq \label{eq:uzchange}
\hat{Z}'^{(1, 0)}_{\QQ}= -\hat{Z}^{(1, 1)}_{\QE} \hat{U} \, . 
\eeq
Needless to say, the above contribution must again be removed by a
suitable change of scheme, in order to abide by the standard $\MSbar$   
renormalization conditions.     
 
We therefore conclude in full generality that a change of basis 
in dimensional regularization is equivalent to a rotation plus a
change of scheme. 
If we discount possible $\mu$-dependent rotations of the operator
basis, it should be clear from the discussion above that the most
general change of basis comprises the three linear transformations
of \Eqs{eq:lineartransformation}, \eq{eq:wchange}, and
\eq{eq:uchange}, as well as a rotation of the evanescent operators,
which will be parameterized by the matrix $\hat{M}$ in what
follows. In total we thus have   
\beq \label{eq:transformations}
\vec{\Qburas} = \hat{R} \left ( \vec{\Qmisiak} + \hat{W}
\vec{\Emisiak} \right ) \, , \hspace{0.5cm} \text{and} \hspace{0.5cm}
\vec{\Eburas} = \hat{M} \left ( \eps \, \hat{U} \vec{\Qmisiak} + \left
[ \hat{1} + \eps \, \hat{U} \hat{W} \right ] \vec{\Emisiak} \right )
\, .       
\eeq
The corresponding residual finite renormalization can be derived with 
simple algebra. Up to second order in $\as$ we find 
\begin{align} \label{eq:zprime}
\hat{Z}'^{(1, 0)}_{\QQ} & = \hat{R} \left [ \hat{W} \hat{Z}^{(1,
0)}_{\EQ} -\left ( \hat{Z}^{(1, 1)}_{\QE} + \hat{W} \hat{Z}^{(1,
1)}_{\EE} - \f{1}{2} \hat{\gamma}^{(0)} \hat{W} \right ) \hat{U}
\right ] \hat{R}^{-1} \, , \non \\
\hat{Z}'^{(2, 0)}_{\QQ} & = \hat{R} \left [ \hat{W} \hat{Z}^{(2,
0)}_{\EQ} - \left ( \hat{Z}^{(2, 1)}_{\QE} + \hat{W} \hat{Z}^{(2,
1)}_{\EE}  - \f{1}{4} \hat{\gamma}^{(1)} \hat{W} - \f{1}{2}
\hat{Z}^{(1, 1)}_{\QE} \hat{Z}^{(1, 0)}_{\EQ} \hat{W}
\right. \right. \non \\  
& \left. \left. - \f{1}{2} \hat{W} \hat{Z}^{(1, 1)}_{\EE} \hat{Z}^{(1,
0)}_{\EQ} \hat{W} - \f{1}{4} \hat{W} \hat{Z}^{(1, 0)}_{\EQ}
\hat{\gamma}^{(0)} \hat{W} + \f{1}{2} \betazero \hat{W} \hat{Z}^{(1, 
0)}_{\EQ} \hat{W} \right ) \hat{U} \right ] \hat{R}^{-1} \, .      
\end{align}

With these expressions at hand, it is now straightforward to deduce
how the ADM and the initial conditions of the Wilson coefficients
transforms under the change of basis as given in
\Eq{eq:transformations}. Up to the NNLO we obtain      
\beq \label{eq:admtransformations}
\begin{split}
\hat{\gamma}'^{(0)} & = \hat{R} \hspace{0.2mm} \hat{\gamma}^{(0)}
\hspace{0.2mm} \hat{R}^{-1} \, , \\[2mm]
\hat{\gamma}'^{(1)} & = \hat{R} \hspace{0.2mm} \hat{\gamma}^{(1)}
\hspace{0.2mm} \hat{R}^{-1} - \left [ \hat{Z}'^{(1, 0)}_{\QQ},
\hat{\gamma}'^{(0)} \right ] - 2 \betazero \hat{Z}'^{(1,0)}_{\QQ} \, ,
\\[2mm] 
\hat{\gamma}'^{(2)} & = \hat{R} \hat{\gamma}^{(2)} \hat{R}^{-1} -
\left [ \hat{Z}'^{(2, 0)}_{\QQ}, \hat{\gamma}'^{(0)} \right ] -
\left [ \hat{Z}'^{(1, 0)}_{\QQ}, \hat{\gamma}'^{(1)} \right ] + 
\left [ \hat{Z}'^{(1, 0)}_{\QQ}, \hat{\gamma}'^{(0)} \right ]
\hat{Z}'^{(1, 0)}_{\QQ} \\  
& - 4 \betazero \hat{Z}'^{(2, 0)}_{\QQ} - 2 \betaone \hat{Z}'^{(1,
0)}_{\QQ} + 2 \betazero \big ( \hat{Z}'^{(1, 0)}_{\QQ} \big )^2 \, , 
\end{split}
\eeq
and 
\beq \label{eq:cprime}
\vec{C}' (\MW) = \left [ \hat{1} + \f{\as (\MW)}{4 \pi} \hat{Z}'^{(1, 
0)}_{\QQ} + \left ( \f{\as (\MW)}{4 \pi} \right )^2 \hat{Z}'^{(2,
0)}_{\QQ} \right ]^T \! \! ( \hat{R}^{-1} \big )^T \vec{C} (\MW) \, .
\eeq 
For what concerns the NLO parts of
\Eqs{eq:zprime}, \eq{eq:admtransformations} and \eq{eq:cprime} our
findings resemble the formulas derived in \cite{Chetyrkin:1997gb}, if
one takes into account that our definition of $\hat{Z}'^{(1,
0)}_{\QQ}$ differs slightly from the residual finite renormalization
matrix used in the latter article. On the other hand, the complete
NNLO relations \Eqs{eq:zprime}, \eq{eq:admtransformations} and
\eq{eq:cprime} have never been presented before.   

After these general considerations, let us discuss in some detail how
the \ads \ given in \Eqs{eq:gamma0new}, \eq{eq:gamma1new} and
\eq{eq:gamma2new} are transformed in going to the ``traditional''
basis of physical operators \cite{Buras:1998ra,  Ciuchini:1993vr,
Buras:1992tc}      
\beq \label{eq:physicaloperatorsburas} 
\begin{split}
\Qburas_1 & = (\bar{s}^\alpha_L \gammadown{\mu_1} c^\beta_L)
(\bar{c}^\beta_L \gammaup{\mu_1} b^\alpha_L) \, , \\  
\Qburas_2 & = (\bar{s}^\alpha_L \gammadown{\mu_1} c^\alpha_L) 
(\bar{c}^\beta_L \gammaup{\mu_1} b^\beta_L) \, , \\      
\Qburas_3 & = (\bar{s}^\alpha_L \gammadown{\mu_1} b^\alpha_L)
\sum\nolimits_q (\bar{q}^\beta_L \gammaup{\mu_1}
q^\beta_L) \, , \\   
\Qburas_4 & = (\bar{s}^\alpha_L \gammadown{\mu_1} b^\beta_L) 
\sum\nolimits_q (\bar{q}^\beta_L \gammaup{\mu_1} 
q^\alpha_L) \, , \\    
\Qburas_5 & = (\bar{s}^\alpha_L \gammadown{\mu_1} b^\alpha_L)
\sum\nolimits_q (\bar{q}^\beta_R \gammaup{\mu_1}
q^\beta_R) \, , \\    
\Qburas_6 & = (\bar{s}^\alpha_L \gammadown{\mu_1} b^\beta_L) 
\sum\nolimits_q (\bar{q}^\beta_R \gammaup{\mu_1} 
q^\alpha_R) \, .   
\end{split}
\eeq
In the above definitions $\alpha$ and $\beta$ denote color indices.  

The one- and two-loop evanescent operators that accompany the
``traditional'' basis can be found by imposing the requirements given
in \cite{Buras:1992tc}. At the one-loop level they are    
\begin{align} \label{eq:oneevanescentoperatorburas}
\Eburas^{(1)}_1 & = (\bar{s}^\alpha_L \gammadown{\mu_1 \mu_2 \mu_3}
c^\beta_L) (\bar{c}^\beta_L \gammaup{\mu_1 \mu_2 \mu_3} b^\alpha_L) -
( 16 - 4 \eps ) \, \Qburas_1 \, , \non \\ 
\Eburas^{(1)}_2 & = (\bar{s}^\alpha_L \gammadown{\mu_1 \mu_2 \mu_3}
c^\alpha_L) (\bar{c}^\beta_L \gammaup{\mu_1 \mu_2 \mu_3} b^\beta_L) -
( 16 - 4 \eps ) \, \Qburas_2 \, , \non \\
\Eburas^{(1)}_3 & = (\bar{s}^\alpha_L \gammadown{\mu_1 \mu_2 \mu_3}
b^\alpha_L) \sum\nolimits_q (\bar{q}^\beta_L
\gammaup{\mu_1 \mu_2 \mu_3} q^\beta_L) - ( 16 - 4 \eps ) \, \Qburas_3
\, , \non \\ 
\Eburas^{(1)}_4 & = (\bar{s}^\alpha_L \gammadown{\mu_1 \mu_2 \mu_3}
b^\beta_L) \sum\nolimits_q (\bar{q}^\beta_L
\gammaup{\mu_1 \mu_2 \mu_3} q^\alpha_L) - ( 16 - 4 \eps ) \, \Qburas_4
\, , \non \\  
\Eburas^{(1)}_5 & = (\bar{s}^\alpha_L \gammadown{\mu_1 \mu_2 \mu_3}
b^\alpha_L) \sum\nolimits_q (\bar{q}^\beta_R
\gammaup{\mu_1 \mu_2 \mu_3} q^\beta_R) - ( 4 + 4 \eps ) \, \Qburas_5
\, , \non \\ 
\Eburas^{(1)}_6 & = (\bar{s}^\alpha_L \gammadown{\mu_1 \mu_2 \mu_3}
b^\beta_L) \sum\nolimits_q (\bar{q}^\beta_R
\gammaup{\mu_1 \mu_2 \mu_3} q^\alpha_R) - ( 4 + 4 \eps ) \, \Qburas_6
\, . 
\end{align}
Following the same procedure, we find the following two-loop
evanescent operators:     
\beq \label{eq:twoevanescentoperatorburas}
\begin{split}
\Eburas^{(2)}_1 & = (\bar{s}^\alpha_L \gammadown{\mu_1 \mu_2 \mu_3 \mu_4
\mu_5} c^\beta_L) (\bar{c}^\beta_L \gammaup{\mu_1 \mu_2 \mu_3 \mu_4
\mu_5} b^\alpha_L) - ( 256 - 224 \eps ) \, \Qburas_1 \, , \\ 
\Eburas^{(2)}_2 & = (\bar{s}^\alpha_L \gammadown{\mu_1 \mu_2 \mu_3 \mu_4
\mu_5} c^\alpha_L) (\bar{c}^\beta_L \gammaup{\mu_1 \mu_2 \mu_3 \mu_4
\mu_5} b^\beta_L) - ( 256 - 224 \eps ) \, \Qburas_2 \, , \\  
\Eburas^{(2)}_3 & = (\bar{s}^\alpha_L \gammadown{\mu_1 \mu_2 \mu_3 \mu_4
\mu_5} b^\alpha_L) \sum\nolimits_q (\bar{q}^\beta_L
\gammaup{\mu_1 \mu_2 \mu_3 \mu_4 \mu_5} q^\beta_L) - ( 256 - 224 \eps
) \, \Qburas_3 \, , \\  
\Eburas^{(2)}_4 & = (\bar{s}^\alpha_L \gammadown{\mu_1 \mu_2 \mu_3 \mu_4
\mu_5} b^\beta_L) \sum\nolimits_q (\bar{q}^\beta_L
\gammaup{\mu_1 \mu_2 \mu_3 \mu_4 \mu_5}  q^\alpha_L) - ( 256 - 224
\eps ) \, \Qburas_4 \, , \\  
\Eburas^{(2)}_5 & = (\bar{s}^\alpha_L \gammadown{\mu_1 \mu_2 \mu_3 \mu_4
\mu_5} b^\alpha_L) \sum\nolimits_q (\bar{q}^\beta_R
\gammaup{\mu_1 \mu_2 \mu_3 \mu_4 \mu_5} q^\beta_R) - ( 16 + 128 \eps )
\, \Qburas_5 \, , \\   
\Eburas^{(2)}_6 & = (\bar{s}^\alpha_L \gammadown{\mu_1 \mu_2 \mu_3 \mu_4
\mu_5} b^\beta_L) \sum\nolimits_q (\bar{q}^\beta_R
\gammaup{\mu_1 \mu_2 \mu_3 \mu_4 \mu_5} q^\alpha_R) - ( 16 + 128 \eps )
\, \Qburas_6 \, . 
\end{split}
\eeq

It turns out that in order to transform the ADM given in
\Eqs{eq:gamma0new}, \eq{eq:gamma1new} and \eq{eq:gamma2new} 
from the initial set of operators to the ``traditional'' basis, we
have to introduce four additional one-loop evanescent operators    
\beq \label{eq:additionaloneevanescentoperatormisiak}
\begin{split} 
\Emisiak^{(1)}_5 & = (\bar{s}_L \gammadown{\mu_1} b_L) 
\sum\nolimits_q (\bar{q} \gammaup{\mu_1} \gamma_5 q) -
\f{5}{3} Q_3 + \f{1}{6} Q_5 \, , \\ 
\Emisiak^{(1)}_6 & = (\bar{s}_L \gammadown{\mu_1} T^a b_L)
\sum\nolimits_q (\bar{q} \gammaup{\mu_1} \gamma_5 T^a
q) - \f{5}{3} Q_4 + \f{1}{6} Q_6 \, , \\  
\Emisiak^{(1)}_7 & = (\bar{s}_L \gammadown{\mu_1 \mu_2 \mu_3} b_L) 
\sum\nolimits_q (\bar{q}  \gammaup{\mu_1 \mu_2 \mu_3}
\gamma_5 q) - \f{32}{3} Q_3 + \f{5}{3} Q_5 \comment{- \f{5}{3}
\Emisiak^{(1)}_5} \, , \\     
\Emisiak^{(1)}_8 & = (\bar{s}_L \gammadown{\mu_1 \mu_2 \mu_3} T^a
b_L) \sum\nolimits_q (\bar{q} \gammaup{\mu_1 \mu_2
\mu_3} \gamma_5 T^a q) - \f{32}{3} Q_4 + \f{5}{3} Q_6 \comment{-
\f{5}{3} \Emisiak^{(1)}_6}  \, , 
\end{split}
\eeq
as well as four additional two-loop evanescent operators
\beq \label{eq:additionaltwoevanescentoperatormisiak} 
\begin{split}
\Emisiak^{(2)}_5 & = (\bar{s}_L \gammadown{\mu_1 \mu_2 \mu_3 \mu_4
\mu_5} b_L) \sum\nolimits_q (\bar{q} \gammaup{\mu_1
\mu_2 \mu_3 \mu_4 \mu_5} \gamma_5 q) - \f{320}{3} Q_3 + \f{68}{3} Q_5 
\comment{- \f{68}{3} \Emisiak^{(1)}_5 \, , \non} \, , \\  
\Emisiak^{(2)}_6 & = (\bar{s}_L \gammadown{\mu_1 \mu_2 \mu_3 \mu_4
\mu_5} T^a b_L) \sum\nolimits_q (\bar{q}
\gammaup{\mu_1 \mu_2 \mu_3 \mu_4 \mu_5}  \gamma_5 T^a q) - \f{320}{3}
Q_4 + \f{68}{3} Q_6 \comment{- \f{68}{3} \Emisiak^{(1)}_6} \, , \\   
\Emisiak^{(2)}_7 & = (\bar{s}_L \gammadown{\mu_1 \mu_2 \mu_3 \mu_4
\mu_5 \mu_6 \mu_7} b_L) \sum\nolimits_q (\bar{q}
\gammaup{\mu_1 \mu_2 \mu_3 \mu_4 \mu_5 \mu_6 \mu_7} \gamma_5 q) -
\f{4352}{3} Q_3 + \f{1040}{3} Q_5 \comment{- \f{1040}{3}
\Emisiak^{(1)}_5 \, , \non} \, , \\   
\Emisiak^{(2)}_8 & = (\bar{s}_L \gammadown{\mu_1 \mu_2 \mu_3 \mu_4
\mu_5 \mu_6 \mu_7} T^a b_L) \sum\nolimits_q (\bar{q}
\gammaup{\mu_1 \mu_2 \mu_3 \mu_4 \mu_5 \mu_6 \mu_7} T^a \gamma_5 q) -
\f{4352}{3} Q_4 + \f{1040}{3} Q_6 \comment{- \f{1040}{3}
\Emisiak^{(1)}_6 \, . \non} \, . 
\end{split}
\eeq
It should be clear, that the evanescent operators
$\Emisiak^{(1)}_5$--$\Emisiak^{(1)}_8$ and 
$\Emisiak^{(2)}_5$--$\Emisiak^{(2)}_8$ are not needed as counterterms 
in the initial basis of operators. However, some linear combinations
of them will become parts of either the physical or the evanescent
operators in the ``traditional'' basis through the change of basis
given by \Eq{eq:transformations}. 

At this point a comment concerning the computation of the
renormalization constants involving the insertions of the additional
evanescent operators is in order. Transforming the three-loop \ads \
from the initial to the ``traditional'' basis requires the knowledge
of one- and two-loop diagrams with insertions of
$\Emisiak^{(1)}_5$--$\Emisiak^{(1)}_8$, which introduces traces with  
$\gamma_5$ into the calculation. In this context we follow
\cite{Misiak:bc}, and avoid anticommutation of $\gamma_5$ in any
fermionic line containing an odd number of $\gamma_5$. Moreover, while
traces containing an odd number of Dirac matrices and a single
$\gamma_5$ do not pose any problem as they vanish algebraically, we do
not evaluate traces containing an even number of Dirac matrices and a
single $\gamma_5$ in $n = 4 - 2 \eps$ dimensions. This brings to life
new evanescent operators --- we will call them trace evanescent in the
following --- that can in general be written as a contraction of a
suitable Dirac structure with one of the following evanescent tensors
\beq \label{eq:evanescenttensors}
\widehat{E}_{\mu_1 \ldots \, \mu_m} =  \mbox{Tr} \left (
\gammadown{\mu_1 \ldots \, \mu_m} \gamma_5 \right ) - 
\widetilde{\mbox{Tr}} \left ( \gammadown{\mu_1 \ldots \, \mu_m} 
\gamma_5 \right ) \, ,  
\eeq
where the four-dimensional traces $\widetilde{\mbox{Tr}} \left (
\gammadown{\mu_1 \ldots \, \mu_m} \gamma_5 \right )$ can be calculated
recursively from the initial value $\widetilde{\mbox{Tr}} \left (
\gamma_5 \right ) = 0$ applying        
\bea \label{eq:fourdimensionalfivetrace}
\widetilde{\mbox{Tr}} \left ( \gammadown{\mu_1 \ldots \, \mu_m} 
\gamma_5 \right ) = \sum_{j = 2}^m \left ( -1 \right )^j
\tilde{g}^{\mu_1 \mu_j} \, \widetilde{\mbox{Tr}} \left (
\gammadown{\mu_2 \ldots \, \mu_{j - 1}  \mu_{j + 1} \ldots \, \mu_m}
\gamma_5  \right ) - \f{i}{6} \eps^{\mu_1 \nu_1 \nu_2 \nu_3} \,
\widetilde{\mbox{Tr}} \left ( \gammadown{\mu_2 \ldots \, \mu_m \nu_1  
\nu_2 \nu_3} \right ) \, . \hspace{1mm}
\eea
Here $\tilde{g}^{\mu_1 \mu_2} \equiv \text{diag} \, (1, -1, -1,
-1)$ denotes the four-dimensional metric tensor, and $\eps^{\mu_1 
\mu_2 \mu_3 \mu_4}$ is the totally antisymmetric Levi-Civita tensor
defined so that $\eps^{0 1 2 3} \equiv 1$. Furthermore, the second
trace in the above equation should also be taken in $n = 4$
dimensions. It can thus be computed recursively from
$\widetilde{\mbox{Tr}} (1) \equiv 4$ using      
\beq \label{eq:fourdimensionaltrace}
\widetilde{\mbox{Tr}} \left ( \gammadown{\mu_1 \ldots \, \mu_m} \right
) = \sum_{j = 2}^m \left ( -1 \right )^j \tilde{g}^{\mu_1 \mu_j} \,
\widetilde{\mbox{Tr}} \left ( \gammadown{\mu_2 \ldots \, \mu_{j - 1}
\mu_{j + 1} \ldots \, \mu_m} \right ) \, .  
\eeq

{%
\begin{figure}[t]
\begin{center}
\vspace{-0.5cm}%
\scalebox{1.8}{\begin{picture}(55,45)(0,-5)\input{gammafive1.tex}\end{picture}}%
\hspace{5mm}%
\scalebox{1.8}{\begin{picture}(55,45)(0,-5)\input{gammafive2.tex}\end{picture}}%
\hspace{5mm}%
\scalebox{1.8}{\begin{picture}(55,45)(0,-5)\input{gammafive3.tex}\end{picture}}%
\hspace{5mm}%
\scalebox{1.8}{\begin{picture}(55,45)(0,-5)\input{gammafive4.tex}\end{picture}}%
\vspace{0.5cm}%
\caption{Typical examples of two-loop 1PI diagrams with an insertion
of $E^{(5)}_1$ involving non-trivial Dirac traces containing $\gamma_5$.}      
\label{fig:gammafive}
\end{center}
\end{figure}
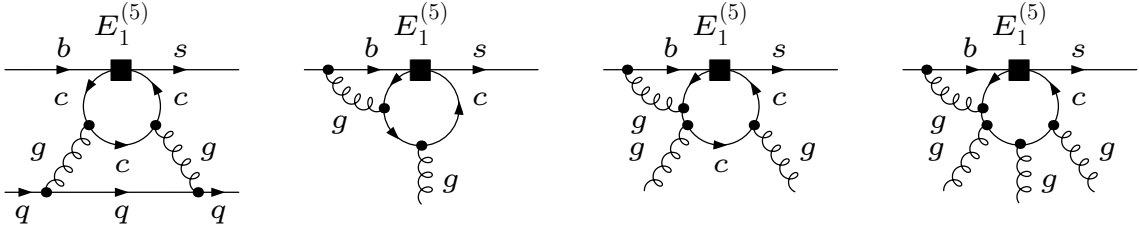
}%

Apparently, the trace evanescent operators originating from
\Eq{eq:evanescenttensors} have to be treated on the same footing as
the regular ones introduced earlier on. The idea of introducing more
and more evanescent operators seems to make the use of an naive
anticommuting $\gamma_5$ in multi-loop calculations involving chiral
operators futile. Fortunately, for the problem at hand this is not the
case, as it turns out that the one-loop insertions of $E_5^{(1)}$ and
$E_6^{(1)}$ needed to find the transformation of the two-loop \ads \
between the initial and the ``traditional'' basis involve only the
trace $\mbox{Tr} (\gammadown{\mu_1 \mu_2} \gamma_5)$, which however is
zero. This observation has been made already in
\cite{Chetyrkin:1997gb}. Furthermore, to find the transformation of  
the three-loop \ads \, matrices, both one-loop insertions of
$E_5^{(1)}$--$E_8^{(1)}$ and two-loop insertions of $E_5^{(1)}$ and
$E_6^{(1)}$ are required. Typical examples of non-vanishing two-loop
1PI diagrams are shown in \Fig{fig:gammafive}. However, also in this
case the number of new evanescent structures is rather small, since
the necessary operator insertions introduce only the non-trivial trace
$\mbox{Tr} (\gammadown{\mu_1 \mu_2 \mu_3 \mu_4} \gamma_5)$. The
complete list of trace evanescent operators relevant to find the
transformation of the three-loop \ads \ between the initial and the
``traditional'' basis is given in \App{app:traceevanescentoperators}.

If possible any regularization prescription should respect all
symmetries of the bare theory, such as gauge and BRST invariance
encoded in the Ward and Slavnov-Taylor identities, or Bose
symmetry. It is interesting to note that all this requirements are not
necessarily fulfilled for an arbitrary choice of trace evanescent
tensors. For example, adding any multiple of $\eps$ times the
four-dimensional traces $\widetilde{\mbox{Tr}} \left (
\gammadown{\mu_1 \ldots \, \mu_m} \gamma_5 \right )$ to the right hand
side of \Eq{eq:evanescenttensors}, in general, spoils the usual
Ward and Slavnov-Taylor identities as well as the Bose symmetry of
some of the resulting 1PI Green's functions even after the correct
subtraction of all subdivergences. In order to verify that neither
problem arises with the definition of trace evanescent operators
adopted in
\Eqsand{eq:onelooptraceevanescentoperators}{eq:twolooptraceevanescentoperators},
we have calculated the pole and finite parts of the off-shell
$\btosccbar$, $\btos$, $\btosgluon$, $\btosgluongluon$ and 
$\btosgluongluongluon$ matrix elements with one-loop insertions of
$E_5^{(1)}$--$E_8^{(1)}$ and two-loop insertions of
$E_5^{(1)}$--$E_6^{(1)}$, and checked explicitly that $i)$ the
resulting operator renormalization constants are independent of the 
external states used in the calculation, and that $ii)$ the
subtracted 1PI Green's functions with two and three external gluons
are cyclic under the interchange of any two gluons as required by Bose
symmetry. Concerning the former issue, let us mention, that in order 
to decompose the finite parts of the two-loop matrix elements of
$E_5^{(1)}$--$E_6^{(1)}$ corresponding to the $\btosgluongluon$ and 
$\btosgluongluongluon$ transitions, one has to enlarge the off-shell
operator basis to contain besides the EOM-vanishing operators
$N_1^{(1)}$ and $N_1^{(2)}$--$N_{10}^{(2)}$ the following four gauge
non-invariant operators of dimension-six:  
\beq \label{eq:anomalousoperators}
\begin{split}
A_1^{(2)} & = \eps_{\mu_1 \mu_2 \mu_3 \mu_4} G^{a \mu_2}
\partial^{\mu_3} G^{a \mu_4} ( \bar{s}_ L \gammaup{\mu_1} b_L ) \, ,
\\     
A_2^{(2)} & = \eps_{\mu_1 \mu_2 \mu_3 \mu_4} d^{a b c} G^{b \mu_2}
\partial^{\mu_3} G^{c \mu_4} ( \bar{s}_ L \gammaup{\mu_1} T^a b_L ) \,
, \\       
A_3^{(2)} & = \eps_{\mu_1 \mu_2 \mu_3 \mu_4} f^{a b c} G^{a \mu_2}
G^{b \mu_3} G^{c \mu_4} ( \bar{s}_ L \gammaup{\mu_1} b_L ) \, , \\
A_4^{(2)} & = \eps_{\mu_1 \mu_2 \mu_3 \mu_4} \left ( d^{a b e} f^{c d
e} + d^{a c e} f^{d b e} + d^{a d e} f^{b c e} \right ) G^{a \mu_2}
G^{b \mu_3} G^{c \mu_4} ( \bar{s}_ L \gammaup{\mu_1} T^a b_L ) \, . 
\end{split}
\eeq
Of course the finite two-loop renormalization between
$E_5^{(1)}$--$E_6^{(1)}$ and the anomalous two- and three-gluon
operators $A_1^{(2)}$--$A_4^{(2)}$ does not affect the residual finite
renormalization matrices $\hat{Z}'^{(1, 0)}_{\QQ}$ and $\hat{Z}'^{(2, 
0)}_{\QQ}$ defined in \Eq{eq:zprime} which are unambiguously fixed by
the corresponding one- and two-loop on-shell $\btosccbar$ and 
$\btosgluon$ matrix elements. In consequence, the presence of the
gauge non-invariant operators $A_1^{(2)}$--$A_4^{(2)}$ leaves the
final result for the \ads \ of the four-quark operators $Q_1$--$Q_6$
in the ``traditional'' scheme unaltered up to the three-loop level.

In case the above considerations might not fully convince a suspicious
reader that our treatment of traces containing an odd number of Dirac
matrices and a single $\gamma_5$ is consistent, it may be worthwhile
to justify it in another independent way. To exclude all possibility
of doubt concerning our regularization scheme, we have computed the
renormalization constants involving the insertions of
$E_5^{(1)}$--$E_8^{(1)}$ using the HV definition of $\gamma_5$ in $n =
4 - 2 \eps$ dimensions, and verified explicitly that, for what
concerns the non-anomalous operators, the latter results coincide with
those obtained by the ``dimensional reduction''-like treatment of
traces containing $\gamma_5$, implemented by
\Eqs{eq:evanescenttensors}, \eq{eq:fourdimensionalfivetrace} and
\eq{eq:fourdimensionaltrace}. For the sake of completeness, let us
also mention, that in order to decompose the finite parts of the
two-loop $\btosgluongluon$ and $\btosgluongluongluon$ matrix elements
with insertions of $E_5^{(1)}$ and $E_6^{(1)}$ computed in the HV
scheme, not four but eight anomalous operators are required. Of
course, the appearance of these additional gauge non-invariant
operators, which correspond to $A_1^{(2)}$--$A_4^{(2)}$ with right
instead of left chiral quark fields, does not alter the final result
for the three-loop $\ord (\as^3)$ \ads \ in the ``traditional''
scheme. 

As regards the relevance of the trace evanescent operators defined in
\Eqsand{eq:onelooptraceevanescentoperators}{eq:twolooptraceevanescentoperators},
let us finally point out, that it would be incorrect to conclude that
they are only needed to find the transformation of the three-loop \ads
\, matrices between the initial and the ``traditional'' basis of  
operators. In fact, their role in the ``traditional'' basis is the
same as the one of the evanescent operators introduced in
\Eq{eq:twoevanescentoperatorburas}, as they, together with the latter
ones, make precise the definition of the renormalization scheme in
that basis at the NNLO. As should be clear from what has been said in
the last section on the cancellation of scheme dependences in general,
the corresponding scheme dependence is canceled by the one of the
two-loop $\ord (\as^2)$ matrix elements, which have to be calculated
using exactly the same definition of trace evanescent operators.     

The renormalization constant matrices entering \Eq{eq:zprime} are
found from one- and two-loop matrix elements of physical and
evanescent operators. We give the relevant ones, as well as the
matrices characterizing the change of basis in
\App{app:changetothetraditionaloperatorbasis}. Our final results for the
residual finite renormalization read  
\beq \label{eq:zprime10explicit}
\hat{Z}'^{(1, 0)}_{\QQ} =
\left (
\begin{array}{cccccc}
{\scriptstyle -\f{7}{3}} & -1 & 0 & 0 & 0 & 0 \\ 
-2 & {\scriptstyle \f{2}{3}} & 0 & 0 & 0 & 0 \\ 
0 & 0 & {\scriptstyle \f{178}{27}} & {\scriptstyle -\f{34}{9}} &
{\scriptstyle -\f{164}{27}} & {\scriptstyle \f{20}{9}} \\  
0 & 0 & {\scriptstyle 1 - \f{1}{9} \nf} & {\scriptstyle -\f{25}{3} + 
\f{1}{3} \nf} & {\scriptstyle -2 - \f{1}{9} \nf} & {\scriptstyle 6 +
\f{1}{3} \nf} \\    
0 & 0 & {\scriptstyle -\f{160}{27}} & {\scriptstyle \f{16}{9}} &
{\scriptstyle \f{146}{27}} & {\scriptstyle -\f{2}{9}} \\  
0 & 0 & {\scriptstyle -2 + \f{1}{9} \nf} & {\scriptstyle 6 - \f{1}{3}
\nf} & {\scriptstyle 3 + \f{1}{9} \nf} & {\scriptstyle -\f{11}{3} -
\f{1}{3} \nf}  
\end{array}
\right ) \, , 
\eeq
and
\beq \label{eq:zprime20explicit}
\hat{Z}'^{(2, 0)}_{\QQ} = 
\left ( 
\begin{array}{cccccc}
{\scriptstyle -\f{200}{9} + \f{7}{54} \nf} & {\scriptstyle \f{68}{3} +
\f{1}{18} \nf} & {\scriptstyle -\f{4}{3}} & 0 & {\scriptstyle
\f{4}{3}} & 0 \\
{\scriptstyle -\f{7}{4} + \f{1}{9} \nf} & {\scriptstyle \f{397}{36} -
\f{1}{27} \nf} & {\scriptstyle -\f{77}{162}} & {\scriptstyle
-\f{35}{54}} & {\scriptstyle \f{77}{162}} & {\scriptstyle \f{35}{54}}
\\   
0 & 0 & {\scriptstyle \f{57253}{2916} - \f{23}{108} \nf} & 
{\scriptstyle -\f{74029}{972} - \f{11}{12} \nf} & {\scriptstyle
-\f{23768}{729} + \f{53}{108} \nf} & {\scriptstyle -\f{10972}{243} -
\f{5}{4} \nf} \\
0 & 0 & {\scriptstyle \f{4541}{54} - \f{1213}{972} \nf} &
{\scriptstyle -\f{165}{2} + \f{493}{324} \nf} & {\scriptstyle
\f{1105}{27} - \f{113}{486} \nf} & {\scriptstyle \f{95}{9} +
\f{305}{162} \nf} \\
0 & 0 & {\scriptstyle -\f{6967}{729} + \f{19}{108} \nf} &
{\scriptstyle \f{17767}{243} + \f{37}{36} \nf} & {\scriptstyle 
\f{42199}{1458} - \f{49}{108} \nf} & {\scriptstyle \f{27677}{486} +
\f{41}{36} \nf} \\
0 & 0 & {\scriptstyle -\f{3461}{54} + \f{1333}{972} \nf} & {\scriptstyle
\f{1085}{18} - \f{421}{324} \nf} & {\scriptstyle -\f{1375}{27} +
\f{53}{486} \nf} & {\scriptstyle \f{43}{3} - \f{341}{162} \nf}  
\end{array} 
\right ) \, . 
\eeq
While the one-loop residual finite renormalization $\hat{Z}'^{(1,
0)}_{\QQ}$ is independent on the choice of trace evanescent operators,
the $\nf$ parts of the two-loop residual finite renormalization
$\hat{Z}'^{(2, 0)}_{\QQ}$ relating the QCD penguin operators in the
unprimed and primed scheme are not. The $\nf$ parts given in 
\Eq{eq:zprime20explicit} correspond to the specific choice of trace
evanescent operators adopted in
\Eqsand{eq:onelooptraceevanescentoperators}{eq:twolooptraceevanescentoperators}.

With these expressions at hand, it is now a matter of simple algebra
to find the ADM and the initial conditions of the Wilson
coefficients in the ``traditional'' basis. Using \Eqs{eq:gamma0new}, 
\eq{eq:gamma1new}, \eq{eq:gamma2new}, \eq{eq:admtransformations}, 
\eq{eq:zprime10explicit} and \eq{eq:zprime20explicit} we obtain for
the regularization- and renormalization scheme independent one-loop
$\ord (\as)$ \ads \ matrix:    
\beq \label{eq:gamma0old}
\hat{\gamma}^{\prime \, (0)} = 
\left (
\begin{array}{cccccc}
-2 & 6 & 0 & 0 & 0 & 0 \\  
6 & -2 & {\scriptstyle -\f{2}{9}} & {\scriptstyle \f{2}{3}} &
{\scriptstyle -\f{2}{9}} & {\scriptstyle \f{2}{3}} \\   
0 & 0 & {\scriptstyle -\f{22}{9}} & {\scriptstyle \f{22}{3}} &
{\scriptstyle -\f{4}{9}} & {\scriptstyle \f{4}{3}} \\   
0 & 0 & {\scriptstyle 6 - \f{2}{9} \nf} & {\scriptstyle -2 + \f{2}{3} 
\nf} & {\scriptstyle -\f{2}{9} \nf} & {\scriptstyle \f{2}{3} \nf} \\  
0 & 0 & 0 & 0 & 2 & -6 \\
0 & 0 & {\scriptstyle -\f{2}{9} \nf} & {\scriptstyle \f{2}{3} \nf} &
{\scriptstyle -\f{2}{9} \nf} & {\scriptstyle -16 + \f{2}{3} \nf} 
\end{array}
\right ) \, .  
\eeq
While the matrix $\hat{\gamma}'^{(0)}$ is renormalization-scheme
independent, $\hat{\gamma}'^{(1)}$ and $\hat{\gamma}'^{(2)}$ are
not. In the $\MSbar$ scheme supplemented by the definition of
evanescent operators given in
\Eqsand{eq:oneevanescentoperatorburas}{eq:twoevanescentoperatorburas}
we obtain for the two-loop $\ord (\as^2)$ \ads \ matrix:         
\beq \label{eq:gamma1old}
\hat{\gamma}^{\prime \, (1)} = 
\left (
\begin{array}{cccccc}
{\scriptstyle -\f{21}{2}} {\scriptstyle -\f{2}{9} \nf} &
{\scriptstyle \f{7}{2}} {\scriptstyle +\f{2}{3} \nf} & {\scriptstyle 
\f{79}{9}} & {\scriptstyle -\f{7}{3}} & {\scriptstyle -\f{65}{9}} &
{\scriptstyle -\f{7}{3}} \\  
{\scriptstyle \f{7}{2}} {\scriptstyle +\f{2}{3} \nf} & {\scriptstyle 
-\f{21}{2}} {\scriptstyle -\f{2}{9} \nf} & {\scriptstyle
-\f{202}{243}} & {\scriptstyle \f{1354}{81}} & {\scriptstyle
-\f{1192}{243}} & {\scriptstyle \f{904}{81}} \\ 
0 & 0 & {\scriptstyle -\f{5911}{486}} {\scriptstyle +\f{71}{9} \nf} &
{\scriptstyle \f{5983}{162}} {\scriptstyle +\f{1}{3} \nf} &
{\scriptstyle -\f{2384}{243}} {\scriptstyle -\f{71}{9} \nf} &
{\scriptstyle \f{1808}{81}} {\scriptstyle -\f{1}{3} \nf} \\ 
0 & 0 & {\scriptstyle \f{379}{18}} {\scriptstyle +\f{56}{243} \nf} &
{\scriptstyle -\f{91}{6}} {\scriptstyle +\f{808}{81} \nf} & 
{\scriptstyle -\f{130}{9}} {\scriptstyle -\f{502}{243} \nf} &
{\scriptstyle -\f{14}{3}} {\scriptstyle +\f{646}{81} \nf} \\
0 & 0 & {\scriptstyle -\f{61}{9} \nf} & {\scriptstyle -\f{11}{3} \nf}
& {\scriptstyle \f{71}{3}} {\scriptstyle +\f{61}{9} \nf} &
{\scriptstyle -99} {\scriptstyle +\f{11}{3} \nf} \\  
0 & 0 & {\scriptstyle -\f{682}{243} \nf} & {\scriptstyle \f{106}{81}
\nf} & {\scriptstyle -\f{225}{2}} {\scriptstyle +\f{1676}{243} \nf} & 
{\scriptstyle -\f{1343}{6}} {\scriptstyle +\f{1348}{81} \nf}
\end{array}
\right ) \, . 
\eeq
The $\nf$ and $\nf^2$ parts of the matrix $\hat{\gamma}'^{(2)}$
describing the mixing of the QCD penguin operators do in addition
depend on the definition of trace evanescent operators given in  
\Eqsand{eq:onelooptraceevanescentoperators}{eq:twolooptraceevanescentoperators}.
In the corresponding scheme we find for the three-loop $\ord (\as^3)$
\ads \ matrix:           
\beq \label{eq:gamma2old}
\begin{split}
\hat{\gamma}'^{(2)} & = 
\left ( 
\begin{array}{ccccccc}
{\scriptstyle \f{19859}{36} - \f{1543}{18} \nf  + \f{298}{81} \nf^2 + 
\f{80}{3} \nf \, \zetathree} & & & & & & \scriptstyle{-\f{9}{4} +
\f{605}{18} \nf - \f{106}{27} \nf^2 - \left ( 672 + 80 \nf \right )  
\zetathree} \\  
{\scriptstyle \f{4741}{12} - \f{11}{2} \nf - \f{82}{27} \nf^2 - \left
    ( 672 + 80 \nf \right ) \zetathree} & & & & & & {\scriptstyle 
-\f{1165}{36} - \f{7}{18} \nf + \f{82}{81} \nf^2 + \f{80}{3}
\nf \, \zetathree} \\      
0 & & & & & & 0 \\  
0 & & & & & & 0 \\
0 & & & & & & 0 \\
0 & & & & & & 0
\end{array}
\right. \\[2mm]
& \hspace{2mm}
\begin{array}{ccc}
{\scriptstyle \f{58231}{1944} + \f{595}{81} \nf} & & {\scriptstyle
  -\f{77239}{648} + \f{53}{27} \nf} \\       
{\scriptstyle -\f{3664721}{52488} + \f{23831}{4374} \nf + \f{280}{81}
\zetathree} & & {\scriptstyle \f{4033865}{17496} - \f{23111}{1458} \nf
- \f{4024}{27} \zetathree} \\
{\scriptstyle \f{22475861}{13122} + \f{1025695}{17496} \nf +
  \f{79}{81} \nf^2 + \left ( \f{560}{81} + \f{80}{3} \nf \right ) 
\zetathree} & & {\scriptstyle \f{12604483}{4374} - \f{1382815}{5832}
\nf - \f{7}{27} \nf^2 - \left ( \f{26192}{27} + 80 \nf \right )
\zetathree} \\    
{\scriptstyle -\f{146039}{81} + \f{8338543}{52488} \nf -
  \f{15961}{4374} \nf^2 - \left ( 672 + \f{6200}{81} \nf \right )  
\zetathree} & & {\scriptstyle \f{31679}{27} - \f{2583223}{17496} \nf 
- \f{6359}{1458} \nf^2 - \f{3304}{81} \nf \, \zetathree} \\    
{\scriptstyle -\f{100832}{81} - \f{17705}{486} \nf - \f{1}{27} \nf^2}
& &  {\scriptstyle -\f{60448}{27} + \f{25841}{81} \nf - \f{23}{9}
  \nf^2} \\   
{\scriptstyle \f{552928}{243} - \f{10779689}{52488} \nf +
\f{18005}{4374} \nf^2 + \f{280}{81} \nf \, \zetathree} & &
{\scriptstyle -\f{128992}{81} + \f{7661297}{17496} \nf - 
\f{31109}{1458} \nf^2 - \f{5464}{27} \nf \, \zetathree}   
\end{array} \\[2mm]
& \left. 
\begin{array}{cc} 
{\scriptstyle -\f{261287}{1944} + \f{163}{81} \nf + \f{20}{3}
\zetathree} & {\scriptstyle -\f{77401}{648} + \f{53}{27} \nf -  
20 \zetathree} \\
{\scriptstyle -\f{2612539}{26244} + \f{38285}{4374} \nf + \f{7228}{81}
  \zetathree} & {\scriptstyle \f{1722187}{8748} - \f{16541}{1458} \nf
  - \f{2044}{27} \zetathree} \\
{\scriptstyle -\f{33248683}{13122} + \f{1049149}{17496} \nf -
\f{49}{27} \nf^2 + \left ( \f{14456}{81} + \f{20}{3} \nf \right )
\zetathree} & {\scriptstyle \f{1098811}{4374} - \f{266077}{5832} \nf +
\f{25}{9} \nf^2 - \left ( \f{4088}{27} + 20 \nf \right ) \zetathree} \\
{\scriptstyle \f{437203}{324} - \f{1360099}{26244} \nf +
  \f{13289}{4374} \nf^2 + \left ( \f{40}{3} + \f{7228}{81} \nf \right
  ) \zetathree} & {\scriptstyle -\f{199123}{108} + \f{1627075}{8748}
  \nf - \f{6233}{1458} \nf^2 - \left ( 40 + \f{2044}{27} \nf \right )
  \zetathree} \\  
{\scriptstyle \f{1826987}{648} - \f{113413}{972} \nf + \f{83}{81}
  \nf^2 - \left ( 378 + 20 \nf \right ) \zetathree} & {\scriptstyle
  -\f{472667}{216} + \f{88933}{324} \nf - \f{11}{27} \nf^2 + \left (
    462 + 60 \nf \right ) \zetathree} \\ 
{\scriptstyle -\f{2091127}{1944}+ \f{509723}{26244} \nf +
\f{27815}{4374} \nf^2 + \left ( 210 + \f{7228}{81} \nf \right )
\zetathree} & {\scriptstyle -\f{1946849}{648} + \f{5194621}{8748} \nf
- \f{9815}{1458} \nf^2 + \left ( 378 + \f{2276}{27} \nf \right ) 
\zetathree}  
\end{array} 
\right ) \, . 
\end{split} 
\eeq
As far as the one- and two-loop self-mixing of the four-quark
operators $\Qburas_1$--$\Qburas_6$, namely $\hat{\gamma}'^{(0)}$ and
$\hat{\gamma}'^{(1)}$ are concerned, our findings agree perfectly with
the results of the direct computations \cite{Ciuchini:1993vr,
Buras:1992tc}. Furthermore, they also agree with the results  
presented in \cite{Chetyrkin:1997gb}, which were obtained by
performing a change of scheme from the unprimed set of operators to
the primed one. On the other hand, the three-loop self-mixing of 
$\Qburas_1$--$\Qburas_6$ described by $\hat{\gamma}'^{(2)}$, is
entirely new. 

Similarly, employing \Eqs{eq:wilsoncoefficientsmisiak},
\eq{eq:cprime}, \eq{eq:zprime10explicit} and \eq{eq:zprime20explicit}
we find for the initial conditions of the Wilson coefficients in the  
``traditional'' basis up to $\ord (\as^2)$:        
\begin{align} \label{eq:wilsoncoefficientsburas}  
  C'_1 (\MW) & = \f{11}{2} \f{\as (\MW)}{4 \pi} + \left( \f{\as
      (\MW)}{4 \pi} \right )^2 \left ( \f{2005}{48} + \f{17}{6} \pi^2
    - \f{1}{2}
    \widetilde{T}_0 (x_t) \right ) \, , \non \\
  C'_2 (\MW) & = 1 - \f{11}{6} \f{\as (\MW)}{4 \pi} - \left( \f{\as
      (\MW)}{4 \pi} \right )^2 \left ( \f{1405}{144} - \f{7}{18} \pi^2
    -
    \f{1}{6} \widetilde{T}_0 (x_t) \right ) \, , \non \\
  C'_3 (\MW) & = -\f{1}{6} \f{\as (\MW)}{4 \pi} \widetilde{E}_0 (x_t)
  + \left( \f{\as (\MW)}{4 \pi} \right )^2 \left ( \f{539}{810} +
    \f{77}{90} \widetilde{E}_0 (x_t) - \f{1}{6} \widetilde{E}_1 (x_t)
    -
    \f{1}{10} \widetilde{G}_1 (x_t) \right ) \, , \non \\
  C'_4 (\MW) & = \f{1}{2} \f{\as (\MW)}{4 \pi} \widetilde{E}_0 (x_t) +
  \left( \f{\as (\MW)}{4 \pi} \right )^2 \left ( \f{49}{54} + \f{7}{6}
    \widetilde{E}_0 (x_t) + \f{1}{2} \widetilde{E}_1 (x_t) - \f{3}{2}
    \widetilde{G}_1 (x_t) \right ) \, , \non \\
  C'_5 (\MW) & = -\f{1}{6} \f{\as (\MW)}{4 \pi} \widetilde{E}_0 (x_t)
  + \left( \f{\as (\MW)}{4 \pi} \right )^2 \left ( \f{308}{405} +
    \f{44}{45} \widetilde{E}_0 (x_t) - \f{1}{6} \widetilde{E}_1 (x_t)
    +
    \f{29}{40} \widetilde{G}_1 (x_t) \right ) \, , \non \\
  C'_6 (\MW) & = \f{1}{2} \f{\as (\MW)}{4 \pi} \widetilde{E}_0 (x_t) +
  \left( \f{\as (\MW)}{4 \pi} \right )^2 \left ( 
      \f{28}{27} + \f{4}{3} \widetilde{E}_0 (x_t) + \f{1}{2}
    \widetilde{E}_1 (x_t) - \f{3}{8} \widetilde{G}_1 (x_t) \right ) \,
  .
\end{align}
Whereas the two-loop $\ord (\as^2)$ corrections to the initial
conditions $C'_1 (\MW)$--$C'_6 (\MW)$ are entirely new, our findings
for the one-loop $\ord (\as)$ corrections agree again perfectly with
all preceding calculations  \cite{Ciuchini:1993vr, Chetyrkin:1997gb,
Buras:1992tc}. Finally, let us mention that the $\ord (\as^2)$ 
corrections to the initial conditions $C'_1 (\MW)$--$C'_6 (\MW)$ do 
not depend on the special choice of trace evanescent operators made in
\Eqsand{eq:onelooptraceevanescentoperators}{eq:twolooptraceevanescentoperators}.
The NNLO analytic formulas for the low-energy Wilson coefficients
relevant for non-leptonic $B$ meson decays in the ``traditional'' 
scheme can be found in
\App{app:wilsoncoefficientsinthetraditionaloperatorbasis}.  

\section{Summary}
\label{sec:summary}

In this paper we have extended the SM analysis of the effective
hamiltonian for non-leptonic $| \Delta F | = 1$ decays to the
NNLO. The main ingredient of this generalization is the three-loop
ADM describing the mixing of the current-current and QCD penguin
operators, which we have computed in an operator basis that allows to
consistently use fully anticommuting $\gamma_5$ in dimensional
regularization to all orders in perturbation theory. The issue of
renormalization scheme dependences, their cancellation in physical
quantities, and the transformation properties of the ADM and the
initial conditions of the Wilson coefficients under a change of scheme
has been discussed thoroughly. In particular, we have elaborated on
the scheme dependence related to the renormalization of the strong 
coupling constant, a feature that, to the best of our knowledge, has
not been studied in the context of the renormalization of effective
field theories so far. As a practical application of our general
considerations, we have derived the explicit NNLO relation between our
and the so-called ``traditional'' renormalization scheme, which
allowed us to calculate indirectly the NNLO ADM and the corresponding 
matching conditions in the latter scheme. Finally, we have solved the
RGE to obtain the analytic expressions for the low-energy Wilson
coefficients relevant for non-leptonic $B$ meson decays through NNLO
in both schemes.   

\subsubsection*{Acknowledgments}
\label{subsec:acknowledgments}

We are grateful to Paolo Gambino for collaboration at an early stage
of this project, for numerous comments and discussions later on, and
for his careful proofreading of the present manuscript. Furthermore,
we are obliged to Mikolaj Misiak for helpful correspondence and for
his critical reading of the final draft of this paper.  In addition,
we would like to thank Matthias Steinhauser for providing us with an
updated version of {\tt MATAD} \cite{Steinhauser:2000ry}, and Guido
Bell and Diego Guadagnoli for pointing out the typos in Eqs. (27) and
(37). Finally, a big thank you goes to Xinqiang Li for bringing the
typos/mistakes in Eqs. (26), (29), (48), (102), and (A.27) to our
attention.  M.~G.\ and U.~H. \ appreciate the warm hospitality of the
Institute T31 for Theoretical Elementary Particle Physics at the
Technical University of Munich where part of this work has been
performed. The work of U.~H.\ is supported by the U.S. Department of
Energy under contract No.~DE-AC02-76CH03000.

\renewcommand{\thesubsection}{A.\arabic{subsection}}
\setcounter{section}{0}  
\renewcommand{\theequation}{A.\arabic{equation}}
\setcounter{equation}{0}  

\section*{Appendix} 

\subsection{Derivation of \boldmath $\hat{S}^{(1)}$ and \boldmath 
$\hat{S}^{(2)}$}    
\label{app:derivationofs1ands2}

In order to derive the explicit expressions for the matrix kernels 
$\hat{S}^{(1)}$ and $\hat{S}^{(2)}$ as given in \Eq{eq:smatrices},
we follow \cite{Buras:1991jm, Buras:1979yt} and compute the partial  
derivative of 
\Eqsand{eq:evolutionmatrix}{eq:asevolutionmatrixexpansion}  
with respect to $\gs$. After some algebra one finds the following
differential equation for $\hat{K} (\gs)$ 
\beq \label{eq:kgdifferntial}
\f{\partial \hat{K} (\gs)}{\partial \gs} + \f{1}{\gs} \left [
\f{\hat{\gamma}^{(0) \, T}}{\betazero}, \hat{K} (\gs) \right ] =   
\left( \f{\hat{\gamma}^T (\gs)}{\beta (\gs)} + \f{1}{\gs}
\f{\hat{\gamma}^{(0) \, T}}{\betazero} \right ) \hat{K} (\gs) \, . 
\eeq
Inserting \Eq{eq:kmatrices} into the last equation we obtain
\beq \label{eq:jcondition}
\begin{split}
\hat{J}^{(1)} + \left [ \f{\hat{\gamma}^{(0) \, T}}{2 \betazero},
\hat{J}^{(1)} \right ] & = -\f{\hat{\gamma}^{(1) \, T}}{2 \betazero} +
\f{\betaone}{2 \betazero^2} \hat{\gamma}^{(0) \, T} \, , \\ 
\hat{J}^{(2)} + \left [ \f{\hat{\gamma}^{(0) \, T}}{4 \betazero},
\hat{J}^{(2)} \right ] & = -\f{\hat{\gamma}^{(2) \, T}}{4 \betazero}
+ \f{\betaone}{4 \betazero^2} \hat{\gamma}^{(1) \, T} + \left (
\f{\betatwo}{4 \betazero^2} - \f{\betaone^2}{4 \betazero^3} \right )
\hat{\gamma}^{(0) \, T} \\ 
& - \left ( \f{\hat{\gamma}^{(1) \, T}}{4 \betazero} - \f{\betaone}{4
\betazero^2} \hat{\gamma}^{(0) \,T} \right ) \hat{J}^{(1)} \, ,    
\end{split}
\eeq
for the parts proportional to $\gs$ and $\gs^3$, respectively. After
diagonalizing these equations with the help of \Eq{eq:magicnumbers} we
find  
\beq \label{eq:smatricesappendix}
\begin{split}
S^{(1)}_{i j} & = \f{\betaone}{\betazero} a_i \delta_{i j} -
\f{G^{(1)}_{i j}}{2 \betazero \left ( 1 + a_i - a_j \right )} \, , \\ 
S^{(2)}_{i j} & = \left ( \f{\betatwo}{2 \betazero} - \f{\betaone^2}{2
\betazero^2} \right ) a_i \delta_{i j} + \sum_k \f{2 \betaone a_i
\delta_{i k} - G^{(1)}_{i k}}{2 \betazero \left ( 2 + a_i - a_j \right
)} S^{(1)}_{k j} + \f{\betaone G^{(1)}_{i j} - \betazero G^{(2)}_{i
j}}{2 \betazero^2 \left ( 2 + a_i - a_j \right )} \, ,    
\end{split}
\eeq
Finally, solving the first equation for $G^{(1)}_{ij}$ and inserting
the result into the second equation, one obtains the expression for
the elements of $\hat{S}^{(2)}$ as given in \Eq{eq:smatrices}.     

\subsection{Trace Evanescent Operators}   
\label{app:traceevanescentoperators}

In the following we specify the exact form of the so-called trace
evanescent operators arising from the one- and two-loop diagrams with 
insertions of $E_5^{(1)}$--$E_8^{(1)}$. At the one-loop level the 
specific structure of only one of them is needed:    
\beq \label{eq:onelooptraceevanescentoperators}
\widehat{E}_1^{(1)} = (\bar{s} \gammadown{\mu_1 \mu_2 \mu_3} \PL
b) \, \mbox{Tr} \left ( \gammaup{\mu_1 \mu_2 \mu_3 \mu_4} \gamma_5 
\right ) - 24 (\bar{s}_L \gammaup{\mu_4} b_L) \, , 
\eeq
while at the two-loop level we encounter eight additional trace
evanescent operators:  
\beq \label{eq:twolooptraceevanescentoperators}
\begin{split}
\widehat{E}_1^{(2)} & = (\bar{s} \gammadown{\mu_1} \PL b)
\sum\nolimits_q ( \bar{q} \gammadown{\mu_2 \mu_3
\mu_4} q ) \, \mbox{Tr} \left ( \gammaup{\mu_1 \mu_2 \mu_3 \mu_4}
\gamma_5 \right ) - 40 (\bar{s}_L \gammadown{\mu_1} b_L)
\sum\nolimits_q ( \bar{q} \gammaup{\mu_1} q ) 
\\[1.5mm]   
& + 4 (\bar{s}_L \gammadown{\mu_1 \mu_2 \mu_3} b_L)
\sum\nolimits_q ( \bar{q} \gammaup{\mu_1 \mu_2 \mu_3} 
q ) \, , \\[3mm]  
\widehat{E}_2^{(2)} & = (\bar{s} \gammadown{\mu_1} \PL
\gammadown{\mu_2 \mu_3} b) \, \mbox{Tr} \left ( \gammaup{\mu_1 \mu_2 
\mu_3 \mu_4} \gamma_5 \right ) - 24 (\bar{s}_L \gammaup{\mu_4}
b_L) \, , \\[2mm] 
\widehat{E}_3^{(2)} & = (\bar{s} \gammadown{\mu_1} \PL
\gammadown{\mu_2 \mu_3} b) \, \mbox{Tr} \left ( \gammaup{\mu_1 \mu_3  
\mu_4 \mu_5} \gamma_5 \right ) + 8  \Big ( \tilde{g}_{\mu_2}^{\; 
\mu_4} \, (\bar{s}_L \gammaup{\mu_5} b_L) - \tilde{g}_{\mu_2}^{\;
\mu_5} \, (\bar{s}_L \gammaup{\mu_4} b_L) \Big ) \, , \\[2mm]
\widehat{E}_4^{(2)} & = (\bar{s} \gammadown{\mu_1} \PL
\gammadown{\mu_2} b) \, \mbox{Tr} \left ( \gammaup{\mu_1 \mu_2 \mu_3
\mu_4} \gamma_5 \right ) + 8  \Big ( (\bar{s}_L \gammaup{\mu_3
\mu_4} b_R) - \tilde{g}^{\mu_3 \mu_4} (\bar{s}_L  b_R) \Big ) \, ,
\\[2mm] 
\widehat{E}_5^{(2)} & = (\bar{s} \gammadown{\mu_1 \mu_2 \mu_3} \PL 
b) \, \mbox{Tr} \left ( \gammaup{\mu_2 \mu_3 \mu_4 \mu_5} \gamma_5
\right ) - 8  \Big ( (\bar{s}_L \gamma_{\mu_1}^{\; \mu_4
\mu_5} b_L) - \tilde{g}^{\mu_4 \mu_5} (\bar{s}_L
\gammadown{\mu_1} b_L) \Big ) \, , \\[2mm]   
\widehat{E}_6^{(2)} & = (\bar{s} \gammadown{\mu_1} \PL b) \,
\mbox{Tr} \left ( \gammaup{\mu_1 \mu_2 \mu_3 \mu_4} \gamma_5 \right ) 
+ 4  \Big ( (\bar{s}_L \gammaup{\mu_2 \mu_3 \mu_4} b_L) \\  
& - \tilde{g}^{\mu_2 \mu_3} \, (\bar{s}_L \gammaup{\mu_4} b_L) +
\tilde{g}^{\mu_2 \mu_4} \, (\bar{s}_L \gammaup{\mu_3} b_L) -
\tilde{g}^{\mu_3 \mu_4} \, (\bar{s}_L \gammaup{\mu_2} b_L) \Big )
\, , \\  
\widehat{E}_7^{(2)} & = (\bar{s} \gammadown{\mu_1 \mu_2 \mu_3} \PL
b) \, \mbox{Tr} \left ( \gammaup{\mu_1 \mu_2 \mu_4 \mu_5} \gamma_5
\right ) + 8  \Big ( (\bar{s}_L \gamma_{\mu_3}^{\; \mu_4
\mu_5} b_L) - 2 \tilde{g}_{\mu_3}^{\; \mu_4} (\bar{s}_L
\gammaup{\mu_5} b_L) \\  
& + 2 \tilde{g}_{\mu_3}^{\; \mu_5} (\bar{s}_L \gammaup{\mu_4}
b_L) - \tilde{g}^{\mu_4 \mu_5} (\bar{s}_L \gammadown{\mu_3}
b_L)\Big ) \, , \\[2mm]   
\widehat{E}_8^{(2)} & = (\bar{s} \gammadown{\mu_1} \PL
\gammadown{\mu_2 \mu_3} b) \, \mbox{Tr} \left ( \gammaup{\mu_1 \mu_2
\mu_4 \mu_5} \gamma_5 \right ) + 8  \Big ( (\bar{s}_L
\gamma_{\mu_3}^{\; \mu_4 \mu_5} b_L) - 2 \tilde{g}_{\mu_3}^{\;
\mu_4} (\bar{s}_L \gammaup{\mu_5} b_L) \\ 
& + 2 \tilde{g}_{\mu_3}^{\; \mu_5} (\bar{s}_L \gammaup{\mu_4}
b_L) - \tilde{g}^{\mu_4 \mu_5} (\bar{s}_L \gammadown{\mu_3} b_L) 
\Big ) \, , 
\end{split}
\eeq
where we have used the definitions $\gamma_{\mu_1 \ldots \mu_j}^{\;
\mu_{j + 1} \ldots \mu_n} \equiv \gamma_{\mu_1} \ldots \gamma_{\mu_j}
\gamma^{\mu_{j + 1}} \ldots \gamma^{\mu_n}$ and $P_L \equiv (1 -
\gamma_5)/2$. Finally, let us recall two important features of the
latter operators. First, their role in the ``traditional'' basis is
the same as the one played by the evanescent operators in
\Eq{eq:twoevanescentoperatorburas}, that is, they are needed to define
the renormalization scheme in that basis at the NNLO. Second, the
$\nf$ and $\nf^2$ parts of the three-loop ADM $\hat \gamma'^{(2)}$
describing the mixing of the QCD penguin operators in the
``traditional'' basis depend on the definition of these operators. The
above definitions correspond to the ``dimensional reduction''-like
treatment of traces with an even number of Dirac matrices and a single
$\gamma_5$, employing  \Eqs{eq:evanescenttensors},
\eq{eq:fourdimensionalfivetrace} and \eq{eq:fourdimensionaltrace}. 

\subsection{Change to the ``Traditional'' Operator Basis}   
\label{app:changetothetraditionaloperatorbasis}

In order to give the explicit expressions for the matrices $\hat{R}$,
$\hat{W}$, $\hat{U}$ and $\hat{M}$ characterizing the change to the
``traditional'' basis, we first have to define the unprimed and primed
set of operators according to \Eq{eq:transformations}. The physical
and evanescent operators in the initial basis are given by 
\beq \label{eq:qandevector}
\begin{split}
\vec{\Qmisiak}^T & = (\Qmisiak_1, \ldots, \Qmisiak_6) \, , \\
\vec{\Emisiak}^T & = (\Emisiak^{(1)}_1, \ldots, \Emisiak^{(1)}_8,
\Emisiak^{(2)}_1, \ldots, \Emisiak^{(2)}_8) \, ,    
\end{split}
\eeq
while the ``traditional'' basis consists of the following two sets of 
operators:  
\beq \label{eq:qprimeandeprimevector}
\begin{split}
\vec{\Qburas}^T & = (\Qburas_1, \ldots, \Qburas_6) \, , \\
\vec{\Eburas}^T & = ( \Eburas^{(1)}_1, \ldots, \Eburas^{(1)}_6,
\Eburas^{(2)}_1, \ldots, \Eburas^{(2)}_6, \Emisiak^{(2)}_3,
\Emisiak^{(2)}_4, \Emisiak^{(2)}_7, \Emisiak^{(2)}_8) \, .    
\end{split}
\eeq 
Needless to say, that $\Emisiak^{(1)}_5$--$\Emisiak^{(1)}_8$ and
$\Emisiak^{(2)}_5$--$\Emisiak^{(2)}_8$ play the role of extra, in
principle unnecessary operators in the initial operator basis. The
same is true for $\Emisiak^{(2)}_3$, $\Emisiak^{(2)}_4$,
$\Emisiak^{(2)}_7$ and $\Emisiak^{(2)}_8$ in the ``traditional''
basis. They are just included for completeness in the above
equations. Although the trace evanescent operators introduced in 
\Eqsand{eq:onelooptraceevanescentoperators}{eq:twolooptraceevanescentoperators}
are needed to make precise the definition of the ``traditional''
renormalization scheme we refrain to include them in
\Eq{eq:qprimeandeprimevector}, since they influence the transformation
between the unprimed and primed set of operators only in an indirect
way, that is, through the finite parts of the two-loop renormalization
constants of some of the evanescent operators included in the initial
basis. This will be explained at the very end of this appendix. 

With this definitions at hand, it is just a matter of simple algebra
to find the explicit expressions for the matrices $\hat{R}$,
$\hat{W}$, $\hat{U}$ and $\hat{M}$. The rotation matrix $\hat{R}$,
which links the physical operators together, is given by    
\beq \label{eq:rmatrix}
\hat{R} = 
\left (
\begin{array}{cccccc}
2 & {\scriptstyle \f{1}{3}} & 0 & 0 & 0 & 0 \\ 
0 & 1 & 0 & 0 & 0 & 0 \\ 
0 & 0 & {\scriptstyle -\f{1}{3}} & 0 & {\scriptstyle \f{1}{12}} & 0 \\
0 & 0 & {\scriptstyle -\f{1}{9}} & {\scriptstyle -\f{2}{3}} &
{\scriptstyle \f{1}{36}} & {\scriptstyle \f{1}{6}} \\  
0 & 0 & {\scriptstyle \f{4}{3}} & 0 & {\scriptstyle -\f{1}{12}} & 0 \\
0 & 0 & {\scriptstyle \f{4}{9}} & {\scriptstyle \f{8}{3}} &
{\scriptstyle -\f{1}{36}} & {\scriptstyle -\f{1}{6}} 
\end{array}
\right ) \, . 
\eeq
The matrix $\hat{W}$ parametrizes a redefinition of the physical
operators $\vec{\Qmisiak}$ by adding some evanescent operators
$\vec{\Emisiak}$ to them. In the case at hand, $\hat{W}$ reads  
\beq \label{eq:wmatrix}
\hat{W} = 
\left (
\begin{array}{ccccccccccccccccccc}
0 & 0 & 0 & 0 & 0 & 0 & 0 & 0 & 0 & 0 & 0 & 0 & 0 & 0 & 0 & 0 \\   
0 & 0 & 0 & 0 & 0 & 0 & 0 & 0 & 0 & 0 & 0 & 0 & 0 & 0 & 0 & 0 \\ 
0 & 0 & 0 & 0 & 0 & 0 & 0 & 0 & 0 & 0 & 0 & 0 & 0 & 0 & 0 & 0 \\ 
0 & 0 & 0 & 0 & 0 & 0 & 0 & 0 & 0 & 0 & 0 & 0 & 0 & 0 & 0 & 0 \\ 
0 & 0 & 0 & 0 & -6 & 0 & 0 & 0 & 0 & 0 & 0 & 0 & 0 & 0 & 0 & 0 \\ 
0 & 0 & 0 & 0 & 0 & -6 & 0 & 0 & 0 & 0 & 0 & 0 & 0 & 0 & 0 & 0
\end{array}
\right ) \, .    
\eeq
On the other hand, $\hat{U}$ describes a redefinition of the
evanescent operators $\vec{\Emisiak}$ by adding some multiples of   
$\eps$ times physical operators $\vec{\Qmisiak}$ to them. The relevant
matrix $\hat{U}$ takes the following form  
\beq \label{eq:umatrix}
\hat{U} = 
\left (
\begin{array}{cccccc}
4 & 0 & 0 & 0 & 0 & 0 \\ 
0 & 4 & 0 & 0 & 0 & 0 \\ 
0 & 0 & -112 & 0 & 16 & 0 \\ 
0 & 0 & 0 & -112 & 0 & 16 \\ 
0 & 0 & {\scriptstyle -\f{10}{9}} & 0 & {\scriptstyle \f{1}{9}} & 0 \\
0 & 0 & 0 & {\scriptstyle -\f{10}{9}} & 0 & {\scriptstyle \f{1}{9}} \\
0 & 0 & {\scriptstyle -\f{136}{9}} & 0 & {\scriptstyle \f{10}{9}} & 0
\\  
0 & 0 & 0 & {\scriptstyle -\f{136}{9}} & 0 & {\scriptstyle \f{10}{9}}
\\  
144 & 0 & 0 & 0 & 0 & 0 \\ 
0 & 144 & 0 & 0 & 0 & 0 \\ 
0 & 0 & 0 & 0 & 0 & 0 \\ 
0 & 0 & 0 & 0 & 0 & 0 \\ 
0 & 0 & {\scriptstyle -\f{2224}{9}} & 0 & {\scriptstyle \f{64}{9}} & 0
\\  
0 & 0 & 0 & {\scriptstyle -\f{2224}{9}} & 0 & {\scriptstyle \f{64}{9}}
\\  
0 & 0 & 0 & 0 & 0 & 0 \\ 
0 & 0 & 0 & 0 & 0 & 0
\end{array}
\right ) \, .   
\eeq
Finally, the matrix $\hat{M}$ represents a simple linear
transformation of the evanescent operators. In our case we find     
\beq \label{eq:mmatrix}
\hat{M} = 
\left (
\begin{array}{ccccccccccccccccccc}
2 & {\scriptstyle \f{1}{3}} & 0 & 0 & 0 & 0 & 0 & 0 & 0 & 0 & 0 & 0 &
0 & 0 & 0 & 0 \\  
0 & 1 & 0 & 0 & 0 & 0 & 0 & 0 & 0 & 0 & 0 & 0 & 0 & 0 & 0 & 0 \\  
0 & 0 & 0 & 0 & 8 & 0 & {\scriptstyle -\f{1}{2}} & 0 & 0 & 0 & 0 & 0 &
0 & 0 & 0 & 0 \\  
0 & 0 & 0 & 0 & {\scriptstyle \f{8}{3}} & 16 & {\scriptstyle
-\f{1}{6}} & -1 & 0 & 0 & 0 & 0 & 0 & 0 & 0 & 0 \\  
0 & 0 & 0 & 0 & -2 & 0 & {\scriptstyle \f{1}{2}} & 0 & 0 & 0 & 0 & 0 &
0 & 0 & 0 & 0 \\  
0 & 0 & 0 & 0 & {\scriptstyle -\f{2}{3}} & -4 & {\scriptstyle
\f{1}{6}} & 1 & 0 & 0 & 0 & 0 & 0 & 0 & 0 & 0 \\  
40 & {\scriptstyle \f{20}{3}} & 0 & 0 & 0 & 0 & 0 & 0 & 2 &
{\scriptstyle \f{1}{3}} & 0 & 0 & 0 & 0 & 0 & 0 \\  
0 & 20 & 0 & 0 & 0 & 0 & 0 & 0 & 0 & 1 & 0 & 0 & 0 & 0 & 0 & 0 \\  
0 & 0 & {\scriptstyle \f{1}{2}} & 0 & 128 & 0 & 0 & 0 & 0 & 0 & 0 & 0
& {\scriptstyle -\f{1}{2}} & 0 & 0 & 0 \\  
0 & 0 & {\scriptstyle \f{1}{6}} & 1 & {\scriptstyle \f{128}{3}} & 256
& 0 & 0 & 0 & 0 & 0 & 0 & {\scriptstyle -\f{1}{6}} & -1 & 0 & 0 \\  
0 & 0 & {\scriptstyle \f{1}{2}} & 0 & -8 & 0 & 0 & 0 & 0 & 0 & 0 & 0 &
{\scriptstyle \f{1}{2}} & 0 & 0 & 0 \\  
0 & 0 & {\scriptstyle \f{1}{6}} & 1 & {\scriptstyle -\f{8}{3}} & -16 &
0 & 0 & 0 & 0 & 0 & 0 & {\scriptstyle \f{1}{6}} & 1 & 0 & 0 \\  
0 & 0 & 0 & 0 & 0 & 0 & 0 & 0 & 0 & 0 & 1 & 0 & 0 & 0 & 0 & 0 \\  
0 & 0 & 0 & 0 & 0 & 0 & 0 & 0 & 0 & 0 & 0 & 1 & 0 & 0 & 0 & 0 \\  
0 & 0 & 0 & 0 & 0 & 0 & 0 & 0 & 0 & 0 & 0 & 0 & 0 & 0 & 1 & 0 \\  
0 & 0 & 0 & 0 & 0 & 0 & 0 & 0 & 0 & 0 & 0 & 0 & 0 & 0 & 0 & 1
\end{array}
\right ) \, .  
\eeq
Parts of the above matrices have already been given explicitly in
\cite{Chetyrkin:1997gb}, where the change of basis from the initial to
the ``traditional'' basis has been performed including the NLO
corrections. If we take into account that the definition of
$\Emisiak^{(1)}_5$--$\Emisiak^{(1)}_8$ adopted in 
\Eq{eq:additionaloneevanescentoperatormisiak} differs slightly from 
the definition of $\Emisiak^{(1)}_5$--$\Emisiak^{(1)}_8$ used in
\cite{Chetyrkin:1997gb}, our results agree with the expressions given
in the latter paper.     

The renormalization constant matrices entering \Eq{eq:zprime} are
found from one- and two-loop matrix elements of physical and
evanescent operators. In the following we will give only the relevant
entries of the necessary renormalization constant matrices, denoting
elements that do not affect the final results for the residual finite 
renormalizations introduced in \Eq{eq:zprime} with a star. For the
finite renormalization between evanescent operators $\vec{\Emisiak}$
and physical operators $\vec{\Qmisiak}$ we get       
\beq \label{eq:Z10EQ}
\hat{Z}^{(1, 0)}_{\EQ} =
\left (
\begin{array}{cccccc}
\star & \star & \star & \star & 0 & 0 \\ 
\star & \star & \star & \star & 0 & 0 \\ 
\star & \star & \star & \star & {\scriptstyle
-\f{1280}{3}} & 320 \\  
\star & \star & \star & \star & {\scriptstyle \f{640}{9}}
& {\scriptstyle \f{1280}{3}} \\  
0 & 0 & {\scriptstyle \f{160}{9}} & {\scriptstyle -\f{128}{9}} &
{\scriptstyle -\f{16}{9}} & {\scriptstyle \f{4}{3}} \\  
 0 & 0 & {\scriptstyle -\f{80}{27}} & {\scriptstyle -\f{476}{27}}
{\scriptstyle -\f{2}{3} \nf} & {\scriptstyle \f{8}{27}} &
{\scriptstyle \f{16}{9}} \\   
\star & \star & \star & \star & {\scriptstyle -\f{160}{9}}
& {\scriptstyle \f{40}{3}} \\  
\star & \star & \star & \star & {\scriptstyle \f{80}{27}}
& {\scriptstyle \f{160}{9}} \\  
\star & \star & \star & \star & 0 & 0 \\  
\star & \star & \star & \star & 0 & 0 \\  
\star & \star & \star & \star & {\scriptstyle
-\f{98560}{3}} & 24640 \\  
\star & \star & \star & \star & {\scriptstyle
\f{49280}{9}} & {\scriptstyle \f{98560}{3}} \\  
\star & \star & \star & \star & {\scriptstyle -\f{256}{9}}
& {\scriptstyle \f{64}{3}} \\  
\star & \star & \star & \star & {\scriptstyle \f{128}{27}}
& {\scriptstyle \f{256}{9}} \\  
\star & \star & \star & \star & {\scriptstyle
\f{154880}{9}} & {\scriptstyle -\f{38720}{3}} \\  
\star & \star & \star & \star & {\scriptstyle
-\f{77440}{27}} & {\scriptstyle -\f{154880}{9}} 
\end{array}
\right ) \, . 
\eeq
For the one-loop mixing of physical operators $\vec{\Qmisiak}$ into
evanescent operators $\vec{\Emisiak}$ we obtain 
\beq \label{eq:Z11QE}
\hat{Z}^{(1, 1)}_{\QE} =
\left (
\begin{array}{cccccccccccccccc}
{\scriptstyle \f{5}{12}} & {\scriptstyle \f{2}{9}} & 0 & 0 & 0 & 0 & 0
& 0 & 0 & 0 & 0 & 0 & 0 & 0 & 0 & 0 \\  
1 & 0 & 0 & 0 & 0 & 0 & 0 & 0 & 0 & 0 & 0 & 0 & 0 & 0 & 0 & 0 \\  
0 & 0 & 0 & 0 & 0 & 0 & 0 & 0 & 0 & 0 & 0 & 0 & 0 & 0 & 0 & 0 \\  
0 & 0 & 0 & 0 & 0 & 0 & 0 & 0 & 0 & 0 & 0 & 0 & 0 & 0 & 0 & 0 \\ 
0 & 0 & 0 & 1 & 0 & 0 & 0 & 0 & 0 & 0 & 0 & 0 & 0 & 0 & 0 & 0 \\ 
0 & 0 & {\scriptstyle \f{2}{9}} & {\scriptstyle \f{5}{12}} & 0 & 0 & 0
& 0 & 0 & 0 & 0 & 0 & 0 & 0 & 0 & 0
\end{array}
\right ) \, . 
\eeq
The one-loop mixing among evanescent operators $\vec{\Emisiak}$ reads
\beq \label{eq:Z11EE}
\hat{Z}^{(1, 1)}_{\EE} =
\left (
\begin{array}{cccccccccccccccc}
\star & \star & \star & \star & \star & \star &
\star & \star & \star & \star & \star & \star &
\star & \star & \star & \star \\  
\star & \star & \star & \star & \star & \star &
\star & \star & \star & \star & \star & \star &
\star & \star & \star & \star \\  
\star & \star & \star & \star & \star & \star &
\star & \star & \star & \star & \star & \star &
\star & \star & \star & \star \\  
\star & \star & \star & \star & \star & \star &
\star & \star & \star & \star & \star & \star &
\star & \star & \star & \star \\  
0 & 0 & 0 & {\scriptstyle \f{1}{6}} & 0 & -10 & 0 & 1 & 0 & 0 & 0 & 0 
& 0 & 0 & 0 & 0 \\  
0 & 0 & {\scriptstyle \f{1}{27}} & {\scriptstyle \f{5}{72}} &
{\scriptstyle -\f{20}{9}} & {\scriptstyle -\f{26}{3}} & {\scriptstyle
\f{2}{9}} & {\scriptstyle \f{5}{12}} & 0 & 0 & 0 & 0 & 0 & 0 & 0 & 0
\\  
\star & \star & \star & \star & \star & \star &
\star & \star & \star & \star & \star & \star &
\star & \star & \star & \star \\  
\star & \star & \star & \star & \star & \star &
\star & \star & \star & \star & \star & \star &
\star & \star & \star & \star \\  
\star & \star & \star & \star & \star & \star &
\star & \star & \star & \star & \star & \star &
\star & \star & \star & \star \\  
\star & \star & \star & \star & \star & \star &
\star & \star & \star & \star & \star & \star &
\star & \star & \star & \star \\  
\star & \star & \star & \star & \star & \star &
\star & \star & \star & \star & \star & \star &
\star & \star & \star & \star \\  
\star & \star & \star & \star & \star & \star &
\star & \star & \star & \star & \star & \star &
\star & \star & \star & \star \\  
\star & \star & \star & \star & \star & \star &
\star & \star & \star & \star & \star & \star &
\star & \star & \star & \star \\  
\star & \star & \star & \star & \star & \star &
\star & \star & \star & \star & \star & \star &
\star & \star & \star & \star \\  
\star & \star & \star & \star & \star & \star &
\star & \star & \star & \star & \star & \star &
\star & \star & \star & \star \\  
\star & \star & \star & \star & \star & \star &
\star & \star & \star & \star & \star & \star &
\star & \star & \star & \star 
\end{array}
\right ) \, .  
\eeq 
At the two-loop order we find for the finite renormalization between
evanescent operators $\vec{\Emisiak}$ and physical operators
$\vec{\Qmisiak}$:
\beq \label{eq:Z20EQ}
\hat{Z}^{(2, 0)}_{\EQ} =
\left (
\begin{array}{cccccc}
\star & \star & \star & \star & \star & \star \\  
\star & \star & \star & \star & \star & \star \\ 
\star & \star & \star & \star & \star & \star \\ 
\star & \star & \star & \star & \star & \star \\ 
0 & 0 & {\scriptstyle \f{21632}{243} - \f{44}{81} \nf} & {\scriptstyle
\f{80092}{243} + \f{143}{27} \nf} & {\scriptstyle -\f{266}{243} +
\f{8}{81} \nf} & {\scriptstyle -\f{1208}{81} - \f{2}{27} \nf} \\  
0 & 0 & {\scriptstyle -\f{77020}{729} + \f{46}{243} \nf} &
{\scriptstyle -\f{114161}{729} - \f{59}{18} \nf} & {\scriptstyle
\f{5182}{729} - \f{1}{243} \nf} & {\scriptstyle \f{21019}{972} -
\f{49}{648} \nf} \\  
\star & \star & \star & \star & \star & \star \\ 
\star & \star & \star & \star & \star & \star \\ 
\star & \star & \star & \star & \star & \star \\ 
\star & \star & \star & \star & \star & \star \\ 
\star & \star & \star & \star & \star & \star \\ 
\star & \star & \star & \star & \star & \star \\ 
\star & \star & \star & \star & \star & \star \\ 
\star & \star & \star & \star & \star & \star \\ 
\star & \star & \star & \star & \star & \star \\ 
\star & \star & \star & \star & \star & \star
\end{array}
\right ) \ . 
\eeq
The two-loop mixing of physical operators $\vec{\Qmisiak}$ into
evanescent operators $\vec{\Emisiak}$ is given by 
{%
\renewcommand{\arraycolsep}{3pt}
\beq \label{eq:Z21QE}
\hat{Z}^{(2, 1)}_{\QE} =
\left (
\begin{array}{cccccccccccccccc}
{\scriptstyle \f{1531}{288}} {\scriptstyle -\f{5}{216} \nf} &
{\scriptstyle -\f{1}{72}} {\scriptstyle -\f{1}{81} \nf} & 0 & 0 & 0 &
0 & 0 & 0 & {\scriptstyle \f{1}{384}} & {\scriptstyle -\f{35}{864}} &
\star & \star & 0 & 0 & \star & \star \\  
{\scriptstyle \f{119}{16}} {\scriptstyle -\f{1}{18} \nf} &
{\scriptstyle \f{8}{9}} & 0 & 0 & 0 & 0 & 0 & 0 & {\scriptstyle
-\f{35}{192}} & {\scriptstyle -\f{7}{72}} & \star & \star & 0 & 0 &
\star & \star \\  
 0 & 0 & {\scriptstyle -\f{7}{72}} & {\scriptstyle -\f{35}{192}} & 0 &
0 & 0 & 0 & 0 & 0 & \star & \star & 0 & 0 & \star & \star \\  
 0 & 0 & {\scriptstyle -\f{35}{864}} & {\scriptstyle \f{1}{384}} & 0 &
0 & 0 & 0 & 0 & 0 & \star & \star & 0 & 0 & \star & \star \\  
 0 & 0 & {\scriptstyle \f{23}{18}} & {\scriptstyle \f{51}{4}}
{\scriptstyle -\f{1}{18} \nf} & 0 & 0 & 0 & 0 & 0 & 0 & \star & \star
& 0 & 0 & \star & \star \\  
 0 & 0 & {\scriptstyle \f{7}{6}} {\scriptstyle -\f{1}{81} \nf} &
{\scriptstyle \f{317}{72}} {\scriptstyle -\f{5}{216} \nf} & 0 & 0 & 0
& 0 & 0 & 0 & \star & \star & 0 & 0 & \star & \star 
\end{array}
\right ) \, . 
\eeq
}%
Finally, the two-loop mixing among evanescent operators
$\vec{\Emisiak}$ reads 
{%
\renewcommand{\arraycolsep}{1pt}
\beq \label{eq:Z21EE}
\hat{Z}^{(2, 1)}_{\EE} =
\left (
\begin{array}{cccccccccccccccc}
\star & \star & \star & \star & \star & \star & \star & \star & \star
& \star & \star & \star & \star & \star & \star & \star \\  
\star & \star & \star & \star & \star & \star & \star & \star & \star
& \star & \star & \star & \star & \star & \star & \star \\  
\star & \star & \star & \star & \star & \star & \star & \star & \star
& \star & \star & \star & \star & \star & \star & \star \\  
\star & \star & \star & \star & \star & \star & \star & \star & \star
& \star & \star & \star & \star & \star & \star & \star \\  
0 & 0 & {\scriptstyle \f{145}{216}} & {\scriptstyle \f{695}{576} -
\f{1}{108} \nf} & {\scriptstyle -\f{157}{9} + 4 \nf} & {\scriptstyle 
-\f{1319}{12} + \f{23}{9} \nf} & {\scriptstyle \f{17}{6}} &
{\scriptstyle \f{133}{12} - \f{1}{18} \nf} & 0 & 0 & \star & \star &
{\scriptstyle -\f{7}{72}} & {\scriptstyle -\f{35}{192}} & \star &
\star \\ 
0 & 0 & {\scriptstyle \f{1703}{2592} - \f{1}{486} \nf} &
{\scriptstyle \f{2035}{1152} - \f{5}{1296} \nf} & {\scriptstyle
-\f{743}{54} + \f{46}{81} \nf} & {\scriptstyle -\f{2819}{36} +
\f{223}{108} \nf} & {\scriptstyle \f{43}{54} - \f{1}{81} \nf} &
{\scriptstyle \f{379}{72} - \f{5}{216} \nf} & 0 & 0 & \star & \star &
{\scriptstyle -\f{35}{864}} & {\scriptstyle \f{1}{384}} & \star &
\star \\ 
\star & \star & \star & \star & \star & \star & \star & \star & \star
& \star & \star & \star & \star & \star & \star & \star \\  
\star & \star & \star & \star & \star & \star & \star & \star & \star
& \star & \star & \star & \star & \star & \star & \star \\  
\star & \star & \star & \star & \star & \star & \star & \star & \star
& \star & \star & \star & \star & \star & \star & \star \\  
\star & \star & \star & \star & \star & \star & \star & \star & \star
& \star & \star & \star & \star & \star & \star & \star \\  
\star & \star & \star & \star & \star & \star & \star & \star & \star
& \star & \star & \star & \star & \star & \star & \star \\  
\star & \star & \star & \star & \star & \star & \star & \star & \star
& \star & \star & \star & \star & \star & \star & \star \\  
\star & \star & \star & \star & \star & \star & \star & \star & \star
& \star & \star & \star & \star & \star & \star & \star \\  
\star & \star & \star & \star & \star & \star & \star & \star & \star
& \star & \star & \star & \star & \star & \star & \star \\  
\star & \star & \star & \star & \star & \star & \star & \star & \star
& \star & \star & \star & \star & \star & \star & \star \\  
\star & \star & \star & \star & \star & \star & \star & \star & \star
& \star & \star & \star & \star & \star & \star & \star 
\end{array}
\right ) \, .  
\eeq
}%
As far as the one-loop renormalization constant matrices are
concerned, let us note, that our results agree with the findings of 
\cite{Chetyrkin:1997gb}, after taking into account that the definition
of $\Emisiak^{(1)}_5$--$\Emisiak^{(1)}_8$ adopted in
\Eq{eq:additionaloneevanescentoperatormisiak} differs slightly from
definition of $\Emisiak^{(1)}_5$--$\Emisiak^{(1)}_8$ used in the
latter article. On the other hand, the two-loop renormalization
constant matrices involving the insertion of $\Emisiak^{(1)}_5$ and
$\Emisiak^{(1)}_6$, are entirely new and have to our knowledge never
been computed before. Finally, let us mention that while the matrices 
$\hat{Z}^{(1, 0)}_{\QE}$, $\hat{Z}^{(1, 1)}_{\QQ}$, $\hat{Z}^{(1,
1)}_{\EE}$, $\hat{Z}^{(2, 1)}_{\QE}$, $\hat{Z}^{(2, 1)}_{\EE}$ do not
depend on the special choice of trace evanescent operators used
throughout the calculation, the $\nf$ parts of $\hat{Z}^{(2,
0)}_{\EQ}$ do depend on the latter choice. The $\nf$ parts of
$\hat{Z}^{(2, 0)}_{\EQ}$ as given in \Eq{eq:Z20EQ} correspond to the  
specific choice of trace evanescent operators adopted in 
\Eqsand{eq:onelooptraceevanescentoperators}{eq:twolooptraceevanescentoperators}.

\subsection{Wilson Coefficients in the ``Traditional'' Operator Basis}
\label{app:wilsoncoefficientsinthetraditionaloperatorbasis} 

In this appendix we shall use the ADM given in \Eqs{eq:gamma0old},
\eq{eq:gamma1old} and \eq{eq:gamma2old} to find the explicit NNLO
expressions for the Wilson coefficients in the ``traditional'' basis:
\beq \label{eq:asexpansionwilsoncoefficientsold}
C'_i (\mub) = C'^{(0)}_i (\mub) + \f{\as (\mub)}{4 \pi} C'^{(1)}_i
(\mub) + \left ( \f{\as (\mub)}{4 \pi} \right )^2 C'^{(2)}_i (\mub) \,
,   
\eeq 
with $i = 1$--$6$. Using the general solution of the RGE given in 
\Eq{eq:asevolutionmatrixexpansion}, we arrive at  
\beq 
\begin{split}
C'^{(0)}_i (\mub) & = \sum^6_{j = 1} c'^{(0)}_{0, i j} \eta^{a'_j} \,
, \\[2mm]  
C'^{(1)}_i (\mub) & = \sum^6_{j = 1} \left ( c'^{(1)}_{0, i j} + 
c'^{(1)}_{1, i j} \eta + e'^{(1)}_{1, i j} \eta \widetilde{E}_0 (x_t) 
\right ) \eta^{a'_j} \, , \\[2mm]  
C'^{(2)}_i (\mub) & = \sum^6_{j = 1} \bigg ( c'^{(2)}_{0, i j} + 
c'^{(2)}_{1, i j} \eta + c'^{(2)}_{2, i j} \eta^2 + \left [ e'^{(2)}_{1,
i j} \eta + e'^{(2)}_{2, i j} \eta^2 \right ] \widetilde{E}_0 (x_t) \\
& + t'^{(2)}_{2, i j} \eta^2 \widetilde{T}_0 (x_t) + e'^{(1)}_{1, i j} 
\eta^2 \widetilde{E}_1 (x_t) + g'^{(2)}_{2, i j} \eta^2 \widetilde{G}_1
(x_t) \bigg ) \eta^{a'_j} \, ,       
\end{split}
\eeq
with 
\begin{align} 
\label{eq:avectorold}
\vec{a}'^T & =
\left ( 
\begin{array}{cccccc}
\f{6}{23} & -\f{12}{23} & 0.4086 & -0.4230 & -0.8994 & 0.1456 
\end{array} 
\right )
\, , \\[2mm]
\label{eq:c00matrixold}
\hat{c}'^{(0)}_0 & = \left ( 
\begin{array}{cccccc}
\f{1}{2} & -\f{1}{2} & 0 & 0 & 0 & 0 \\[1mm]
\f{1}{2} & \f{1}{2} & 0 & 0 & 0 & 0 \\[1mm]
-\f{1}{14} & \f{1}{6} & 0.0510 & -0.1403 & -0.0113 & 0.0054 \\[1mm]
-\f{1}{14} & -\f{1}{6} & 0.0984 & 0.1214 & 0.0156 & 0.0026 \\[1mm]
0 & 0 & -0.0397 & 0.0117 & -0.0025 & 0.0304 \\[1mm]
0 & 0 & 0.0335 & 0.0239 & -0.0462 & -0.0112 
\end{array} 
\right ) \, , \\[2mm]
\label{eq:c10matrixold}
\hat{c}'^{(1)}_0 & = \left ( 
\begin{array}{cccccc}
0.8136 & 0.7142 & 0 & 0 & 0 & 0 \\
0.8136 & -0.7142 & 0 & 0 & 0 & 0 \\
-0.0766 & -0.1455 & -0.8848 & 0.4137 & -0.0114 & 0.1722 \\
-0.2353 & -0.0397 & 0.4920 & -0.2758 & 0.0019 & -0.1449 \\
0.0397 & 0.0926 & 0.7342 & -0.1261 & -0.1209 & -0.1085  \\
-0.1190 & -0.2778 & -0.5544 & 0.1915 & -0.2744 & 0.3568
\end{array} 
\right ) \, , \\[2mm]
\label{eq:c11matrixold}
\hat{c}'^{(1)}_1 & = \left ( 
\begin{array}{cccccc}
1.0197 & 2.9524 & 0 & 0 & 0 & 0 \\
1.0197 & -2.9524 & 0 & 0 & 0 & 0 \\
-0.1457 & -0.9841 & 0.3299 & 1.2188 & 0.1463 & -0.0328 \\
-0.1457 & 0.9841 & 0.6370 & -1.0547 & -0.2032 & -0.0160 \\
0 & 0 & -0.2568 & -0.1014 & 0.0320 & -0.1847 \\
0 & 0 & 0.2168 & -0.2074 & 0.6000 & 0.0679
\end{array} 
\right ) \, , \\[2mm]
\label{eq:e11matrixold}
\hat{e}'^{(1)}_1 & = \left (  
\begin{array}{cccccc}
0 & 0 & 0 & 0 & 0 & 0 \\
0 & 0 & 0 & 0 & 0 & 0 \\
0 & 0 & 0.1494 & -0.3725 & 0.0738 & -0.0173 \\
0 & 0 & 0.2885 & 0.3224 & -0.1025 & -0.0084 \\
0 & 0 & -0.1163 & 0.0310 & 0.0162 & -0.0975 \\
0 & 0 & 0.0982 & 0.0634 & 0.3026 & 0.0358
\end{array} 
\right ) \, , \\[2mm]
\label{eq:c20matrixold}
\hat{c}'^{(2)}_0 & = \left ( 
\begin{array}{cccccc}
7.9372 & 23.1398 & 0 & 0 & 0 & 0 \\ 
19.9372 & -23.1398 & 0 & 0 & 0 & 0 \\  
-1.8694 & -20.8668 & 5.1370 & 15.9499 & 3.1041 &  1.5552 \\  
2.7077 &  3.4331 & -1.0616 & -0.5150 & -1.8471 & 0.6903 \\ 
-0.1694 & 11.7230 & -5.3883 & -11.6376 & -0.9016 & 0.1902 \\   
-7.6029 & -7.4283 & 9.8588 & 1.6456 & 1.9006 & 1.6776
\end{array} 
\right ) \, , \\[2mm]
\label{eq:c21matrixold}
\hat{c}'^{(2)}_1 & = \left ( 
\begin{array}{cccccc}
1.6593 & -4.2175 & 0 & 0 & 0 & 0 \\
1.6593 & 4.2175 & 0 & 0 & 0 & 0 \\
-0.1561 & 0.8591 & -5.7288 & -3.5925 & 0.1475 & -1.0445 \\
-0.4798 & 0.2344 & 3.1857 & 2.3950 & -0.0252 & 0.8789 \\
0.0809 & -0.5467 & 4.7534 & 1.0956 & 1.5714 & 0.6584 \\
-0.2428 & 1.6402 & -3.5892 & -1.6632 & 3.5653 & -2.1649
\end{array} 
\right ) \, , \\[2mm]
\label{eq:c22matrixold}
\hat{c}'^{(2)}_2 & = \left ( 
\begin{array}{cccccc}
16.3114 & 24.9044 & 0 & 0 & 0 & 0 \\
16.3114 & -24.9044 & 0 & 0 & 0 & 0 \\
-2.3302 & -8.3015 & 0.4831 & 15.7207 & 1.4758 & 0.1230 \\
-2.3302 & 8.3015 & 0.9329 & -13.6035 & -2.0495 & 0.0598 \\
0 & 0 & -0.3760 & -1.3081 & 0.3232 & 0.6922 \\ 
0 & 0 & 0.3174 & -2.6752 & 6.0523 & -0.2543
\end{array} 
\right ) \, , \\[2mm]
\label{eq:e21matrixold}
\hat{e}'^{(2)}_1 & = \left (  
\begin{array}{cccccc}
0 & 0 & 0 & 0 & 0 & 0 \\
0 & 0 & 0 & 0 & 0 & 0 \\
0 & 0 & -2.5948 & 1.0981 & 0.0744 & -0.5514 \\
0 & 0 & 1.4429 & -0.7320 & -0.0127 & 0.4640 \\
0 & 0 & 2.1530 & -0.3349 & 0.7925 & 0.3476 \\
0 & 0 & -1.6257 & 0.5084 & 1.7980 & -1.1429
\end{array} 
\right ) \, , \\[2mm]
\label{eq:e22matrixold}
\hat{e}'^{(2)}_2 & = \left (  
\begin{array}{cccccc}
0 & 0 & 0 & 0 & 0 & 0 \\
0 & 0 & 0 & 0 & 0 & 0 \\
0 & 0 & 1.0180 & 1.7567 & 0.2494 & -0.1948 \\
0 & 0 & 1.9657 & -1.5201 & -0.3463 & -0.0948 \\
0 & 0 & -0.7924 & -0.1462 & 0.0546 & -1.0965 \\
0 & 0 & 0.6689 & -0.2989 & 1.0227 & 0.4029 
\end{array} 
\right ) \, , \\[2mm]
\label{eq:t22matrixold}
\hat{t}'^{(2)}_2 & = \left ( 
\begin{array}{cccccc}
-\f{1}{6} & -\f{1}{3} & 0 & 0 & 0 & 0 \\[1mm]
-\f{1}{6} & \f{1}{3} & 0 & 0 & 0 & 0 \\[1mm]
\f{1}{42} & \f{1}{9} & -0.0100 & -0.1172 & -0.0047 & -0.0029 \\[1mm]
\f{1}{42} & -\f{1}{9} & -0.0193 & 0.1015 & 0.0066 & -0.0014 \\[1mm]
0 & 0 & 0.0078 & 0.0098 & -0.0010 & -0.0165 \\[1mm]
0 & 0 & -0.0066 & 0.0200 & -0.0194 & 0.0061
\end{array} 
\right ) \, , \\[2mm]
\label{eq:g22matrixold}
\hat{g}'^{(2)}_2 & = \left (  
\begin{array}{cccccc}
0 & 0 & 0 & 0 & 0 & 0 \\
0 & 0 & 0 & 0 & 0 & 0 \\
0 & 0 & -0.5842 & 0.3877 & 0.0445 & 0.0520 \\
0 & 0 & -1.1280 & -0.3355 & -0.0618 & 0.0253 \\
0 & 0 & 0.4547 & -0.0323 & 0.0097 & 0.2928 \\
0 & 0 & -0.3839 & -0.0660 & 0.1824 & -0.1076
\end{array} 
\right ) \, .  
\end{align}
As far as the LO and NLO corrections parameterized by 
$\hat{c}'^{(0)}_0$, $\hat{c}'^{(1)}_0$, $\hat{c}'^{(1)}_1$ and
$\hat{e}'^{(1)}_1$ are concerned our results agree perfectly with the
findings of \cite{Buras:1991jm, Ciuchini:1993vr,
Buras:1993dy}. Contrariwise, the resummation of the NNLO logarithms is
entirely new, and the corresponding matrices $\hat{c}'^{(2)}_0$,
$\hat{c}'^{(2)}_1$, $\hat{c}'^{(2)}_2$, $\hat{e}'^{(2)}_1$,
$\hat{e}'^{(2)}_2$, $\hat{t}'^{(2)}_2$ and $\hat{g}'^{(2)}_2$ have
never been computed before. Finally, let us mention that while the
matrices $\hat{c}'^{(2)}_1$, $\hat{c}'^{(2)}_2$, $\hat{e}'^{(2)}_1$,
$\hat{e}'^{(2)}_2$, $\hat{t}'^{(2)}_2$ and $\hat{g}'^{(2)}_2$ do not
depend on the special choice of trace evanescent operators used
throughout the calculation, the entries of $\hat{c}'^{(2)}_0$
describing the evolution of the Wilson coefficients of the QCD penguin
operators do depend on the latter choice. The numbers given in
\Eq{eq:c20matrixold} correspond to the specific choice of trace
evanescent operators adopted in
\Eqsand{eq:onelooptraceevanescentoperators}{eq:twolooptraceevanescentoperators}.
Recalling what has been said earlier on the cancellation of scheme
dependences in general, it should be clear, that the associated scheme
dependence is canceled by the one of the two-loop $\ord (\as^2)$
matrix elements, which have to be calculated using the same definition
of trace evanescent operators.

\end{document}

%% file: oneloopeom1.tex
\unitlength=1bp%

\begin{feynartspicture}(55,70)(1,1.3)

\FADiagram{}
\FAProp(0.,10.)(10.,10.)(0.,){/Straight}{1}
\FALabel(5.,11.07)[b]{${\scriptscriptstyle b}$}
\FAProp(20.,10.)(10.,10.)(0.,){/Straight}{-1}
\FALabel(15.,11.07)[b]{${\scriptscriptstyle s}$}
\FAProp(10.,10.)(10,3.5)(1.,){/Straight}{1}
\FALabel(4.65,8.)[t]{${\scriptscriptstyle c}$}
\FAProp(10.,3.5)(10.,10.)(1.,){/Straight}{1}
\FALabel(15.,8.)[t]{${\scriptscriptstyle c}$}
\FAProp(10.,3.5)(10.,-1.5)(0.,){/Cycles}{0}
\FALabel(12.5,-0.5)[b]{${\scriptscriptstyle g}$}
\FALabel(10.,10.)[c]{${\scriptscriptstyle \blacksquare}$}
\FALabel(10.,12.07)[b]{$\scriptscriptstyle Q_{\scalebox{0.4}{2}}$}
\FAVert(10.,3.5){0}

\end{feynartspicture}

%% file: oneloopeom2.tex
\unitlength=1bp%

\begin{feynartspicture}(55,70)(1,1.3)

\FADiagram{}
\FAProp(0.,10.)(10.,10.)(0.,){/Straight}{1}
\FALabel(5.,11.07)[b]{${\scriptscriptstyle b}$}
\FAProp(20.,10.)(10.,10.)(0.,){/Straight}{-1}
\FALabel(15.,11.07)[b]{${\scriptscriptstyle s}$}
\FAProp(7.15,5.25)(12.85,5.25)(0.6,){/Straight}{1}
\FAProp(10.,10.)(7.15,5.25)(0.6,){/Straight}{1}
\FAProp(12.85,5.25)(10.,10.)(0.6,){/Straight}{1}
\FALabel(10.,2.25)[t]{${\scriptscriptstyle c}$}
\FALabel(15.,8.)[t]{${\scriptscriptstyle c}$}
\FALabel(4.75,8.)[t]{${\scriptscriptstyle c}$}
\FALabel(10.,10.)[c]{${\scriptscriptstyle \blacksquare}$}
\FALabel(10.,12.07)[b]{$\scriptscriptstyle Q_{\scalebox{0.4}{2}}$}
\FAProp(7.15,5.25)(3.5,-0.5)(0.,){/Cycles}{0}
\FALabel(2.85,2.25)[b]{${\scriptscriptstyle g}$}
\FAProp(12.85,5.25)(16.5,-0.5)(0.,){/Cycles}{0}
\FALabel(17.45,2.25)[b]{${\scriptscriptstyle g}$}
\FAVert(7.15,5.25){0}
\FAVert(12.85,5.25){0}

\end{feynartspicture}

%% file: oneloopeom3.tex
\unitlength=1bp%

\begin{feynartspicture}(55,70)(1,1.3)

\FADiagram{}
\FAProp(0.,10.)(10.,10.)(0.,){/Straight}{1}
\FALabel(5.,11.07)[b]{${\scriptscriptstyle b}$}
\FAProp(20.,10.)(10.,10.)(0.,){/Straight}{-1}
\FALabel(15.,11.07)[b]{${\scriptscriptstyle s}$}
\FAProp(7.15,5.25)(12.85,5.25)(0.6,){/Straight}{0}
\FAProp(10.,10.)(7.15,5.25)(0.6,){/Straight}{1}
\FAProp(12.85,5.25)(10.,10.)(0.6,){/Straight}{1}
\FALabel(15.,8.)[t]{${\scriptscriptstyle c}$}
\FALabel(4.75,8.)[t]{${\scriptscriptstyle c}$}
\FALabel(10.,10.)[c]{${\scriptscriptstyle \blacksquare}$}
\FALabel(10.,12.07)[b]{$\scriptscriptstyle Q_{\scalebox{0.4}{2}}$}
\FAProp(7.15,5.25)(3.5,-0.5)(0.,){/Cycles}{0}
\FALabel(2.85,2.25)[b]{${\scriptscriptstyle g}$}
\FAProp(12.85,5.25)(16.5,-0.5)(0.,){/Cycles}{0}
\FALabel(17.45,2.25)[b]{${\scriptscriptstyle g}$}
\FAProp(10.,3.65)(10.,-1.5)(0.,){/Cycles}{0}
\FALabel(12.5,-0.5)[b]{${\scriptscriptstyle g}$}
\FAVert(7.15,5.25){0}
\FAVert(12.85,5.25){0}
\FAVert(10.,3.65){0}

\end{feynartspicture}

%% file: twoloopeom1.tex
\unitlength=1bp%

\begin{feynartspicture}(55,70)(1,1.3)

\FADiagram{}
\FAProp(0.,10.)(2.,10.)(0.,){/Straight}{0}
\FAProp(2.,10.)(10.,10.)(0.,){/Straight}{1}
\FALabel(5.75,11.07)[b]{${\scriptscriptstyle b}$}
\FAProp(20.,10.)(10.,10.)(0.,){/Straight}{-1}
\FALabel(15.,11.07)[b]{${\scriptscriptstyle s}$}
\FAProp(10.,10.)(6.8,6.7)(0.425,){/Straight}{1}
\FAProp(6.8,6.7)(10,3.5)(0.425,){/Straight}{1}
\FAProp(10.,3.5)(10.,10.)(1.,){/Straight}{1}
\FALabel(15.,8.)[t]{${\scriptscriptstyle c}$}
\FAProp(10.,3.5)(10.,-1.5)(0.,){/Cycles}{0}
\FALabel(12.5,-0.5)[b]{${\scriptscriptstyle g}$}
\FALabel(10.,10.)[c]{${\scriptscriptstyle \blacksquare}$}
\FALabel(10.,12.07)[b]{$\scriptscriptstyle Q_{\scalebox{0.4}{2}}$}
\FAVert(10.,3.5){0}
\FAProp(2.,10.)(6.8,6.7)(0.4,){/Cycles}{0}
\FALabel(3.,4.5)[b]{${\scriptscriptstyle g}$}
\FAVert(6.8,6.7){0}
\FAVert(2.,10.){0}

\end{feynartspicture}

%% file: twoloopeom2.tex
\unitlength=1bp%

\begin{feynartspicture}(55,70)(1,1.3)

\FADiagram{}
\FAProp(0.,10.)(2.,10.)(0.,){/Straight}{0}
\FAProp(2.,10.)(10.,10.)(0.,){/Straight}{1}
\FALabel(5.75,11.07)[b]{${\scriptscriptstyle b}$}
\FAProp(20.,10.)(10.,10.)(0.,){/Straight}{-1}
\FALabel(15.,11.07)[b]{${\scriptscriptstyle s}$}
\FAProp(7.15,5.25)(12.85,5.25)(0.6,){/Straight}{1}
\FAProp(10.,10.)(6.8,6.7)(0.45,){/Straight}{1}
\FAProp(6.8,6.7)(7.15,5.25)(0.3,){/Straight}{0}
\FAProp(12.85,5.25)(10.,10.)(0.6,){/Straight}{1}
\FALabel(10.,2.25)[t]{${\scriptscriptstyle c}$}
\FALabel(15.,8.)[t]{${\scriptscriptstyle c}$}
\FALabel(10.,10.)[c]{${\scriptscriptstyle \blacksquare}$}
\FALabel(10.,12.07)[b]{$\scriptscriptstyle Q_{\scalebox{0.4}{2}}$}
\FAProp(2.,10.)(6.8,6.7)(0.4,){/Cycles}{0}
\FALabel(3.,4.5)[b]{${\scriptscriptstyle g}$}
\FAProp(7.15,5.25)(3.5,-0.5)(0.,){/Cycles}{0}
\FALabel(2.85,2.25)[b]{${\scriptscriptstyle g}$}
\FAProp(12.85,5.25)(16.5,-0.5)(0.,){/Cycles}{0}
\FALabel(17.45,2.25)[b]{${\scriptscriptstyle g}$}
\FAVert(6.8,6.7){0}
\FAVert(2.,10.){0}
\FAVert(7.15,5.25){0}
\FAVert(12.85,5.25){0}

\end{feynartspicture}

%% file: twoloopeom3.tex
\unitlength=1bp%

\begin{feynartspicture}(55,70)(1,1.3)

\FADiagram{}
\FAProp(0.,10.)(2.,10.)(0.,){/Straight}{0}
\FAProp(2.,10.)(10.,10.)(0.,){/Straight}{1}
\FALabel(5.75,11.07)[b]{${\scriptscriptstyle b}$}
\FAProp(20.,10.)(10.,10.)(0.,){/Straight}{-1}
\FALabel(15.,11.07)[b]{${\scriptscriptstyle s}$}
\FAProp(7.15,5.25)(12.85,5.25)(0.6,){/Straight}{0}
\FAProp(10.,10.)(6.8,6.7)(0.45,){/Straight}{1}
\FAProp(6.8,6.7)(7.15,5.25)(0.3,){/Straight}{0}
\FAProp(12.85,5.25)(10.,10.)(0.6,){/Straight}{1}
\FALabel(15.,8.)[t]{${\scriptscriptstyle c}$}
\FALabel(10.,10.)[c]{${\scriptscriptstyle \blacksquare}$}
\FALabel(10.,12.07)[b]{$\scriptscriptstyle Q_{\scalebox{0.4}{2}}$}
\FAProp(2.,10.)(6.8,6.7)(0.4,){/Cycles}{0}
\FALabel(3.,4.5)[b]{${\scriptscriptstyle g}$}
\FAProp(7.15,5.25)(3.5,-0.5)(0.,){/Cycles}{0}
\FALabel(2.85,2.25)[b]{${\scriptscriptstyle g}$}
\FAProp(12.85,5.25)(16.5,-0.5)(0.,){/Cycles}{0}
\FALabel(17.45,2.25)[b]{${\scriptscriptstyle g}$}
\FAProp(10.,3.65)(10.,-1.5)(0.,){/Cycles}{0}
\FALabel(12.5,-0.5)[b]{${\scriptscriptstyle g}$}
\FAVert(6.8,6.7){0}
\FAVert(2.,10.){0}
\FAVert(7.15,5.25){0}
\FAVert(12.85,5.25){0}
\FAVert(10.,3.65){0}

\end{feynartspicture}

%% file: brstexact1.tex
\unitlength=1bp%

\begin{feynartspicture}(55,70)(1,1.3)

\FADiagram{}
\FAProp(0.,10.)(10.,10.)(0.,){/Straight}{1}
\FALabel(5.,11.07)[b]{${\scriptscriptstyle b}$}
\FAProp(20.,10.)(10.,10.)(0.,){/Straight}{-1}
\FALabel(15.,11.07)[b]{${\scriptscriptstyle s}$}
\FAProp(7.15,5.25)(12.85,5.25)(0.6,){/Straight}{1}
\FAProp(10.,10.)(7.15,5.25)(0.6,){/Straight}{1}
\FAProp(12.85,5.25)(10.,10.)(0.6,){/Straight}{1}
\FALabel(10.,2.25)[t]{${\scriptscriptstyle c}$}
\FALabel(15.,8.)[t]{${\scriptscriptstyle c}$}
\FALabel(4.75,8.)[t]{${\scriptscriptstyle c}$}
\FALabel(10.,10.)[c]{${\scriptscriptstyle \blacksquare}$}
\FALabel(10.,12.07)[b]{$\scriptscriptstyle Q_{\scalebox{0.4}{2}}$}
\FAProp(7.15,5.25)(3.5,-0.5)(0.,){/Cycles}{0}
\FALabel(2.85,2.25)[b]{${\scriptscriptstyle g}$}
\FAProp(12.85,5.25)(16.5,-0.5)(0.,){/Cycles}{0}
\FALabel(17.45,2.25)[b]{${\scriptscriptstyle g}$}
\FAProp(0.,-0.5)(3.5,-0.5)(0.,){/GhostDash}{0}
\FAProp(3.5,-0.5)(16.5,-0.5)(0.,){/GhostDash}{0}
\FAProp(16.5,-0.5)(20.,-0.5)(0.,){/GhostDash}{0}
\FALabel(1.5,-3.73)[b]{${\scriptscriptstyle u^{\scalebox{0.4}{a}}}$}
\FALabel(10.,-3.73)[b]{${\scriptscriptstyle u^{\scalebox{0.4}{a}}}$}
\FALabel(18.25,-3.73)[b]{${\scriptscriptstyle u^{\scalebox{0.4}{a}}}$}
\FAVert(7.15,5.25){0}
\FAVert(12.85,5.25){0}
\FAVert(3.5,-0.5){0}
\FAVert(16.5,-0.5){0}

\end{feynartspicture}

%% file: brstexact2.tex
\unitlength=1bp%

\begin{feynartspicture}(55,70)(1,1.3)

\FADiagram{}
\FAProp(0.,10.)(10.,10.)(0.,){/Straight}{1}
\FALabel(5.,11.07)[b]{${\scriptscriptstyle b}$}
\FAProp(20.,10.)(10.,10.)(0.,){/Straight}{-1}
\FALabel(15.,11.07)[b]{${\scriptscriptstyle s}$}
\FAProp(10.,10.)(10,3.5)(1.,){/Straight}{1}
\FALabel(4.65,8.)[t]{${\scriptscriptstyle c}$}
\FAProp(10.,3.5)(10.,10.)(1.,){/Straight}{1}
\FALabel(15.,8.)[t]{${\scriptscriptstyle c}$}
\FAProp(10.,3.5)(10.,-1.5)(0.,){/Cycles}{0}
\FALabel(12.5,-0.5)[b]{${\scriptscriptstyle g}$}
\FALabel(10.,10.)[c]{${\scriptscriptstyle \blacksquare}$}
\FALabel(10.,12.07)[b]{$\scriptscriptstyle Q_{\scalebox{0.4}{2}}$}
\FAVert(10.,3.5){0}

\end{feynartspicture}

%% file: brstexact3.tex
\unitlength=1bp%

\begin{feynartspicture}(55,70)(1,1.3)

\FADiagram{}
\FAProp(0.,10.)(4.,10.)(0.,){/Straight}{1}
\FAProp(4.,10.)(16.,10.)(0.,){/Straight}{1}
\FAProp(16.,10.)(20.,10.)(0.,){/Straight}{1}
\FALabel(1.,11.07)[b]{${\scriptscriptstyle b}$}
\FALabel(10.,11.07)[b]{${\scriptscriptstyle s}$}
\FALabel(18.1,11.07)[b]{${\scriptscriptstyle s}$}
\FAProp(4.,10.)(4.,-0.5)(0.,){/Cycles}{0}
\FALabel(2.,3.5)[b]{${\scriptscriptstyle g}$}
\FALabel(4.,10.)[c]{${\scriptscriptstyle \blacksquare}$}
\FALabel(5.,12.07)[b]{$\scriptscriptstyle N^{\scalebox{0.4}{(1)}}_{\scalebox{0.4}{1}}$}
\FAProp(16.,10.)(16.,-0.5)(0.,){/Cycles}{0}
\FALabel(18.5,3.5)[b]{${\scriptscriptstyle g}$}
\FAProp(0.,-0.5)(4.,-0.5)(0.,){/GhostDash}{0}
\FAProp(4.,-0.5)(16.,-0.5)(0.,){/GhostDash}{0}
\FAProp(16.,-0.5)(20.,-0.5)(0.,){/GhostDash}{0}
\FALabel(2.,-3.73)[b]{${\scriptscriptstyle u^{\scalebox{0.4}{a}}}$}
\FALabel(10.,-3.73)[b]{${\scriptscriptstyle u^{\scalebox{0.4}{a}}}$}
\FALabel(18.,-3.73)[b]{${\scriptscriptstyle u^{\scalebox{0.4}{a}}}$}
\FAVert(16.,10.){0}
\FAVert(16.,-0.5){0}
\FAVert(4.,-0.5){0}

\end{feynartspicture}

%% file: brstexact4.tex
\unitlength=1bp%

\begin{feynartspicture}(55,70)(1,1.3)

\FADiagram{}
\FAProp(0.,10.)(10.,10.)(0.,){/Straight}{1}
\FALabel(5.,11.07)[b]{${\scriptscriptstyle b}$}
\FAProp(20.,10.)(10.,10.)(0.,){/Straight}{-1}
\FALabel(15.,11.07)[b]{${\scriptscriptstyle s}$}
\FAProp(10.,10.)(10,3.5)(1.,){/GhostDash}{0}
\FALabel(4.35,8.)[t]{${\scriptscriptstyle u^{\scalebox{0.4}{a}}}$}
\FAProp(10.,3.5)(10.,10.)(1.,){/GhostDash}{0}
\FALabel(15.7,8.)[t]{${\scriptscriptstyle u^{\scalebox{0.4}{a}}}$}
\FAProp(10.,3.5)(10.,-1.5)(0.,){/Cycles}{0}
\FALabel(12.5,-0.5)[b]{${\scriptscriptstyle g}$}
\FALabel(10.,10.)[c]{${\scriptscriptstyle \blacksquare}$}
\FALabel(10.,12.07)[b]{$\scriptscriptstyle B_{\scalebox{0.4}{1}}^{\scalebox{0.4}{(2)}}$} 
\FAVert(10.,3.5){0}

\end{feynartspicture}

%% file: matching1.tex
\unitlength=1bp%

\begin{feynartspicture}(55,70)(1,1.3)

\FADiagram{}
\FAProp(0.,15.)(5.,14.5)(0.,){/Straight}{1}
\FAProp(5.,14.5)(15.,14.5)(0.,){/Straight}{1}
\FAProp(15.,14.5)(20.,15.)(0.,){/Straight}{1}
\FALabel(2.54577,17.5377)[t]{$\scriptscriptstyle b$}
\FALabel(10.,17.5377)[t]{$\scriptscriptstyle b$}
\FALabel(17.34577,16.9377)[t]{$\scriptscriptstyle c$}
\FAProp(0.,5.)(5.,5.5)(0.,){/Straight}{1}
\FAProp(5.,5.5)(15.,5.5)(0.,){/Straight}{1}
\FAProp(15.,5.5)(20.,5.)(0.,){/Straight}{1}
\FALabel(2.35423,4.13769)[t]{$\scriptscriptstyle c$}
\FALabel(10.,4.13769)[t]{$\scriptscriptstyle c$}
\FALabel(17.34577,4.13769)[t]{$\scriptscriptstyle s$}
\FAProp(15.,14.5)(15.,5.5)(0.,){/Sine}{1}
\FALabel(14.33,10.)[r]{$\scriptscriptstyle W$}
\FAProp(5.,14.5)(5.,11.5)(0.,){/Cycles}{0}
\FAProp(5.,8.5)(5.,5.5)(0.,){/Cycles}{0}
\FALabel(4.03,12.9)[r]{$\scriptscriptstyle g$}
\FALabel(4.03,7.)[r]{$\scriptscriptstyle g$}
\FAProp(5.,11.5)(5.,8.5)(1.,){/Straight}{1}
\FAProp(5.,8.5)(5.,11.5)(1.,){/Straight}{1}
\FALabel(8.43,10.)[r]{$\scriptscriptstyle t$}
\FALabel(2.73,10.)[r]{$\scriptscriptstyle t$}
\FAVert(5.,14.5){0}
\FAVert(5.,5.5){0}
\FAVert(5.,11.5){0}
\FAVert(5.,8.5){0}
\FAVert(15.,14.5){0}
\FAVert(15.,5.5){0}

\end{feynartspicture}

%% file: matching2.tex
\unitlength=1bp%

\begin{feynartspicture}(55,70)(1,1.3)

\FADiagram{}
\FAProp(0.,15.)(5.,14.5)(0.,){/Straight}{1}
\FAProp(5.,14.5)(10.,14.5)(0.,){/Straight}{1}
\FAProp(10.,14.5)(15.,14.5)(0.,){/Straight}{1}
\FAProp(15.,14.5)(20.,15.)(0.,){/Straight}{1}
\FALabel(2.54577,17.5377)[t]{$\scriptscriptstyle b$}
\FALabel(7.34577,16.9377)[t]{$\scriptscriptstyle c$}
\FALabel(12.34577,16.9377)[t]{$\scriptscriptstyle c$}
\FALabel(17.34577,16.9377)[t]{$\scriptscriptstyle c$}
\FAProp(0.,5.)(5.,5.5)(0.,){/Straight}{1}
\FAProp(5.,5.5)(15.,5.5)(0.,){/Straight}{1}
\FAProp(15.,5.5)(20.,5.)(0.,){/Straight}{1}
\FALabel(2.35423,4.13769)[t]{$\scriptscriptstyle c$}
\FALabel(10.,4.13769)[t]{$\scriptscriptstyle s$}
\FALabel(17.34577,4.13769)[t]{$\scriptscriptstyle s$}
\FAProp(15.,14.5)(15.,5.5)(0.,){/Cycles}{0}
\FALabel(11.34577,10.4)[r]{$\scriptscriptstyle g$}
\FAProp(5.,14.5)(5.,5.5)(0.,){/Sine}{1}
\FALabel(9.03,10.)[r]{$\scriptscriptstyle W$}
\FAProp(15.,10.)(10.,14.5)(0.,){/Cycles}{0}
\FALabel(18.23,12.4)[r]{$\scriptscriptstyle g$}
\FALabel(18.23,7.5)[r]{$\scriptscriptstyle g$}
\FAVert(5.,14.5){0}
\FAVert(5.,5.5){0}
\FAVert(15.,14.5){0}
\FAVert(15.,5.5){0}
\FAVert(10.,14.5){0}
\FAVert(15.,10.){0}

\end{feynartspicture}

%% file: matching3.tex
\unitlength=1bp%

\begin{feynartspicture}(55,70)(1,1.3)

\FADiagram{}
\FAProp(0.,15.)(5.,14.5)(0.,){/Straight}{1}
\FAProp(5.,14.5)(10.,14.5)(0.,){/Sine}{1}
\FAProp(10.,14.5)(15.,14.5)(0.,){/Straight}{1}
\FAProp(15.,14.5)(20.,15.)(0.,){/Straight}{1}
\FALabel(2.54577,17.5377)[t]{$\scriptscriptstyle b$}
\FALabel(7.34577,17.2377)[t]{$\scriptscriptstyle W$}
\FALabel(12.34577,16.9377)[t]{$\scriptscriptstyle s$}
\FALabel(17.34577,16.9377)[t]{$\scriptscriptstyle s$}
\FAProp(0.,5.)(5.,5.5)(0.,){/Straight}{1}
\FAProp(5.,5.5)(15.,5.5)(0.,){/Straight}{1}
\FAProp(15.,5.5)(20.,5.)(0.,){/Straight}{1}
\FALabel(2.35423,4.13769)[t]{$\scriptscriptstyle q$}
\FALabel(10.,4.13769)[t]{$\scriptscriptstyle q$}
\FALabel(17.34577,4.13769)[t]{$\scriptscriptstyle q$}
\FAProp(15.,14.5)(15.,5.5)(0.,){/Cycles}{0}
\FALabel(3.5,7.5)[r]{$\scriptscriptstyle g$}
\FAProp(5.,14.5)(5.,10.)(0.,){/Straight}{1}
\FALabel(3.5,12.5)[r]{$\scriptscriptstyle t$}
\FAProp(5.,10.)(5.,5.)(0.,){/Cycles}{0}
\FALabel(18.53,10.)[r]{$\scriptscriptstyle g$}
\FAProp(5.,10.)(10.,14.5)(0.,){/Straight}{1}
\FALabel(9.,10.5)[r]{$\scriptscriptstyle t$}
\FAVert(5.,14.5){0}
\FAVert(5.,5.5){0}
\FAVert(15.,14.5){0}
\FAVert(15.,5.5){0}
\FAVert(10.,14.5){0}
\FAVert(5.,10.){0}

\end{feynartspicture}

%% file: matching4.tex
\unitlength=1bp%

\begin{feynartspicture}(55,70)(1,1.3)

\FADiagram{}
\FAProp(0.,15.)(3.25,14.75)(0.,){/Straight}{1}
\FAProp(3.25,14.75)(6.5,14.5)(0.,){/Straight}{1}
\FALabel(1.56888,15.9154)[b]{$\scriptscriptstyle b$}
\FALabel(5.06888,15.7154)[b]{$\scriptscriptstyle b$}
\FAProp(20.,15.)(13.5,14.5)(0.,){/Straight}{-1}
\FALabel(16.6311,15.8154)[b]{$\scriptscriptstyle s$}
\FAProp(10.,8.5)(10.,3.5)(0.,){/Cycles}{0}
\FALabel(13.73,6.)[r]{$\scriptscriptstyle g$}
\FAProp(6.5,14.5)(13.5,14.5)(0.,){/Sine}{1}
\FALabel(10.,15.57)[b]{$\scriptscriptstyle W$}
\FAProp(6.5,14.5)(8.25,11.5)(0.,){/Straight}{1}
\FAProp(8.25,11.5)(10.,8.5)(0.,){/Straight}{1}
\FALabel(6.5089,9.599)[tl]{$\scriptscriptstyle t$}
\FALabel(4.79114,10.599)[tr]{$\scriptscriptstyle g$}
\FAProp(3.25,14.75)(8.25,11.5)(.5,){/Cycles}{0}
\FALabel(12.6089,11.199)[tl]{$\scriptscriptstyle t$}
\FAProp(13.5,14.5)(10.,8.5)(0.,){/Straight}{-1}
\FAVert(6.5,14.5){0}
\FAVert(13.5,14.5){0}
\FAVert(10.,8.5){0}
\FAVert(3.25,14.75){0}
\FAVert(8.25,11.5){0}

\end{feynartspicture}

%% file: mixing1.tex
\unitlength=1bp%

\begin{feynartspicture}(55,70)(1,1.3)

\FADiagram{}
\FAProp(3.,-1.5)(5.5,1.)(0.,){/Straight}{1}
\FAProp(5.5,1.)(7.5,3.)(0.,){/Straight}{0}
\FAProp(7.5,3.)(10.,5.5)(0.,){/Straight}{0}
\FALabel(3.,1.25)[c]{${\scriptscriptstyle c}$}
\FAProp(10.,5.5)(14.5,10.)(0.,){/Straight}{1}
\FALabel(13.5,5.5)[c]{$\scriptscriptstyle Q_{\scalebox{0.4}{2}}$}
\FAProp(14.5,10.)(17.,12.5)(0.,){/Straight}{1}
\FALabel(17.,9.5)[c]{${\scriptscriptstyle c}$}
\FAProp(17.,-1.5)(14.5,1.)(0.,){/Straight}{-1}
\FAProp(14.5,1.)(10.,5.5)(0.,){/Straight}{-1}
\FAProp(10.,5.5)(7.5,8.)(0.,){/Straight}{0}
\FALabel(17.,1.5)[c]{${\scriptscriptstyle s}$}
\FAProp(7.5,8.)(5.5,10.)(0.,){/Straight}{0}
\FAProp(5.5,10.)(3.,12.5)(0.,){/Straight}{-1}
\FALabel(3.,9.5)[c]{${\scriptscriptstyle b}$}
\FALabel(10.,5.5)[c]{${\scriptscriptstyle \blacksquare}$}
\FALabel(10.,1.5)[c]{$\scriptscriptstyle g$}
\FAProp(7.5,8.)(7.5,3.)(0.5,){/Cycles}{0}
\FALabel(4.5,5.5)[c]{$\scriptscriptstyle g$}
\FAProp(5.5,10.)(14.5,10.)(-0.5,){/Cycles}{0}
\FALabel(10.,9.5)[c]{$\scriptscriptstyle g$}
\FAProp(5.5,1.)(14.5,1.)(0.5,){/Cycles}{0}
\FAVert(7.5,8.){0}
\FAVert(7.5,3.){0}
\FAVert(5.5,10.){0}
\FAVert(14.5,10.){0}
\FAVert(5.5,1.){0}
\FAVert(14.5,1.){0}

\end{feynartspicture}

%% file: mixing2.tex
\unitlength=1bp%

\begin{feynartspicture}(55,70)(1,1.3)

\FADiagram{}
\FAProp(0.,10.)(4.,10.)(0.,){/Straight}{1}
\FAProp(4.,10.)(12.,10.)(0.,){/Straight}{1}
\FAProp(12.,10.)(16.,10.)(0.,){/Straight}{1}
\FAProp(16.,10.)(20.,10.)(0.,){/Straight}{1}
\FALabel(1.,11.07)[b]{${\scriptscriptstyle b}$}
\FALabel(8.1,11.07)[b]{${\scriptscriptstyle s}$}
\FALabel(14.,11.07)[b]{${\scriptscriptstyle s}$}
\FALabel(18.1,11.07)[b]{${\scriptscriptstyle s}$}
\FAProp(4.,10.)(4,3.5)(1.,){/Straight}{1}
\FALabel(2.7,7.25)[t]{${\scriptscriptstyle c}$}
\FAProp(4.,3.5)(7.2,6.7)(0.425,){/Straight}{1}
\FAProp(7.2,6.7)(4.,10.)(0.425,){/Straight}{1}
\FALabel(7.75,4.)[t]{${\scriptscriptstyle c}$}
\FAProp(4.,3.5)(4.,-0.5)(0.,){/Cycles}{0}
\FALabel(11.,4.75)[b]{${\scriptscriptstyle g}$}
\FALabel(4.,10.)[c]{${\scriptscriptstyle \blacksquare}$}
\FALabel(4.,12.07)[b]{$\scriptscriptstyle Q_{\scalebox{0.4}{2}}$}
\FAProp(12.,10.)(7.2,6.7)(-0.4,){/Cycles}{0}
\FALabel(2.,0.75)[b]{${\scriptscriptstyle g}$}
\FAProp(16.,10.)(16.,-0.5)(0.,){/Cycles}{0}
\FALabel(18.5,3.75)[b]{${\scriptscriptstyle g}$}
\FAProp(0.,-0.5)(4.,-0.5)(0.,){/Straight}{1}
\FAProp(4.,-0.5)(16.,-0.5)(0.,){/Straight}{1}
\FAProp(16.,-0.5)(20.,-0.5)(0.,){/Straight}{1}
\FALabel(2.,-3.13)[b]{${\scriptscriptstyle q}$}
\FALabel(10.,-3.13)[b]{${\scriptscriptstyle q}$}
\FALabel(18.,-3.13)[b]{${\scriptscriptstyle q}$}
\FAVert(7.2,6.7){0}
\FAVert(12.,10.){0} 
\FAVert(4.,3.5){0}
\FAVert(16.,10.){0}
\FAVert(16.,-0.5){0}
\FAVert(4.,-0.5){0}

\end{feynartspicture}

%% file: mixing3.tex
\unitlength=1bp%

\begin{feynartspicture}(55,70)(1,1.3)

\FADiagram{}
\FAProp(0.,10.)(2.,10.)(0.,){/Straight}{0}
\FAProp(2.,10.)(10.,10.)(0.,){/Straight}{1}
\FALabel(5.75,11.07)[b]{${\scriptscriptstyle b}$}
\FAProp(20.,10.)(10.,10.)(0.,){/Straight}{-1}
\FALabel(15.,11.07)[b]{${\scriptscriptstyle s}$}
\FAProp(7.15,5.25)(12.85,5.25)(0.6,){/Straight}{1}
\FAProp(10.,10.)(6.8,6.7)(0.45,){/Straight}{1}
\FAProp(6.8,6.7)(7.15,5.25)(0.3,){/Straight}{0}
\FAProp(12.85,5.25)(10.,10.)(0.6,){/Straight}{1}
\FALabel(10.,2.25)[t]{${\scriptscriptstyle c}$}
\FALabel(15.,8.)[t]{${\scriptscriptstyle c}$}
\FALabel(10.,10.)[c]{${\scriptscriptstyle \blacksquare}$}
\FALabel(10.,12.07)[b]{$\scriptscriptstyle Q_{\scalebox{0.4}{2}}$}
\FAProp(2.,10.)(6.8,6.7)(0.4,){/Cycles}{0}
\FALabel(3.,4.5)[b]{${\scriptscriptstyle g}$}
\FAProp(7.15,5.25)(3.5,-0.5)(0.,){/Cycles}{0}
\FALabel(2.85,2.25)[b]{${\scriptscriptstyle g}$}
\FAProp(12.85,5.25)(16.5,-0.5)(0.,){/Cycles}{0}
\FALabel(17.45,2.25)[b]{${\scriptscriptstyle g}$}
\FAProp(0.,-0.5)(3.5,-0.5)(0.,){/Straight}{1}
\FAProp(3.5,-0.5)(16.5,-0.5)(0.,){/Straight}{1}
\FAProp(16.5,-0.5)(20.,-0.5)(0.,){/Straight}{1}
\FALabel(1.5,-3.13)[b]{${\scriptscriptstyle q}$}
\FALabel(10.,-3.13)[b]{${\scriptscriptstyle q}$}
\FALabel(18.25,-3.13)[b]{${\scriptscriptstyle q}$}
\FAVert(6.8,6.7){0}
\FAVert(2.,10.){0}
\FAVert(7.15,5.25){0}
\FAVert(12.85,5.25){0}
\FAVert(3.5,-0.5){0}
\FAVert(16.5,-0.5){0}

\end{feynartspicture}

%% file: mixing4.tex
\unitlength=1bp%

\begin{feynartspicture}(55,70)(1,1.3)

\FADiagram{}
\FAProp(0.,10.)(2.,10.)(0.,){/Straight}{0}
\FAProp(2.,10.)(10.,10.)(0.,){/Straight}{1}
\FALabel(5.75,11.07)[b]{${\scriptscriptstyle b}$}
\FAProp(18.,10.)(10.,10.)(0.,){/Straight}{-1}
\FAProp(20.,10.)(18.,10.)(0.,){/Straight}{0}
\FALabel(14.,11.07)[b]{${\scriptscriptstyle s}$}
\FAProp(10.,10.)(6.8,6.7)(0.425,){/Straight}{1}
\FAProp(6.8,6.7)(10,3.5)(0.425,){/Straight}{1}
\FALabel(14.,4.)[t]{${\scriptscriptstyle c}$}
\FAProp(10.,3.5)(13.2,6.7)(0.425,){/Straight}{1}
\FAProp(13.2,6.7)(10.,10.)(0.425,){/Straight}{1}
\FALabel(5.75,4.)[t]{${\scriptscriptstyle c}$}
\FAProp(10.,3.5)(10.,-1.5)(0.,){/Cycles}{0}
\FALabel(12.5,-0.5)[b]{${\scriptscriptstyle g}$}
\FALabel(10.,10.)[c]{${\scriptscriptstyle \blacksquare}$}
\FALabel(10.,12.07)[b]{$\scriptscriptstyle Q_{\scalebox{0.4}{2}}$}
\FAProp(2.,10.)(6.8,6.7)(0.4,){/Cycles}{0}
\FALabel(3.,4.5)[b]{${\scriptscriptstyle g}$}
\FAProp(18.,10.)(13.2,6.7)(-0.4,){/Cycles}{0}
\FALabel(17.,4.5)[b]{${\scriptscriptstyle g}$}
\FAVert(6.8,6.7){0}
\FAVert(2.,10.){0}
\FAVert(13.2,6.7){0}
\FAVert(18.,10.){0} 
\FAVert(10.,3.5){0}

\end{feynartspicture}

%% file: gammafive1.tex
\unitlength=1bp%

\begin{feynartspicture}(55,70)(1,1.3)

\FADiagram{}
\FAProp(0.,10.)(10.,10.)(0.,){/Straight}{1}
\FALabel(5.,11.07)[b]{${\scriptscriptstyle b}$}
\FAProp(20.,10.)(10.,10.)(0.,){/Straight}{-1}
\FALabel(15.,11.07)[b]{${\scriptscriptstyle s}$}
\FAProp(7.15,5.25)(12.85,5.25)(0.6,){/Straight}{1}
\FAProp(10.,10.)(7.15,5.25)(0.6,){/Straight}{1}
\FAProp(12.85,5.25)(10.,10.)(0.6,){/Straight}{1}
\FALabel(10.,2.25)[t]{${\scriptscriptstyle c}$}
\FALabel(15.,8.)[t]{${\scriptscriptstyle c}$}
\FALabel(4.75,8.)[t]{${\scriptscriptstyle c}$}
\FALabel(10.,10.)[c]{${\scriptscriptstyle \blacksquare}$}
\FALabel(10.,12.07)[b]{$\scriptscriptstyle E^{\scalebox{0.4}{(5)}}_{\scalebox{0.4}{1}}$}
\FAProp(7.15,5.25)(3.5,-0.5)(0.,){/Cycles}{0}
\FALabel(2.85,2.25)[b]{${\scriptscriptstyle g}$}
\FAProp(12.85,5.25)(16.5,-0.5)(0.,){/Cycles}{0}
\FALabel(17.45,2.25)[b]{${\scriptscriptstyle g}$}
\FAProp(0.,-0.5)(3.5,-0.5)(0.,){/Straight}{1}
\FAProp(3.5,-0.5)(16.5,-0.5)(0.,){/Straight}{1}
\FAProp(16.5,-0.5)(20.,-0.5)(0.,){/Straight}{1}
\FALabel(1.5,-3.13)[b]{${\scriptscriptstyle q}$}
\FALabel(10.,-3.13)[b]{${\scriptscriptstyle q}$}
\FALabel(18.25,-3.13)[b]{${\scriptscriptstyle q}$}
\FAVert(7.15,5.25){0}
\FAVert(12.85,5.25){0}
\FAVert(3.5,-0.5){0}
\FAVert(16.5,-0.5){0}

\end{feynartspicture}

%% file: gammafive2.tex
\unitlength=1bp%

\begin{feynartspicture}(55,70)(1,1.3)

\FADiagram{}
\FAProp(0.,10.)(2.,10.)(0.,){/Straight}{0}
\FAProp(2.,10.)(10.,10.)(0.,){/Straight}{1}
\FALabel(5.75,11.07)[b]{${\scriptscriptstyle b}$}
\FAProp(20.,10.)(10.,10.)(0.,){/Straight}{-1}
\FALabel(15.,11.07)[b]{${\scriptscriptstyle s}$}
\FAProp(10.,10.)(6.8,6.7)(0.425,){/Straight}{1}
\FAProp(6.8,6.7)(10,3.5)(0.425,){/Straight}{1}
\FAProp(10.,3.5)(10.,10.)(1.,){/Straight}{1}
\FALabel(15.,8.)[t]{${\scriptscriptstyle c}$}
\FAProp(10.,3.5)(10.,-1.5)(0.,){/Cycles}{0}
\FALabel(12.5,-0.5)[b]{${\scriptscriptstyle g}$}
\FALabel(10.,10.)[c]{${\scriptscriptstyle \blacksquare}$}
\FALabel(10.,12.07)[b]{$\scriptscriptstyle E^{\scalebox{0.4}{(5)}}_{\scalebox{0.4}{1}}$}
\FAVert(10.,3.5){0}
\FAProp(2.,10.)(6.8,6.7)(0.4,){/Cycles}{0}
\FALabel(3.,4.5)[b]{${\scriptscriptstyle g}$}
\FAVert(6.8,6.7){0}
\FAVert(2.,10.){0}

\end{feynartspicture}

%% file: gammafive3.tex
\unitlength=1bp%

\begin{feynartspicture}(55,70)(1,1.3)

\FADiagram{}
\FAProp(0.,10.)(2.,10.)(0.,){/Straight}{0}
\FAProp(2.,10.)(10.,10.)(0.,){/Straight}{1}
\FALabel(5.75,11.07)[b]{${\scriptscriptstyle b}$}
\FAProp(20.,10.)(10.,10.)(0.,){/Straight}{-1}
\FALabel(15.,11.07)[b]{${\scriptscriptstyle s}$}
\FAProp(7.15,5.25)(12.85,5.25)(0.6,){/Straight}{1}
\FAProp(10.,10.)(6.8,6.7)(0.45,){/Straight}{1}
\FAProp(6.8,6.7)(7.15,5.25)(0.3,){/Straight}{0}
\FAProp(12.85,5.25)(10.,10.)(0.6,){/Straight}{1}
\FALabel(10.,2.25)[t]{${\scriptscriptstyle c}$}
\FALabel(15.,8.)[t]{${\scriptscriptstyle c}$}
\FALabel(10.,10.)[c]{${\scriptscriptstyle \blacksquare}$}
\FALabel(10.,12.07)[b]{$\scriptscriptstyle E^{\scalebox{0.4}{(5)}}_{\scalebox{0.4}{1}}$}
\FAProp(2.,10.)(6.8,6.7)(0.4,){/Cycles}{0}
\FALabel(3.,4.5)[b]{${\scriptscriptstyle g}$}
\FAProp(7.15,5.25)(3.5,-0.5)(0.,){/Cycles}{0}
\FALabel(2.85,2.25)[b]{${\scriptscriptstyle g}$}
\FAProp(12.85,5.25)(16.5,-0.5)(0.,){/Cycles}{0}
\FALabel(17.45,2.25)[b]{${\scriptscriptstyle g}$}
\FAVert(6.8,6.7){0}
\FAVert(2.,10.){0}
\FAVert(7.15,5.25){0}
\FAVert(12.85,5.25){0}

\end{feynartspicture}

%% file: gammafive4.tex
\unitlength=1bp%

\begin{feynartspicture}(55,70)(1,1.3)

\FADiagram{}
\FAProp(0.,10.)(2.,10.)(0.,){/Straight}{0}
\FAProp(2.,10.)(10.,10.)(0.,){/Straight}{1}
\FALabel(5.75,11.07)[b]{${\scriptscriptstyle b}$}
\FAProp(20.,10.)(10.,10.)(0.,){/Straight}{-1}
\FALabel(15.,11.07)[b]{${\scriptscriptstyle s}$}
\FAProp(7.15,5.25)(12.85,5.25)(0.6,){/Straight}{0}
\FAProp(10.,10.)(6.8,6.7)(0.45,){/Straight}{1}
\FAProp(6.8,6.7)(7.15,5.25)(0.3,){/Straight}{0}
\FAProp(12.85,5.25)(10.,10.)(0.6,){/Straight}{1}
\FALabel(15.,8.)[t]{${\scriptscriptstyle c}$}
\FALabel(10.,10.)[c]{${\scriptscriptstyle \blacksquare}$}
\FALabel(10.,12.07)[b]{$\scriptscriptstyle E^{\scalebox{0.4}{(5)}}_{\scalebox{0.4}{1}}$}
\FAProp(2.,10.)(6.8,6.7)(0.4,){/Cycles}{0}
\FALabel(3.,4.5)[b]{${\scriptscriptstyle g}$}
\FAProp(7.15,5.25)(3.5,-0.5)(0.,){/Cycles}{0}
\FALabel(2.85,2.25)[b]{${\scriptscriptstyle g}$}
\FAProp(12.85,5.25)(16.5,-0.5)(0.,){/Cycles}{0}
\FALabel(17.45,2.25)[b]{${\scriptscriptstyle g}$}
\FAProp(10.,3.65)(10.,-1.5)(0.,){/Cycles}{0}
\FALabel(12.5,-0.5)[b]{${\scriptscriptstyle g}$}
\FAVert(6.8,6.7){0}
\FAVert(2.,10.){0}
\FAVert(7.15,5.25){0}
\FAVert(12.85,5.25){0}
\FAVert(10.,3.65){0}

\end{feynartspicture}